\documentclass[reprint,prd,showpacs,aps,amsmath,amssymb,nofootinbib,floatfix,floats,
superscriptaddress,twocolumn,eqsecnum]{revtex4-1} 

\pdfoutput=1
\allowdisplaybreaks[2]

\usepackage{times}

\usepackage[usenames, dvipsnames]{color}
\usepackage[colorlinks, pdfborder={0 0 0}, plainpages=false]{hyperref}
\usepackage{breakurl}
\usepackage{amsmath, amssymb, amsfonts}
\usepackage{xspace} % Sensible space treatment at end of simple macros
\usepackage{dcolumn}% Align table columns on decimal point
\usepackage{bm}% bold math
\usepackage{multirow} % Multiple rows in tables
\usepackage{graphics,psfrag}
\usepackage{graphicx,psfrag}
\usepackage{natbib}

\usepackage{xfrac}
\usepackage{wasysym}
\usepackage{tikz}
\usepackage{tabularx}
\usepackage{booktabs}

\usepackage{ulem}

\allowdisplaybreaks	

\def\ben{\begin{equation}}
\def\een{\end{equation}}

   \let\d=\delta 
   
  \let\n=\nu   

\def\be{\begin{equation}}
\def\ee{\end{equation}}
\def\beq{\begin{equation}}
\def\eeq{\end{equation}}
\def\ba{\begin{array}}
\def\ea{\end{array}}

\def\dalemb#1#2{{\vbox{\hrule height .#2pt
       \hbox{\vrule width.#2pt height#1pt \kern#1pt
               \vrule width.#2pt}
       \hrule height.#2pt}}}
\def\square{\mathord{\dalemb{6.8}{7}\hbox{\hskip1pt}}}
\newcommand{\bea}{\begin{eqnarray}}
\newcommand{\eea}{\end{eqnarray}}

\def\({\left(}
\def\){\right)}

\renewcommand{\v}[1]{\ensuremath{\mathbf{#1}}} % for vectors
\newcommand{\gv}[1]{\ensuremath{\mbox{\boldmath$ #1 $}}} 
\newcommand{\uv}[1]{\ensuremath{\mathbf{\hat{#1}}}} % for unit vector
\renewcommand{\d}[2]{\frac{d #1}{d #2}} % for derivatives
\newcommand{\dd}[2]{\frac{d^2 #1}{d #2^2}} % for double derivatives
\newcommand{\pd}[2]{\frac{\partial #1}{\partial #2}} 
\newcommand{\grad}[1]{\gv{\nabla} #1} % for gradient
\let\baraccent=\= % rename builtin command \= to \baraccent
\renewcommand{\=}[1]{\stackrel{#1}{=}} % for putting numbers above =
\newcommand{\Ord}{\mathcal{O}}

\makeatletter
\def\x@multispan#1{%
  \global\let\@tempa\@empty
  \@multicnt#1\relax
  \loop\ifnum\@multicnt>\@ne
  \xdef\@tempa{\@tempa\kern\dimen@i\hfill&\omit}%
   \advance\@multicnt\m@ne
  \repeat
  \@tempa\kern\dimen@i\hfill}

\long\def\xmulticolumn#1#2#3{%
 \omit
 \begingroup
   \def\@addamp{\if@firstamp \@firstampfalse \else
                \@preamerr 5\fi}%
  \@mkpream{#2}\@addtopreamble\@empty
  \endgroup
  \def\@sharp{#3}%
 \setbox\z@\hbox{{\@preamble}}%
\global\dimen@i\wd\z@
\global\divide\dimen@i#1\relax
 \ignorespaces
\x@multispan{#1}}
\makeatother

\AtBeginDocument{
\heavyrulewidth=.08em
\lightrulewidth=.05em
\cmidrulewidth=.03em
\belowrulesep=.65ex
\belowbottomsep=0pt
\aboverulesep=.4ex
\abovetopsep=0pt
\cmidrulesep=\doublerulesep
\cmidrulekern=.5em
\defaultaddspace=.5em
}

\begin{document}
\newcolumntype{L}[1]{>{\raggedright\arraybackslash}p{#1}}
\newcolumntype{C}[1]{>{\centering\arraybackslash}p{#1}}
\newcolumntype{R}[1]{>{\raggedleft\arraybackslash}p{#1}}

\title{Modeling dynamical scalarization with a resummed post-Newtonian expansion}
\author{Noah Sennett}
\email{nsennett@umd.edu}
\affiliation{Department of Physics, University of Maryland, College Park, MD 20742, USA}
\affiliation{Max Planck Institute for Gravitational Physics (Albert Einstein Institute), Am M\"{u}hlenberg 1, Potsdam-Golm 14776, Germany}
\author{Alessandra Buonanno}
\affiliation{Max Planck Institute for Gravitational Physics (Albert Einstein Institute), Am M\"{u}hlenberg 1, Potsdam-Golm 14776, Germany}
\affiliation{Department of Physics, University of Maryland, College Park, MD 20742, USA}
\date{\today}

\begin{abstract}
Despite stringent constraints set by astrophysical observations, there remain viable scalar-tensor theories that could be distinguished from general relativity with gravitational-wave detectors. A promising signal predicted in these alternative theories is dynamical scalarization, which can dramatically affect the evolution of neutron-star binaries near merger. Motivated by the successful treatment of spontaneous scalarization, we develop a formalism that partially resums the post-Newtonian expansion to capture dynamical scalarization in a mathematically consistent manner. We calculate the post-Newtonian order corrections to the equations of motion and scalar mass of a binary system. Through comparison with quasi-equilibrium configuration calculations, we verify that this new approximation scheme can accurately predict the onset and magnitude of dynamical scalarization. 
\end{abstract}
\maketitle

\section{Introduction}

The detection of gravitational-wave (GW) event GW150914 by Advanced LIGO heralds a new era of experimental relativity \cite{detection}. Every test of the past hundred years has indicated that gravity behaves as predicted by general relativity (GR). Until now, the best constraints have come from solar-system experiments \cite{Will2014} and binary-pulsar observations \cite{Wex2014,Berti2015}. These measurements probe the mildly-relativistic, strong-field regime of gravity generated by objects with velocities $v/c\lesssim 10^{-3}$ and gravitational fields $\Phi_\text{Newt}/c^2 \lesssim 10^{-1}$ (see Table 4 of Ref. \cite{Will2014} for a summary of model-independent constraints).

For the first time, these constraints can be extended through the direct observation of strong, dynamical gravitational fields. In particular, GW detectors can track the coalescence of compact objects in binary systems, a process in which the objects are highly-relativistic and strongly self-gravitating, with $v/c \sim0.5$ and $\Phi_\text{Newt}/c^2\sim0.5$. We expect to observe several GWs per year \cite{Aasi2013,Abbott2016} with the upcoming global network of detectors comprised of Advanced LIGO \cite{Abbott2016b}, advanced Virgo \cite{aVIRGO}, and KAGRA \cite{KAGRA}.

These ground-based GW detectors will be able to observe binaries of solar-mass objects for thousands of orbital cycles before merger. Significant effort has gone into the development of techniques to test GR with these measurements; see Refs. \cite{YunesLR,GairLR} and references therein. During the first stage of a binary's coalescence (the early inspiral), the waveform measured by the detector is well described within the stationary-phase approximation. The waveform generated in GR for the early inspiral can be approximated by
\begin{align}
h_\text{GR}(\bm{\theta};f)&=\frac{\mathcal{A}(\bm\theta)}{D}f^{-7/6} e^{i \psi(\bm\theta;f)}, \label{eq:GRwaveform}
\end{align}
where $f$ is the observed frequency, $D$ is the distance to the binary, and $\mathcal{A}$ and $\psi$ are the amplitude and phase of the GW, respectively, dependent on the intrinsic (e.g. chirp mass, component spins, etc.) and extrinsic (e.g. sky position, time of coalescence, etc.) parameters of the binary, represented collectively by $\bm\theta$.

Using this signal as a baseline, one can parameterize any non-GR waveform in the early inspiral as
\begin{align}
h(\bm\theta;f)&=h_\text{GR}(\bm\theta;f)\left(1+\delta \mathcal{A}(\bm\theta,\bm \zeta;f)\right)e^{i \delta \psi(\bm{\theta}, \bm\zeta;f)}\label{eq:ppE}
\end{align}
where $\bm\zeta$ represents the parameters that characterize the alternative theory \cite{Arun2006b,Yunes2009}. Then, given a GW detection, Bayesian inference can be used to estimate $\delta \mathcal{A}$ and $\delta \psi$ \cite{DelPozzo2011, Cornish2011,Li2012a, Li2012b,Sampson2013b,Sampson2013}.  Typically one expands these functions in powers of the frequency $f$ (and its logarithm $\log f$), then performs a hypothesis test to constrain the corresponding expansion coefficients. This approach can be used either to search for generic deviations from GR by treating these coefficients independently or to test a specific alternative theory against GR by relating the coefficients to the underlying physical parameters $\bm\zeta$. Using a parameterized waveform that also included the merger and ringdown signal, both types of tests were done for GW150914 in Ref. \cite{testingGR}: with the former, the authors constrained the higher-order expansion coefficients in $\delta \psi$, and with the latter, they placed a lower bound on the Compton wavelength $\lambda_g$ of the graviton in a hypothetical massive gravity theory \cite{Will1997} ($\lambda_g$ is signified by $\bm\zeta$ in our notation).

However, this type of analysis rests on the assumption that $\delta \mathcal{A}$ and $\delta \psi$ admit expansions in powers of $f$. There exist certain alternative theories of gravity where this assumption of analyticity breaks down due to phase transitions or resonant effects \cite{Sampson2013}. Fortunately, several complementary tests were performed in Ref. \cite{testingGR} to verify that GW150914 is indeed consistent with GR. Still, our ability to model non-analytic features in waveforms is essential in case future events do not match the predictions of GR as closely.

The task of modeling a non-analytic deviation $\delta\psi$ in a generic, theory-independent way is intractable. Instead, previous work has focused on modeling specific non-GR phenomena predicted in particular alternative theories of gravity. We continue this effort here, focusing on dynamical scalarization (DS), an effect that can arise in neutron-star binaries in certain scalar-tensor (ST) theories of gravity \cite{Barausse2013,Shibata2014}. Previous efforts to model this effect have simply grafted models of DS onto independently developed analytic approximations of the inspiral \cite{Sampson2013,Palenzuela2014,Sampson2014a}. In this paper, we propose a new perturbative formalism that incorporates DS from first principles. Our aim is to lay the groundwork for a model whose accuracy can be improved iteratively in a way that is more straightforward and self-consistent than previous methods.

The paper is organized as follows. In Sec. \ref{sec:STPN}, we examine the relationship between DS and the better understood phenomenon of spontaneous scalarization. From this discussion, we motivate a resummation of the post-Newtonian formalism to incorporate DS, which is then developed in Sec. \ref{sec:MRPNformalism}. We derive the equations of motion for a neutron-star binary up to next-to-leading order in Secs. \ref{sec:NZfields} and \ref{sec:EOM}. In Secs. \ref{sec:FZfields} and \ref{sec:ScalarMass}, we calculate its scalar mass (a measure of the system's scalarization) at the same order. As a test of its validity, in Sec. \ref{sec:Discussion}, we compare our model with numerical quasi-equilibrium configurations of neutron stars \cite{Taniguchi2014} and previous analytical models \cite{Palenzuela2014}. We provide a summary in Sec. \ref{sec:Conclusions} and outline the future work needed to produce waveforms with our model.

\section{Non-perturbative phenomena in scalar-tensor gravity}\label{sec:STPN}

Scalar-tensor theories of gravity are amongst the most natural and well-motivated alternatives to GR. We consider the class of theories detailed in Ref. \cite{Damour1992}, in which a massless scalar field couples non-minimally to the metric, effectively allowing a spin-0 polarization of the graviton. These theories are described by the action 
\begin{align}
S&=\int d^4 x \frac{c^3\sqrt{-g}}{16 \pi G}\left[\phi R-\frac{\omega(\phi)}{\phi} g^{\mu \nu} \nabla_\mu \phi \nabla_\nu \phi\right]+S_m[g_{\mu \nu},\Xi],\label{eq:Action}
\end{align}
where $\Xi$ represents all of the matter degrees of freedom in the theory. Note that in the limit that $\omega\rightarrow\infty$, the scalar field relaxes to a constant value, and the theory reduces to GR with the modified gravitational constant $G_{\text{eff}}=G/\phi$; we refer to this extreme as the GR limit.

The form of the action in Eq. (\ref{eq:Action}) is known as the ``Jordan frame'' action. Alternatively, the action can be cast into the ``Einstein frame'' by performing a conformal transformation $\tilde{g}_{\mu \nu}\equiv \phi g_{\mu \nu}$ as
\begin{align}
S=&\int d^4 x \frac{c^3\sqrt{-\tilde{g}}}{16 \pi G}\left[\tilde{R}-2 \tilde{g}^{\mu \nu} \nabla_\mu \tilde \varphi \nabla_\nu \tilde\varphi\right]\nonumber \\
&+S_m \left[e^{-\int 2d\tilde\varphi/\sqrt{3+2\omega(\tilde\varphi)}}\tilde{g}_{\mu \nu},\Xi\right],\label{eq:EinsteinS}
\end{align}
where we have introduced the scalar field
\begin{align}
\tilde\varphi\equiv \int d\phi\frac{\sqrt{3+2\omega(\phi)}}{2\phi}.
\end{align}
From Eq. (\ref{eq:EinsteinS}), we see that the coupling of the scalar field to matter (through the metric $\tilde{g}_{\mu\nu}$) is characterized by 
\begin{align}
a=(3+2\omega)^{-1/2}.
\end{align}

Measurable phenomena absent in GR arise in theories whose coupling is linear in $\tilde\varphi$
\begin{align}
a=\frac{B\tilde\varphi}{2}.
\end{align}
This coupling can be expressed in terms of Jordan frame variables as
\begin{align}
\frac{1}{\omega(\phi)+3/2}&=B \log \phi\label{eq:coupling},
\end{align}
and imposes the relation between $\phi$ and $\tilde\varphi$
\begin{align}
\phi=\exp(B \tilde{\varphi}^2/2).\label{eq:PhiVarphi}
\end{align}

Damour and Esposito-Far\`{e}se discovered an instability in the scalar field triggered by the presence of relativistic matter in theories with $B>0$ \cite{Esposito-Farese1993}.\footnote{Based off the work of Refs. \cite{Damour1993,Damour1993b,Damour1996}, the authors of Ref. \cite{Sampson2014a} recently discussed a related instability in this theory that would cause the scalar field to grow rapidly over cosmological timescales throughout the Universe. Consequently, the scalar field today would be so large that its presence would have already been detected by solar-system experiments. The addition of a potential $V(\phi)$ or slight modification of $\omega(\phi)$ could ameliorate this issue while preserving the neutron star phenomena discussed in this paper (for example, see Ref. \cite{Ramazanoglu2016}). As is done in the literature, we ignore here this cosmological problem.} For sufficiently large $B$, compact neutron stars were found to undergo a phase transition now known as spontaneous scalarization. Spontaneously scalarized stars are expected to behave differently than their (un-scalarized) GR counterparts (see Refs. \cite{DeDeo2003,Sotani2004,Damour1998,DeDeo2004,Doneva2014} for examples of such deviations).

Observation of a scalarized star would be a smoking gun for modifying gravity; in turn, our lack of evidence for such stars places constraints on this class of ST theories \cite{Freire2012}. Because scalarization arises from the non-linear interaction between strong gravitational fields and matter, it is unconstrained by weak-field experiments and GW150914. However, pulsar timing measurements have ruled out nearly all theories that can sustain spontaneous scalarization \cite{Freire2012}.

Dynamical scalarization is a similar phenomenon revealed by recent numerical-relativity simulations that is not ruled out by binary-pulsar observations  \cite{Barausse2013,Shibata2014}. In a binary system, neutron stars too diffuse to spontaneously scalarize in isolation were found to scalarize collectively. Despite the name, DS has also been found in recent quasi-equilibrium calculations \cite{Taniguchi2014}; the phenomenon is caused by the proximity of the neutron stars rather than their dynamical evolution. The onset of DS produces an abrupt change in the stars' motion, generating sharp features in the GW signal produced by the binary.

Gravitational-wave detectors may be able to extend the current constraints on ST theories by searching for DS \cite{Sampson2014a}. This endeavor hinges on our ability to accurately and efficiently model GWs from binaries that undergo DS.  Such effects have been modeled by using a Heaviside function for $\delta \psi$ in Eq. (\ref{eq:ppE}) \cite{Sampson2013, Sampson2014a} or augmenting the post-Newtonian (PN) evolution of the binary with a semi-analytic feedback model \cite{Palenzuela2014}. This work follows a general strategy similar to that of Ref. \cite{Palenzuela2014}. However, we adopt a top-down approach to incorporate DS into the PN formalism in hopes of creating a model that is more consistent and streamlined conceptually (see Appendix \ref{sec:Palenzuela} for a detailed analysis of the results of Ref. \cite{Palenzuela2014}).

The PN expansion is an effective tool for analytically approximating the evolution of binary systems of interest to ground-based GW detectors. In this approach, one expands solutions to the Einstein equations about flat space in the small parameter $\epsilon\sim  G M/r c^2\sim(v/c)^2<1$, where $M, r, v$ represent the characteristic mass, distance, and velocity scales in the problem, respectively. In ST theories, this expansion is done about the Minkowski metric $\eta_{\mu \nu}$ and background field $\phi_0$ (assumed to be constant and homogeneous over the time and distance scales of the evolution of a binary system). We refer to the $\epsilon^{n+1}$ corrections to these quantities as the ``$n$PN'' fields.
We define \textit{non-perturbative phenomena} as behavior found in the full gravitational theory that cannot be recovered at any finite PN order. 

In the remainder of this section, we argue that DS is a non-perturbative phenomenon. First, we review the analytic treatment of spontaneous scalarization, describing the way in which the phenomenon has been identified as non-perturbative and then incorporated into the PN expansion in a rigorous manner. We then perform a similar analysis for DS and present a quantitative argument that the phenomenon is non-perturbative. Finally, we describe how the analytic treatment of spontaneous scalarization could be adapted to incorporate DS into the PN formalism.

\subsection{Spontaneous scalarization: single neutron star}

In ST theories, static, spherically symmetric spacetimes are characterized by three parameters: the asymptotic field $\phi_0$, the Arnowitt-Deser-Misner (ADM) mass $m$, and the scalar charge $\alpha$ \cite{Damour1992,Just}. These parameters can be extracted from the asymptotic behavior of the metric and scalar field
\begin{align}
\lim_{|\v{x}|\rightarrow \infty}g_{ij}&=\left(1+\frac{2 G m}{|\v{x}| c^2}\right)\delta_{ij} +\Ord\left(|\v{x}|^{-2}\right),\label{eq:Isolated1}\\
\lim_{|\v{x}|\rightarrow \infty}\phi&=\phi_0+\frac{2 G \mu_0 m \alpha}{|\v{x}| c^2}+\Ord\left(|\v{x}|^{-2}\right),\label{eq:Isolated2}
\end{align}
where we have defined
\begin{align}
\mu_0\equiv\frac{1}{\sqrt{3+2\omega(\phi_0)}}=\sqrt{\frac{B \log \phi_0}{2}}.
\end{align}

\begin{figure}
\includegraphics[width=\columnwidth,clip=true, trim=0 5 1 15]{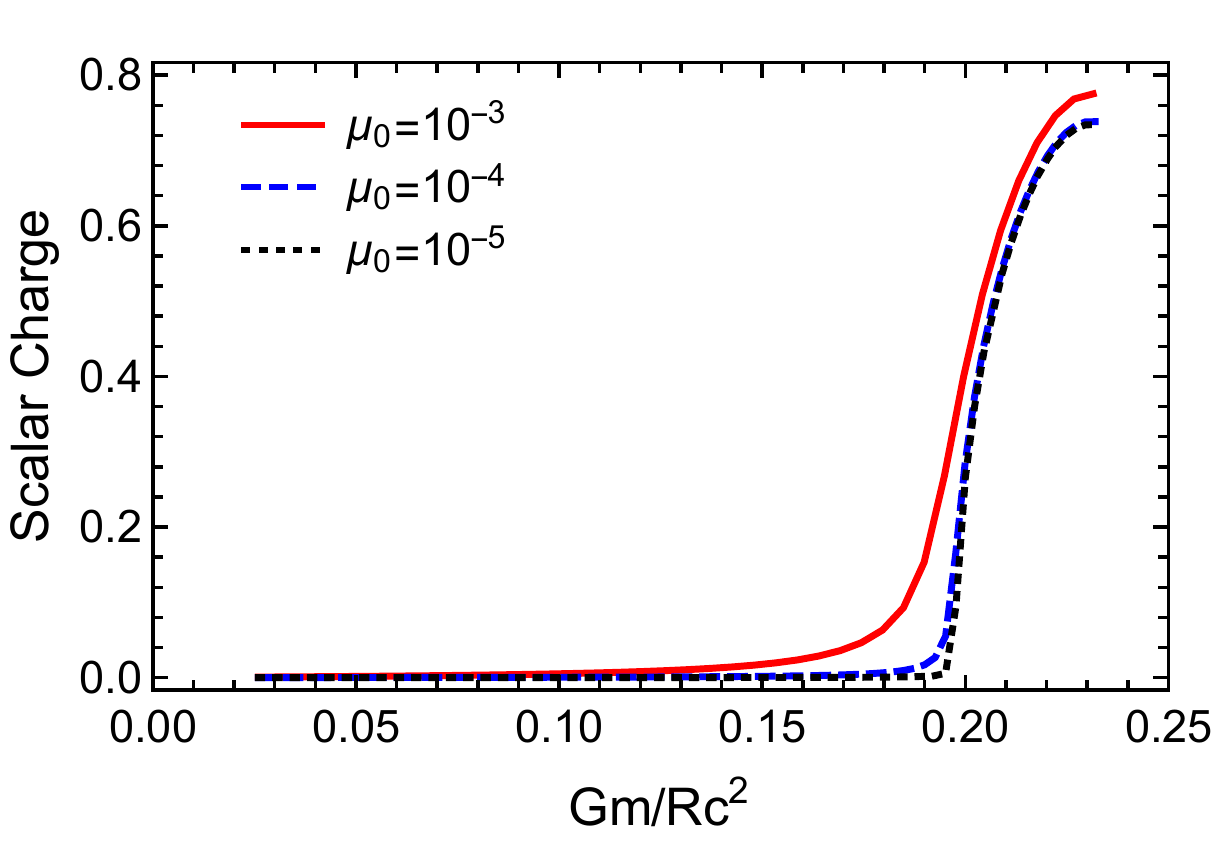}
\caption{The scalar charge of an isolated non-spinning neutron star as a function of its compactness in the limit that $\mu_0$ approaches zero (the GR limit). We use the theory parameter $B=9$ with a piecewise polytropic fit to the APR4 equation of state detailed in Ref. \cite{Read2009}.}\label{fig:Spontaneous}
\end{figure}

It was shown in Ref. \cite{Damour1992} that the scalar charge of an isolated star can be written in the PN expansion as
\begin{align}
&\alpha=\mu_0\left[1+A_1 \left(\frac{G m}{R c^2}\right)+A_2 \left(\frac{G m}{R c^2}\right)^2+\cdots\right],  \label{eq:SpontaneousScalarization}
\end{align}
where $R$ is the radius of the body, and the coefficients $A_i$ are of order unity. Because $\mu_0$ vanishes in the GR limit, one finds that the right hand side of Eq. (\ref{eq:SpontaneousScalarization}), truncated at any finite order, must vanish as well. However, as first discovered in Ref. \cite{Esposito-Farese1993}, exactly solving the geometry numerically shows that a sufficiently compact body can sustain an appreciable scalar charge even when $\mu_0=0$ (corresponding to the GR limit $\phi_0=1$).\footnote{Because of the additional prefactor of $\mu_0$, the $|\v{x}|^{-1}$ term in Eq. (\ref{eq:Isolated2}) vanishes even for spontaneously scalarized stars in the GR limit. The dramatic effect of spontaneous scalarization is more easily seen through $\tilde\varphi$ [given in Eq. (\ref{eq:PhiVarphi})], which can be approximated as $\tilde\varphi=\tilde{\varphi}_0+\frac{G m \alpha}{|\v{x}| c^2}+\Ord\left(|\v{x}|^{-2}\right),$ where $\phi_0=\phi(\tilde{\varphi}_0)$.}  Thus, we would describe this scalarization as non-perturbative (in the sense defined above). Figure \ref{fig:Spontaneous} depicts the sharp growth in scalar charge in the limit $\mu_0\rightarrow0$ as one increases the compactness of a neutron star. For this figure and all that follow, we use a piecewise polytropic fit \cite{Read2009} to the APR4 equation of state given in Ref. \cite{Akmal1998}.

The tension between the analytic and numerical results suggests that the PN expansion must break down beyond some compactness $G m/R c^2$ for this class of ST theories. The scalar charge is non-analytic at this critical compactness, at which point the isolated body undergoes a phase transition. Analogous to ferromagnetism, the derivative of the charge diverges when $\mu_0$ approaches zero, indicating that this transition is of second order. Beyond the critical point, the vanishing of $\mu_0$ in the GR limit is compensated by the divergence of the bracketed sum in Eq. (\ref{eq:SpontaneousScalarization}). The only astrophysical objects that could reach this critical compactness are neutron stars and black holes. However, no-hair theorems protect isolated black holes from developing a scalar charge \cite{Hawking1972}. We focus exclusively on neutron stars for the remainder of this paper.

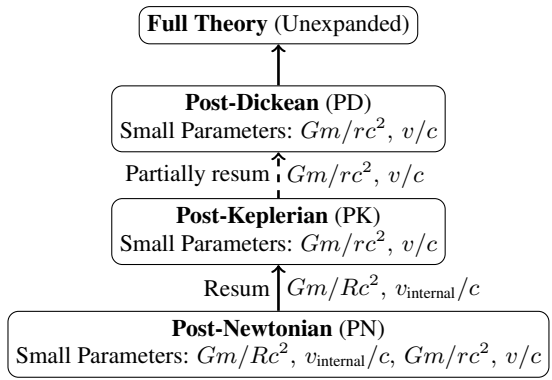
\begin{figure}
\begin{tikzpicture}[rec/.style={draw, rectangle, rounded corners,align=center,minimum width=3cm}]
\node (Full) at (0,0) [rec,below] {\textbf{Full Theory} (Unexpanded)};
\node (MRPN) at (0,-1.5) [rec] {\textbf{Post-Dickean} (PD)\\Small Parameters: $Gm/rc^2,\,v/c$};
\node (PK) at (0,-3) [rec] {\textbf{Post-Keplerian} (PK)\\Small Parameters: $Gm/rc^2,\,v/c$};
\node (PN) at (0,-4.5) [rec] {\textbf{Post-Newtonian} (PN)\\Small Parameters: $Gm/Rc^2,\,v_{\text{internal}}/c,\,Gm/rc^2,\,v/c$};
\draw[->,line width=1pt] (PN) to (PK);
\draw[->,dashed,line width=1pt] (PK) to (MRPN);
\draw[->,line width=1 pt] (MRPN) to (Full);
\node at (0,-3.75) [left] {Resum};
\node at (0,-3.75) [right] {$Gm/Rc^2,\,v_{\text{internal}}/c$};
\node at (0,-2.25) [left] {Partially resum};
\node at (0,-2.25) [right] {$Gm/rc^2,\,v/c$};
\end{tikzpicture}
\caption{Analytic approximations of ST theories. Starting from the PN expansion about the Minkowski metric $\eta_{\mu \n}$ and the background field $\phi_0$, one resums all expansions in the compactness $Gm/Rc^2$ and associated internal velocity $v_\text{Internal}/c$ to capture spontaneous scalarization. Recombining these expansions produces the PK approximation. To capture dynamical scalarization, one resums the PK expansion in $Gm/rc^2$ and $v/c$. Fully recombining these expansions reproduces the original (unexpanded) theory. Instead, one partially resums the PK expansions to generate the PD approximation.}\label{fig:Approximations}
\end{figure}

In anticipation of our discussion of dynamical scalarization, we briefly review how spontaneous scalarization is incorporated into analytic models of binary pulsars. A binary system of non-spinning stars is characterized by two length scales: the characteristic size of the bodies $R$ and their separation $r$. As in the case of an isolated body, the individual stars can spontaneously scalarize if they exceed some critical compactness, at which point the PN expansion no longer accurately predicts the evolution of the binary. Damour and Esposito-Far\`{e}se developed the ``post-Keplerian'' (PK) expansion to accommodate such systems \cite{Damour1992,Damour1996b} (not to be confused with the ``parameterized post-Keplerian'' formalism for modeling binary pulsars in generic alternative theories \cite{Damour1992a}). In the PK approach, one expands only in $G m/r c^2$, leaving quantities dependent on $R$ unexpanded (e.g the scalar charge $\alpha$). Equivalently, one can recombine the sum in powers of $Gm/Rc^2$ in the PN expansion to produce the PK expansion. The relationship between the PN and PK expansions is summarized in the bottom two panels of Fig. \ref{fig:Approximations} (the remaining panels are discussed in Sec. \ref{sec:DS}).  Spontaneous scalarization is captured by explicitly including all of the terms in Eq. (\ref{eq:SpontaneousScalarization}) at each order in the PK expansion.

\subsection{Dynamical scalarization: neutron-star binaries}\label{sec:DS}

Despite its successful application to binary pulsars, the PK approximation does not predict dynamical scalarization. The asymptotic scalar field for a binary system has been computed recently to 1.5PK order in Ref. \cite{Lang2014a}.\footnote{In the literature, the distinction between the PN and PK expansions is often overlooked; the PK expansion (i.e., the approximation in which power series in $Gm/Rc^2$ have been resummed) is often referred to as the ``PN expansion,'' (for example Refs. \cite{Will2014,Mirshekari2013,Lang2014a}). To avoid confusion, we have taken care to distinguish the two in Sec. \ref{sec:STPN} when discussing spontaneous scalarization. Because both the PN and PK expansions fail to capture dynamical scalarization, starting from Sec. \ref{sec:MRPNformalism}, we continue the popular conflation of these two approximation schemes, referring to the expansions collectively as ``PN.''} For a system containing neutron stars too diffuse to spontaneously scalarize individually, the PK prediction of the total scalar charge remains small as the binary coalesces. However, numerical-relativity calculations indicate that the scalar charge can greatly increase beyond this estimate as the two neutron stars draw close \cite{Barausse2013,Shibata2014,Taniguchi2014}. We postpone a quantitative comparison between these analytic and numerical predictions until Sec. \ref{sec:Discussion} (see Fig. \ref{fig:ScalarMassPlot1}); we must first formulate a precise measure of the scalarization of a binary system. Akin to spontaneous scalarization, we suspect that the mismatch between analytic and numerical results stems from a breakdown of the PK expansion. We posit that DS is a non-perturbative phenomenon, and hence the PK expansion needs to be suitably modified to capture it.

To support this intuition, we carefully examine how the mass and scalar charge of a star depend on the nearby scalar field. For an isolated body, these are the relations $m(\phi_0)$ and $\alpha(\phi_0)$ where $\phi_0,\, m,$ and $\alpha$ are defined in Eqs. (\ref{eq:Isolated1}) and (\ref{eq:Isolated2}). As shown in Appendix A of Ref. \cite{Damour1992}, the scalar charge is related to the mass by
\begin{align}
\alpha_A(\phi_0)&=\mu_0\left(1-2\d{\log m_A}{\log \phi_0}\right).
\end{align}
The dependence of the mass on $\phi_0$ can only be found by numerically solving the Tolman-Oppenheimer-Volkoff (TOV) equations modified for ST gravity with a given equation of state \cite{Esposito-Farese1993}.

The mass and scalar charge of each neutron star in a binary system can be similarly determined provided that the system is well-separated ($R/r\ll1$). Working at leading order in $R/r$, each star can be treated as an isolated body immersed in the scalar field produced by its partner \cite{Damour1992}. At a distance $|\v{x}|=d\sim \sqrt{R r}$ from each star (``far'' from the star relative to $R$), the metric and scalar field will behave as in Eqs. (\ref{eq:Isolated1}) and (\ref{eq:Isolated2}) with $\phi_0$ replaced by the background field produced by the other star. As above, we numerically solve the modified TOV equations to relate the mass $m$ and scalar charge $\alpha$ to this background scalar field. Because we work in the limit $d/r=\sqrt{R/r} \rightarrow 0$, this matching occurs effectively at each star relative to $r$, the smallest distance scale relevant to GW generation.\footnote{In this paper, we ignore all effects that arise from the finite size of the neutron stars. Such effects could influence the dynamics of a binary system of scalarized stars at 1PK order \cite{Esposito-Farese2011}.} In this limit, the TOV equations provide us with the dependence on the mass and charge on the \emph{local} scalar field for each body in a binary system, i.e. the functions $m(\phi)$ and $\alpha(\phi)$ where $\phi$ is evaluated \emph{at} the star.

To analytically model these relations using the PK approximation, one must expand $m$ and $\alpha$ about the background field $\phi_0$, where now $\phi_0$ is the value taken very far from the binary system at $|\v{x}|\gg r$. Because the analytic form of the function $m(\phi)$ is unknown, Eardley \cite{Eardley1975} proposed the agnostic expansion
\begin{align}
\begin{split}
m_A(\phi)&=m_A^{(0)}\left[1+s_A \Psi+\frac{1}{2}\left(s_A^2-s_A+s_A'\right)\Psi^2+\cdots\right],\label{eq:Mexpand}
\end{split}
\end{align}
where
\begin{align}
m_A^{(0)}&\equiv m_A(\phi_0),\\
s_A&\equiv \left(\d{\log m_A}{\log \phi}\right)_{\phi=\phi_0} ,\\
s'_A&\equiv \left(\dd{\log m_A}{(\log \phi)}\right)_{\phi=\phi_0} ,\\
\Psi&\equiv\frac{\phi-\phi_0}{\phi_0}\propto \frac{G m}{r c^2}. \label{eq:PsiDef}
\end{align}

\begin{figure}
\includegraphics[width=\columnwidth,clip=true, trim=0 0 0 0]{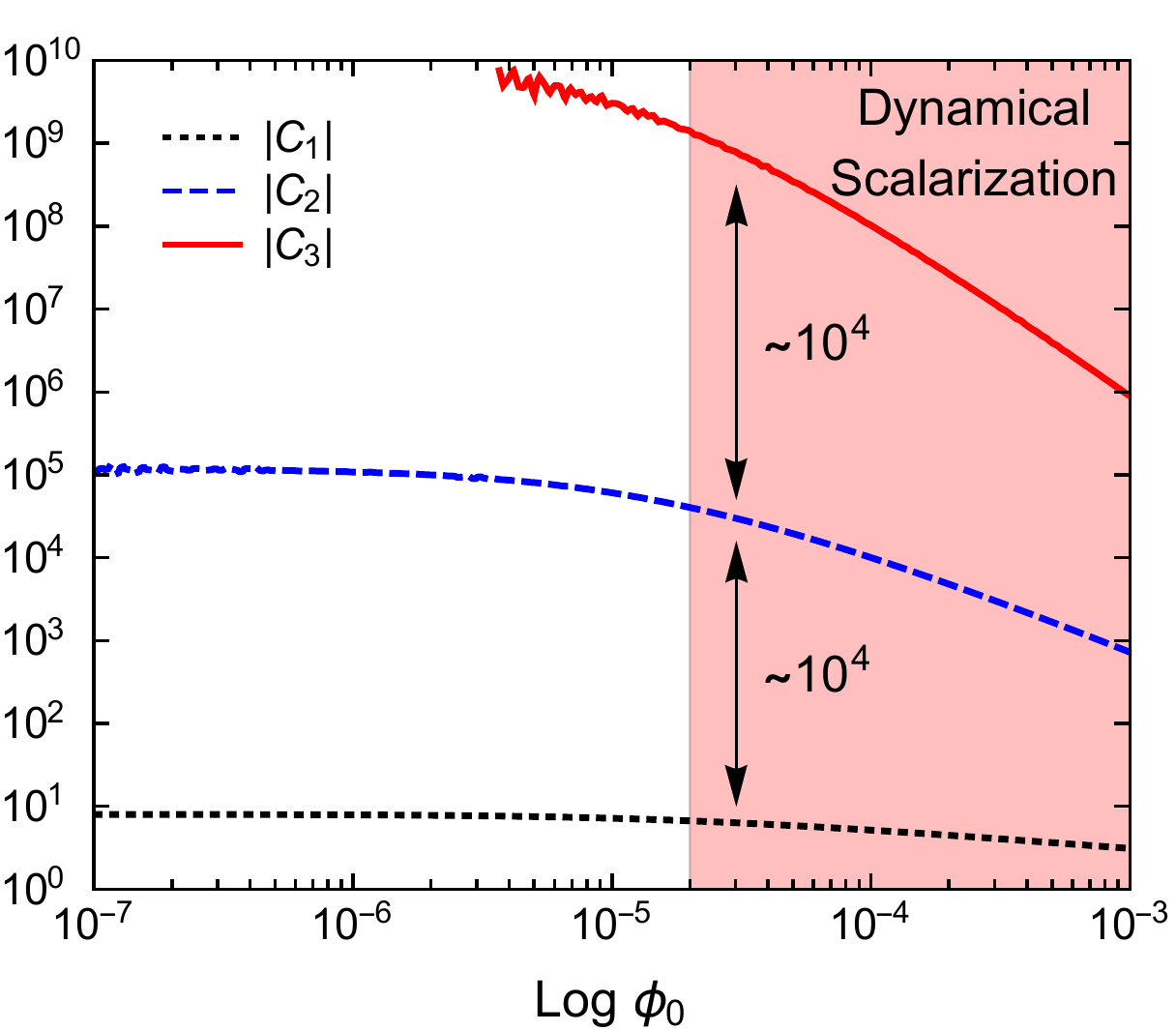}
\caption{Magnitude of the coefficients of the expansion $m(\phi)=m^{(0)}\left(1+C_1 \Psi +C_2 \Psi^2+\cdots\right)$ across the typical scalar field values achieved during the evolution of a compact binary. Values shown here are for an isolated body with $m(\phi_0)=1.35 M_\odot$ and APR4 equation of state with $B=9$. Interpolation errors dominate the computation of $C_3$ for small values of $\log \phi_0$; we omit these regions of the curve.}\label{fig:Sensitivities}
\end{figure}

We plot the magnitude of the coefficients in Eq. (\ref{eq:Mexpand}) in Fig. \ref{fig:Sensitivities} across a range of scalar field values reached during the coalescence of a binary neutron star system \cite{Barausse2013,Shibata2014}. Using the model of Ref. \cite{Palenzuela2014}, we estimate that DS occurs when the field at each body reaches a value of 
\begin{align}
\Psi \sim 10^{-4},\label{eq:PsiMagnitude}
\end{align}
depicted as the pink region in the figure.

For neutron stars with realistic, piecewise polytropic equations of states (e.g. fits to APR4 and H4 defined in Ref. \cite{Read2009}), we find that for $\Psi$ near this maximal value,
\begin{align}
\bigg|\frac{C_{n+1}}{C_n}\bigg|\sim 10^3 - 10^5,\label{eq:Sgrowth}
\end{align}
for $n=1,2$, where $C_i$ is the coefficient of the $i$-th PN correction in Eq. (\ref{eq:Mexpand}).

Comparing Eqs. (\ref{eq:PsiMagnitude}) and (\ref{eq:Sgrowth}), we see that the rapid growth of the expansion coefficients in Eq. (\ref{eq:Mexpand}) can overpower the ``smallness'' of our expansion parameter $\Psi$. In particular, the relative contribution of each term on the right hand side of Eq. (\ref{eq:Mexpand}) does not diminish as one moves to increasingly higher order. These symptoms indicate that $m(\phi)$ may not be analytic in this regime, and thus, the PK expansion would break down at this point in the binary's evolution. Inspired by the treatment of spontaneous scalarization, we posit that the best way to work around this restriction is to resum the expansion in Eq. (\ref{eq:Mexpand}). The hierarchy of these expansions is outlined in Fig. \ref{fig:Approximations}.

Unfortunately, such a prescription is not as straightforward as the case for spontaneous scalarization. To capture spontaneous scalarization, one simply ``unexpands'' all expansions in $Gm/Rc^2$ in the PN approximation [i.e. those of the form as in Eq. (\ref{eq:SpontaneousScalarization})], leaving only expansions in $Gm/rc^2$ and the corresponding orbital velocity $v/c$. Completely resumming \textit{these} expansions would reproduce the full ST theory. Instead, we need to choose certain quantities dependent on $Gm/rc^2$ to resum, and leave the rest expanded. Based on the discussion above, we suspect that the best quantities to keep unexpanded are the mass $m(\phi)$ and its derivatives (including the scalar charge $\alpha(\phi)$). However, \emph{a priori}, there is no clear indication of precisely ``what to resum.'' We need to incorporate the flexibility of this choice into our model.

\section{The post-Dickean expansion}\label{sec:MRPNformalism}

\subsection{Action and field equations}

We refer to our method of resumming the PN expansion as the ``post-Dickean'' (PD) approach --- named after Robert Dicke, one of several pioneers of ST gravity \cite{Scherrer,Thiry,Jordan,Fierz,Brans} who made many important contributions to experimental relativity throughout his career. For notational convenience, we introduce an auxiliary field $\xi$ that is related to $\phi$ in a small neighborhood of each particle's worldline.\footnote{Formally, the matching of $\xi$ and $\phi$ is done at the boundary of the \textit{body zone}, defined at a distance $d\sim \sqrt{Rr}$ from each body. As justified in Appendix A of Ref. \cite{Damour1992}, in the limit that $d/r\rightarrow 0$, we can represent each body as a point particle; in this limit, the matching of the two field variables is done exactly on each body's worldline.} This new field is used to demarcate the resummed variables ($m$ and its derivatives). We explicitly constrain $\xi$ in the matter action via the Lagrange multipliers $\lambda_A$
\begin{align}
\begin{split}
S_m&\equiv c^2\sum_A \int d^4 x\int d \tau_A \delta^{(4)}\left(x-\gamma_A(\tau_A)\right)\\
&\times\left(m_A(\phi,\xi)+\lambda_A(\tau_A)\left(F(\phi)-\xi\right)\right),
\end{split}
\end{align}
where the arbitrary functions $m(\phi,\xi)$ and $F(\phi)$ encode our choice of how to resum the mass and scalar charge, respectively.

With this expression, the action in Eq. (\ref{eq:Action}) gives rise to the field equations
\begin{align}
F \left(\phi(\gamma_A(\tau_A))\right)&=\xi(\gamma_A(\tau_A))\label{eq:JordanConstraint},\\
u_A^\sigma\nabla_\sigma\left( m(\phi,\xi)u_A^\alpha\right)&=-\frac{D m}{D \phi}\partial^\alpha \phi\label{eq:JordanEOM}
\end{align}
\begin{align}
\begin{split}
R_{\mu \nu}-\frac{1}{2}R g_{\mu\nu}=&\frac{\omega(\phi)}{\phi^2} \left(\nabla_\mu\phi \nabla_\nu \phi -\frac{1}{2}g_{\mu \nu} g^{\alpha \beta}\nabla_\alpha \phi \nabla_\beta \phi\right)\\
&+\frac{1}{\phi}\left(\nabla_\mu \nabla_\nu \phi-g_{\mu \nu} \square \phi\right)+\frac{8 \pi G}{\phi c^4} T_{\mu \nu},\label{eq:JordanMetric}
\end{split}
\end{align}
\begin{align}
\begin{split}
\square \phi=&\frac{1}{3+2\omega(\phi)}\left(\frac{8 \pi G}{c^4} T-\frac{16 \pi G}{c^4}\phi \frac{D T}{D \phi}\right.\\
&\left.-\d{\omega}{\phi}g^{\alpha \beta}\nabla_\alpha \phi \nabla_\beta \phi\right),\label{eq:JordanScalar}
\end{split}
\end{align}
where we have defined
\begin{align}
\frac{D}{D\phi}&\equiv \pd{}{\phi}+\d{F}{\phi}\pd{}{\xi},\\
\begin{split}\label{eq:StressEnergy}
T^{\mu \nu}&\equiv\frac{2c}{\sqrt{-g}}\frac{\delta S_m}{\delta g_{\mu \nu}}\\
&=c^3\sqrt{-g}\sum_A\int d \tau_A m_A(\phi,\xi) u_A^\mu u_A^\nu \delta^{(4)}(x-\gamma_A(\tau_A)),
\end{split}
\end{align}
where $\gamma_A$, $\tau_A$, and $u_A^\mu=\d{\gamma_A^\mu}{\tau_A}$ are the worldline, proper time, and four velocity of particle $A$, respectively.

In this paper, we focus on only a few, natural choices for $m$ and $F$, given in Table \ref{table:ResummationSchemes}. Physically, the choice of $m^{(\text{RJ})}$ corresponds to resumming the mass measured in the Jordan frame, while the choice and $m^{(\text{RE})}$ corresponds to resumming the Einstein-frame mass
\begin{align}
m^{(E)}(\phi)=\frac{m(\phi)}{\sqrt{\phi}}.
\end{align}
The choices of $F^{(\phi)}$ and $F^{(\tilde\varphi)}$ respectively equate the auxiliary field $\xi$ to $\phi$ and $\tilde\varphi$, defined in Eq. (\ref{eq:PhiVarphi}). We refer to the joint selection of $m$ and $F$ as the \textit{resummation scheme}. The PD parameterization also encompasses the (non-resummed) PN expansion; this limit is reached with the choice of $m^{(\text{PN})}$ given in the table (recall that here ``PN'' is used to refer collectively to the post-Newtonian and -Keplerian approximations).

\begin{table}[t]\centering
\caption{Resummation schemes discussed in this paper. We abbreviate $m(\phi,\xi)$ with $m$ and $F(\phi)$ with $F$.}\label{table:ResummationSchemes}
\begin{ruledtabular}
\begin{tabular}{ C{.15\columnwidth} C{.4\columnwidth} C{.4\columnwidth}}
\multicolumn{1}{r}{}
& \multicolumn{1}{c}{$F^{(\phi)}$}
& \multicolumn{1}{c}{$F^{(\tilde\varphi)}$}\\
\cmidrule{2-3}
\multirow{2}{*}{$m^{(\text{RJ})}$} & $m=m(\xi)$ & $m=m(\xi)$\\
 & $F=\phi$ &$F=\sqrt{2\log \phi/B}$ \vspace{1em}\\
\multirow{2}{*}{$m^{(\text{RE})}$} &$m=(\phi/\xi)^{1/2}m(\xi)$  & $m=\phi^{1/2}e^{-B\xi^2/4}m(\xi)$ \\
 & $F=\phi$ & $F=\sqrt{2\log \phi/B}$\\
\cmidrule{1-3}
\multirow{2}{*}{$m^{(\text{PN})}$}&\multicolumn{2}{c}{\multirow{2}{*}{$m=m(\phi)$}}\\
& \multicolumn{2}{c}{}\\
\end{tabular}
\end{ruledtabular}
\end{table}

\subsection{Relaxed field equations}
To solve Eqs. (\ref{eq:JordanMetric}) and (\ref{eq:JordanScalar}), we employ a technique known as direction integration of the relaxed Einstein equations, originally developed in GR in Refs. \cite{Epstein1975,Wagoner1976,Will1996,Pati2000,Pati2002} and then extended to ST gravity in Refs. \cite{Mirshekari2013,Lang2014a,Lang2014b}. The remainder of this section closely follows the framework presented in Sec. II.B of Ref. \cite{Mirshekari2013}. We define
\begin{align}
\mathfrak{g}^{\mu \nu}&\equiv\sqrt{-g}g^{\mu \nu}\label{eq:Gothicg},\\
H^{\mu \alpha \nu \beta}&\equiv \mathfrak{g}^{\mu \nu}\mathfrak{g}^{\alpha \beta}-\mathfrak{g}^{\alpha \nu}\mathfrak{g}^{\mu \beta}\label{eq:SuperPotential}.
\end{align}
As in general relativity, the following identity holds:
\begin{align}
{H^{\mu \alpha \nu \beta}}_{,\alpha \beta}&=(-g)(2 R_{\mu \nu}-R g_{\mu\nu}+\frac{16 \pi G}{c^4} t_{\text{LL}}^{\mu \nu}),\label{eq:SPidentity}
\end{align}
where $t_{\text{LL}}^{\mu \nu}$ is the Landau-Lifshitz pseudotensor.\\

We assume that far from any sources, the metric reduces to the Minkowski metric $\eta_{\mu \nu}$ and that the scalar field approaches a constant value $\phi_0$. Let $\varphi\equiv \phi/\phi_0$ be the normalized scalar field. We introduce the conformally transformed metric
\begin{align}
\tilde{g}_{\mu \nu}&\equiv \varphi g_{\mu \nu},\label{eq:Defgtilde}
\end{align}
the gravitational field
\begin{align}
\tilde{h}^{\mu \nu}\equiv \eta^{\mu \nu}-\sqrt{-\tilde{g}}\tilde{g}^{\mu \nu},\label{eq:Defh}
\end{align}
and the ``conformal gothic metric''
\begin{align}
\tilde{\mathfrak{g}}^{\mu \nu}\equiv\sqrt{-\tilde{g}}\tilde{g}^{\mu \nu}.
\end{align}
 We impose the Lorentz gauge condition
\begin{align}
\partial_\nu \tilde{h}^{\mu \nu}&=0.\label{eq:LorentzGauge}
\end{align}

Substituting Eqs. (\ref{eq:Gothicg})--(\ref{eq:Defh}) into the gauge condition (\ref{eq:LorentzGauge}), the field equation (\ref{eq:JordanMetric}) is rewritten as
\begin{align}
\square_\eta \tilde{h}^{\mu \n}&= -\frac{16 \pi G}{c^2} \tau^{\mu \nu},\label{eq:DIRE1}
\end{align}
where $\square_\eta$ is the Minkowski space d'Alembertian and
\begin{align}
\tau^{\mu \nu}\equiv& (-g) \frac{\varphi}{\phi_0 c^2} T^{\mu \nu}+\frac{c^2}{16 \pi G}\left(\Lambda^{\mu \nu}+\Lambda_S^{\mu \nu}\right)\label{eq:TauDef},\\
\begin{split}
\Lambda^{\mu \nu}\equiv& \frac{16 \pi G}{c^4}\left[(-g) t^{\mu \nu}_\text{LL}\right](\tilde{\mathfrak{g}}^{\mu \nu})+\partial_\beta \tilde{h}^{\mu \alpha}\partial_\alpha \tilde{h}^{\nu \beta}\\
&-\tilde{h}^{\alpha \beta}\partial_\alpha \partial_\beta \tilde{h}^{\mu \nu},\label{eq:LambdaDef1}
\end{split}\\
\Lambda_S^{\mu \nu}\equiv&\frac{3+2\omega}{\varphi^2}\partial_\alpha\varphi \partial_\beta \varphi \left(\tilde{\mathfrak{g}}^{\mu \alpha}\tilde{\mathfrak{g}}^{\nu \beta}-\frac{1}{2}\tilde{\mathfrak{g}}^{\mu \nu}\tilde{\mathfrak{g}}^{\alpha \beta}\right),\label{eq:LambdaDef2}
\end{align}
where the notation $[(-g)t_{\text{LL}}^{\mu \nu}](\tilde{\mathfrak{g}}^{\mu \nu})$ indicates that the Landau-Lifshitz pseudotensor should be calculated using $\tilde{\mathfrak{g}}$ rather than the physical metric $g$. Similarly, the scalar field equation (\ref{eq:JordanScalar}) can be recast into the form
\begin{align}
\square_\eta \varphi&= -\frac{8 \pi G}{c^2} \tau_s,\label{eq:DIRE2}
\end{align}
with 
\begin{align}
\begin{split}
\tau_s\equiv& -\frac{1}{3+2 \omega}\sqrt{-g}\frac{\varphi}{\phi_0 c^2}\left(T-2\phi \frac{D T}{D \phi}\right)-\frac{c^2}{8\pi G}\tilde{h}^{\alpha \beta}\partial_\alpha \partial_\beta \varphi\\
&+\frac{c^2}{16 \pi G}\d{}{\varphi}\left[\log\left(\frac{3+2\omega}{\varphi^2}\right)\right]\partial_\alpha \varphi \partial_\beta\varphi\tilde{\mathfrak{g}}^{\alpha\beta}\label{eq:TauSDef}.
\end{split}
\end{align}
The differential equations (\ref{eq:DIRE1}) and (\ref{eq:DIRE2}) can be solved formally using the standard flat-space Green's function; we only consider retarded solutions, i.e. those with no incoming radiation
\begin{align}
\begin{split}
\tilde{h}^{\mu \nu}(t,\v{x})&=\frac{4 G}{c^2}\int d^3x' \frac{\tau^{\mu \nu}(t-|\v{x}-\v{x'}|,\v{x}')}{|\v{x}-\v{x}'|},\label{eq:hTildeDef}
\end{split}\\
\begin{split}
\varphi(t,\v{x})&=1+\frac{2G}{c^2}\int d^3x' \frac{\tau_s(t-|\v{x}-\v{x'}|,\v{x}')}{|\v{x}-\v{x}'|}, \label{eq:VarphiDef}
\end{split}
\end{align}
where the integration constant is explicitly added to enforce the asymptotic boundary condition on the scalar field. By construction, the constraint equation Eq. (\ref{eq:JordanConstraint}) acts as an additional boundary condition on the scalar field along the worldline of each body; this constraint distinguishes our work from the PN solutions found in Refs. \cite{Mirshekari2013,Lang2014a,Lang2014b}.

We approximate the formal solutions given in Eqs. (\ref{eq:hTildeDef}) and (\ref{eq:VarphiDef}) with an expansion in terms of $\epsilon\sim (v/c)^2 \sim Gm/rc^2$. However, to capture the strong-field effects behind dynamical scalarization, \textit{we expand only the metric $g_{\mu \nu}$ and scalar field $\phi$, leaving $\xi$ unexpanded.} Note that $\xi$ appears only in the function $m_A(\phi,\xi)$ in Eqs. (\ref{eq:DIRE1}) and (\ref{eq:DIRE2}). Thus, by not expanding $\xi$, we effectively resum the variable mass found in the PN treatment. The constraint equation Eq. (\ref{eq:JordanConstraint}) is used to solve $\xi$ exactly on each worldline at a given order in $\epsilon$.

\section{Structure of the near-zone fields}\label{sec:NZfields}

The resummation detailed above only enters through the sources, i.e. the stress-energy tensor $T^{\mu\nu}$ and its derivatives. As such, we adopt the same techniques used for the PN calculation of the metric and scalar field in Refs. \cite{Mirshekari2013,Lang2014a,Lang2014b}. We summarize this approach below, leaving our results in terms of $T^{\mu\nu}$ and its derivatives. For more detail, see Secs. III and IV of Ref. \cite{Mirshekari2013}.

The integration in Eqs. (\ref{eq:hTildeDef}) and (\ref{eq:VarphiDef}) is done over the flat space past null cone $\mathcal{C}$ emanating from the point $(t,\v{x})$. We divide this three-dimensional hypersurface into two regions. For matter sources of characteristic size $\mathcal{S}$, we define the \emph{near zone} as the worldtube with $|\v{x}|<\mathcal{R}$ where $\mathcal{R}\sim \mathcal{S}/v$ is the characteristic wavelength of the emitted gravitational radiation. The \emph{radiation zone} is the region outside of the near zone, that is, $|\v{x}|>\mathcal{R}$. We demarcate the intersection of $\mathcal{C}$ with the near zone as $\mathcal{N}$ and the intersection of $\mathcal{C}$ with the radiation zone as $\mathcal{C}-\mathcal{N}$.

We focus first on finding the metric and scalar field in the near zone, as these determine the equations of motion of the binary system through Eq. (\ref{eq:JordanEOM}). Following Refs. \cite{Pati2000,Mirshekari2013}, we establish the following notation
\begin{gather}
\begin{split}
N\equiv \tilde{h}^{00}, \qquad K^i \equiv \tilde{h}^{0i},\\
B^{ij}\equiv \tilde{h}^{ij}, \qquad B\equiv \tilde{h}^{ii}.\label{eq:MetricDef}
\end{split}
\end{gather}
To post-Newtonian order, we express the metric in terms of these fields using Eqs. (\ref{eq:Defgtilde}) and (\ref{eq:Defh})
\begin{align}
\begin{split}
g_{00}=&-1+\left(\frac{1}{2} N +\Psi\right)\\
&+\left(\frac{1}{2} B -\frac{3}{8} N^2 -\frac{1}{2} N \Psi -\Psi^2\right)+\Ord\left(\frac{1}{c^6}\right),\end{split}\label{eq:g00}\\
g_{0i}=&-K^i+\Ord\left(\frac{1}{c^5}\right),\\
g_{ij}=& \delta_{ij}\left[1+\left(\frac{1}{2}N-\Psi\right)\right]+\Ord\left(\frac{1}{c^4}\right),
\end{align}
where $\Psi$ was defined in Eq. (\ref{eq:PsiDef}).

At the point $(t,\v{x})$ in the near zone, the near-zone contribution to the integrals in Eqs. (\ref{eq:hTildeDef}) and (\ref{eq:VarphiDef}) can be expanded in powers of $|\v{x}-\v{x}'|$
\begin{widetext}
\begin{align}
N_\mathcal{N}(t,\v{x})&= \frac{4 G}{c^2} \int_\mathcal{M} \frac{\tau^{00}(t,\v{x}')}{|\v{x}-\v{x}'|}d^3 x'+\frac{2 G}{c^4} \partial_t^2 \int_\mathcal{M} \tau^{0 0}(t,\v{x}')|\v{x}-\v{x}'|d^3 x'+N_{\partial \mathcal{M}}+\Ord\left(\frac{1}{c^6}\right),\label{eq:NZfield1}\\
K^i_\mathcal{N}(t,\v{x})&=\frac{4 G}{c^2}\int_\mathcal{M} \frac{\tau^{0i}(t,\v{x}')}{|\v{x}-\v{x}'|}d^3 x'+K^i_{\partial \mathcal{M}}+\Ord\left(\frac{1}{c^5}\right),\\
B^{ij}_\mathcal{N}(t,\v{x})&=\frac{4 G}{c^2} \int_\mathcal{M} \frac{\tau^{ij}(t,\v{x}')}{|\v{x}-\v{x}'|}d^3 x'+B^{ij}_{\partial \mathcal{M}}+\Ord\left(\frac{1}{c^4}\right),\\
\Psi_\mathcal{N}(t,\v{x})&= \frac{2 G}{c^2} \int_\mathcal{M} \frac{\tau_s(t,\v{x}')}{|\v{x}-\v{x}'|}d^3 x'- \frac{2 G}{c^3}\partial_t \int_\mathcal{M} \tau_s(t,\v{x}') d^3 x'+ \frac{G}{c^4}\partial_t^2 \int_\mathcal{M} \tau_s(t,\v{x}')|\v{x}-\v{x}'|d^3 x'+\Ord\left(\frac{1}{c^6}\right)\label{eq:NZfield4},
\end{align}
\end{widetext}
where $\mathcal{M}$ is a constant-time hypersurface which covers the near zone and we have used Eq. (\ref{eq:LorentzGauge}) to eliminate the first order correction in Eq. (\ref{eq:NZfield1}). There will also be a contribution to the fields at $(t,\v{x})$ from the radiation zone, but these only enter at higher order \cite{Mirshekari2013}. The boundary terms $N_{\partial \mathcal{M}},K^i_{\partial \mathcal{M}},B^{ij}_{\partial \mathcal{M}}$ depend on the value of $\mathcal{R}$. Because the left hand side of Eqs. (\ref{eq:NZfield1})--(\ref{eq:NZfield4}) should not depend on the arbitrarily chosen boundary between the near and radiation zones, we argue (as in Ref. \cite{Mirshekari2013}) that these terms are exactly cancelled by the contributions from the radiation zone. This cancellation was shown explicitly in GR in Refs. \cite{Will1996,Pati2000}.

All that remains is to expand the sources $\tau^{\mu \nu}$ and $\tau_s$. We first define the densities
%\begin{align}
%\begin{split}
%c^2 \sigma&\equiv T^{00}+T^{ii},\qquad c^2 \sigma^i\equiv T^{0 i},\\
%c^2\sigma^{ij}&\equiv T^{ij},\qquad c^2 \sigma_s\equiv-T+2\phi \frac{DT}{D\phi}.\label{eq:SigmaDef}
%\end{split}
%\end{align}
\begin{align}
\sigma&\equiv (T^{00}+T^{ii})c^{-2},\label{eq:SigmaDefstart}\\
\sigma^i&\equiv T^{0 i}c^{-2},\label{eq:SigmaDef2}\\
\sigma^{ij} &\equiv T^{ij}c^{-2},\\
\sigma_s &\equiv-\frac{T}{c^2}+\frac{2\phi}{c^2} \frac{DT}{D\phi}.\label{eq:SigmaDefend}
\end{align}
We expand Eqs. (\ref{eq:TauDef}) and (\ref{eq:TauSDef}) to post-Newtonian order
\begin{align}
\begin{split}
\tau^{00}=&\frac{1}{\phi_0}\left[\sigma-\sigma^{ii}+\frac{G}{\phi_0 c^2}\left(4 \sigma U-\frac{7}{8\pi} (\nabla U)^2\right)\right.\\
&\left.- \frac{G {\mu_0}^2}{\phi_0 c^2}\left(6 \sigma U_s -\frac{1}{8\pi} (\nabla U_s)^2\right)\right],
\end{split}\\
\tau^{0i}=&\frac{\sigma^i}{\phi_0},\\
\tau^{ii}=&\frac{1}{\phi_0}\left[\sigma^{ii}-\frac{1}{8\pi}\frac{G}{\phi_0 c^2}(\nabla U)^2 -\frac{1}{8\pi} \frac{G {\mu_0}^2}{\phi_0 c^2}(\nabla U_s)^2\right]\\
\begin{split}
\tau_s=&\frac{{\mu_0}^2}{\phi_0}\left[\sigma_s+2\frac{G}{\phi_0 c^2} \sigma_s U+\frac{G (B-2{\mu_0}^2)}{\phi_0 c^2} \sigma_s U_s \vphantom{\frac{1}{2 \pi}}\right.\\
&\left.-\frac{1}{8 \pi} \frac{G (B+4{\mu_0}^2)}{\phi_0 c^2}(\nabla U_s)^2\right]\label{eq:TauSExpansion},
\end{split}
\end{align}
where we have introduced the potentials
\begin{align}
U&\equiv \int_\mathcal{M} \frac{\sigma(t,\v{x}')}{|\v{x}-\v{x}'|} d^3 x',\\
U_s&\equiv \int_\mathcal{M} \frac{\sigma_s(t,\v{x}')}{|\v{x}-\v{x}'|} d^3 x'.
\end{align}
Plugging these expressions back into Eqs. (\ref{eq:g00})--(\ref{eq:NZfield4}), the 1PD metric and scalar field are given by
\begin{widetext}
\begin{align}
\begin{split}
g_{00}=&-1+\frac{2 G}{\phi_0 c^2} U+\frac{2 G \mu_0}{\phi_0 c^2} U_s-\frac{2 G}{c^3}\dot{M}_s-\frac{2 G^2}{\phi_0^2 c^4}U^2+\frac{G^2 \mu_0(B-4 \mu_0)}{2\phi_0^2 c^4}U_s^2-\frac{G^2 \mu_0}{\phi_0^2 c^4}U U_s\label{eq:1PSsolStart}\\
&+\frac{4 G^2 \mu_0}{\phi_0^2 c^4}\Phi^s_2-\frac{12 G^2 \mu_0}{\phi_0^2 c^4} \Phi_{2s}+\frac{G^2 \mu_0(B-8\mu_0)}{\phi_0^2 c^4}\Phi^s_{2s}+\frac{G}{\phi_0 c^4} \ddot{X}+\frac{G \mu_0}{\phi_0 c^4} \ddot{X}_s+\Ord\left(\frac{1}{c^6}\right),
\end{split}\\
g_{0i}=&-\frac{4G}{\phi_0 c^2}V^i+\Ord\left(\frac{1}{c^5}\right),\\
g_{ij}=&\delta_{ij}\left[1+\frac{2 G}{\phi_0 c^2} U-\frac{2 G \mu_0}{\phi_0 c^2}U_s\right]+\Ord\left(\frac{1}{c^4}\right),\\
\begin{split}
\phi=&\phi_0+\frac{2 G \mu_0 U_s}{c^2}-\frac{2 G}{c^3}\dot{M}_s+\frac{G \mu_0}{ c^2}\left[\frac{G (B+4\mu_0)}{2\phi_0 c^2} U_s^2+4\frac{G}{\phi_0 c^2} \Phi^s_2+\frac{G (B-8\mu_0)}{\phi_0 c^2}\Phi^s_{2s}+\ddot{X}_s\right]+\Ord\left(\frac{1}{c^6}\right),
\end{split}\label{eq:Phi1PS}
\end{align}
\end{widetext}
with the additional potentials
\begin{align}
M_s&\equiv\int \sigma_s(t,\v{x}')d^3 x',\\
V^i&\equiv\int\frac{\sigma^i(t,\v{x}')}{|\v{x}-\v{x}'|} d^3 x',\\
\Phi^s_2&\equiv\int\frac{\sigma_s(t,\v{x}')U(t,\v{x}')}{|\v{x}-\v{x}'|} d^3 x',\\
\Phi_{2s}&\equiv\int\frac{\sigma(t,\v{x}')U_s(t,\v{x}')}{|\v{x}-\v{x}'|} d^3 x',\\
\Phi^s_{2s}&\equiv\int\frac{\sigma_s(t,\v{x}')U_s(t,\v{x}')}{|\v{x}-\v{x}'|} d^3 x',\\
X&\equiv\int \sigma(t,\v{x}') |\v{x}-\v{x}'| d^3 x',\\
X_s&\equiv\int \sigma_s(t,\v{x}') |\v{x}-\v{x}'| d^3 x',
\end{align}
\section{Two-body Equations of motion} \label{sec:EOM}
\subsection{Newtonian order} \label{sec:0PS}
We now apply these calculations to a binary system whose stress-energy tensor is given by Eq. (\ref{eq:StressEnergy}). To highlight the novel aspects of the PD approach, we explicitly work out the leading-order equations of motion here before calculating their higher-order corrections in the following section. In keeping with PN conventions, we describe the leading order as Newtonian and the next-to-leading order as post-Dickean or ``1PD.''

At Newtonian order, the densities defined in Eqs. (\ref{eq:SigmaDefstart}) and (\ref{eq:SigmaDef2}) are given by
\begin{align}
\sigma&=\sum_A m_A(\phi,\xi)\delta^{(3)}(x-x_A)+\Ord\left(\frac{1}{c^2}\right),\\
\sigma_s&=\sum_A m_A(\phi,\xi)\frac{\alpha_A(\phi,\xi)}{\mu_0}\delta^{(3)}(x-x_A)+\Ord\left(\frac{1}{c^2}\right).
\end{align}
where we have introduced the scalar charge of each body
\begin{align}
\alpha_A(\phi,\xi)&\equiv\left(\frac{B \log \phi}{2}\right)^{1/2} \left(1-2 \phi\frac{ D \log m_A}{D \phi}\right).\label{eq:ChargeDef}
\end{align}
Our definition of the scalar charge is the natural generalization of the expression used in Ref. \cite{Damour1992}; with no resummation, i.e. $m(\phi,\xi)=m(\phi)$, one recovers the definition
\begin{align}
\alpha_A=-\d{\log m_A^{(E)}}{\tilde{\varphi}},
\end{align}
where $\tilde \varphi$ is defined in Eq. (\ref{eq:PhiVarphi}).

Evaluating Eq. (\ref{eq:Phi1PS}) at Newtonian order, the scalar field for a 2-body system is given by
\begin{align}
\begin{split}
\phi&=\phi_0+2 \frac{G {\mu_0}^2U_s}{c^2},\\
&=\phi_0+\frac{2 G m_1 \mu_0 \alpha_1}{c^2 r_1}+\Ord\left(\frac{1}{c^3}\right)+\left(1\rightleftharpoons 2\right),
\end{split}
\end{align}
where we have adopted the shorthand
\begin{align}
\begin{split}
m_A\equiv &m_A(\phi(x_A),\xi(x_A)),\qquad \alpha_A\equiv \alpha_A(\phi(x_A),\xi(x_A)),\\
&r_A\equiv |\v{x}-\v{x}_A|,\qquad \v{n}_A\equiv (\v{x}-\v{x}_A)/r_A. \label{eq:ShorthandDef}
\end{split}
\end{align}
Because $m_A$ and $\alpha_A$ depend on $\phi$, these quantities must be expanded around the background field $\phi_0$. We suppress these expansions (given in Appendix \ref{sec:Charge}) throughout the remainder of this paper for notational convenience, denoting with the shorthand in Eq. (\ref{eq:ShorthandDef}) that the mass and charge should be expanded and truncated at the appropriate PD order.

On each worldline, we \emph{exactly} solve (i.e. not perturbatively) Eq. (\ref{eq:JordanConstraint}), ignoring the divergent terms that arise from self-interactions of each body
\begin{align}
\xi(x_1)&=\begin{cases}\phi_0+\frac{2 G m_2 \mu_0\alpha_2}{c^2 r},& \text{if } F(\phi)=\phi\\
\tilde\varphi_0+\frac{G m_2 \alpha_2}{c^2 r},& \text{if } F(\phi)=\sqrt{\frac{2 \log \phi}{B}} \end{cases}\label{eq:NewtonianXi}\\
\xi(x_2)&=\left(1\leftrightharpoons 2\right),
\end{align}
where $r\equiv|\v{x}_1-\v{x}_2|$ is the orbital separation of the binary and
\begin{align}
\tilde\varphi_0\equiv\frac{2 \mu_0}{B}=\sqrt{\frac{2 \log\phi_0}{B}}.
\end{align}
Note that this system of equations cannot be solved analytically, as $m_A$ and $\alpha_A$ depend on $\xi$ along each worldline. This final step is analogous to the feedback model proposed in Ref. \cite{Palenzuela2014}; with the choice of $F^{(\tilde\varphi)}$ given in Table \ref{table:ResummationSchemes}, we exactly reproduce this model.

Plugging in the expressions for the metric and scalar field into Eq. (\ref{eq:JordanEOM}), we find the Newtonian equations of motion
\begin{align}
a_1^i&=-\frac{G m_2 \left(1+\alpha_1 \alpha_2\right)}{\phi_0 r^2}n^i,\label{eq:NewtonianEOM1}\\
a^i_{2}&=\left(1\rightleftharpoons 2\right),\label{eq:NewtonianEOM2}
\end{align}
where $\v{n}\equiv(\v{x}_1-\v{x}_2)/r$. The mass $m_A$ and scalar charge $\alpha_A$ depend on the choice of resummation scheme; their leading order piece is given in Appendix \ref{sec:Charge}.

\subsection{Post-Dickean order}

To find the equations of motion of the binary to next order in $c^{-2}$, we expand the stress-energy tensor and evaluate the potentials introduced in Sec. \ref{sec:NZfields} (see Appendix \ref{sec:1PSpotentials}).

On each worldline, we plug the above potentials into Eq. (\ref{eq:Phi1PS}) and numerically solve Eq. (\ref{eq:JordanConstraint})

\begin{widetext}
\begin{align}
\xi(x_1)&=\begin{cases}\begin{aligned}&\phi_0+\frac{2 G \mu_0 m_2 \alpha_2}{\phi_0 r c^2}+\frac{G m_2 \alpha_2}{\phi_0 r c^4}\left[-\mu_0(\v{v}_2\cdot\v{n})^2+\alpha_2\left(\frac{B}{2}+2 {\mu_0}^2\right)\frac{G m_2}{\phi_0 r}\right.\\
&\left.-\left({2 \mu_0}^2\alpha_1+\mu_0\left(3+\alpha_1\alpha_2\right)\right)\frac{G m_1}{\phi_0 r}\right],\end{aligned}& \text{if } F(\phi)=\phi\\
\tilde\varphi_0+\frac{G m_2 \alpha_2}{\phi_0 r c^2}+\frac{G m_2 \alpha_2}{\phi_0 r c^4}\left[-\frac{1}{2}(\v{v}_2\cdot\v{n})^2-\left(\frac{3}{2}+\mu_0 \alpha_1 +\frac{1}{2}\alpha_1\alpha_2\right)\frac{G m_1}{\phi_0 r}\right],&\text{if } F(\phi)=\sqrt{\frac{2 \log \phi}{B}}
\end{cases}\label{eq:Xi1PS}\\
\xi(x_2)&=\left(1 \rightleftharpoons 2\right).\label{eq:Xi1PS2}
\end{align}

 Substituting Eqs. (\ref{eq:1PSsolStart})-(\ref{eq:Phi1PS}) into Eq. (\ref{eq:JordanEOM}), we find the following equation of motions for each particle
\begin{align}
\begin{split}
a^i_{(1)}=&-\frac{G m_2\left(1+\alpha_1\alpha_2\right)}{\phi_0 r^2}n^i+\frac{G m_2}{\phi_0 r^2 c^2}n^i\left[-\left(1-\alpha_1\alpha_2\right)v_1^2-2 (v_2^2-2 \v{v_1}\cdot\v{v_2})\vphantom{\frac{G}{r}}\right.\\
&\left. +\frac{3}{2}\left(1+\alpha_1\alpha_2\right)\left(\v{v}_2\cdot\v{n}\right)^2+4\left(1+\alpha_1\alpha_2\right)\frac{G m_2}{\phi_0 r}+\left(5+\mu_0\alpha_1\right)\left(1+\alpha_1\alpha_2 \right)\frac{G m_1}{\phi_0 r}\right]\\
&+\frac{G m_2}{\phi_0 r^2 c^2}(v_1-v_2)^i\left[4\left(\v{v}_1\cdot\v{n}\right)-\left(3-\alpha_1\alpha_2\right)\left(\v{v}_2\cdot\v{n}\right)\right],
\end{split}\label{eq:1PSEOM1}\\
a^i_{(2)}&=\left(1\rightleftharpoons 2\right),\label{eq:1PSEOM2}
\end{align}
where $m$ and $\alpha$ themselves receive post-Dickean corrections dependent on the resummation scheme used (see Appendix \ref{sec:Charge}).

For reference later, the 1PN equations of motion (with no resummation of the mass) are recovered with the choice $m^{(\text{PN})}$
\begin{align}
\begin{split}
a^i_{1\,\text{(PN)}}=&-\frac{G \bar m_2\left(1+\bar\alpha_1\bar\alpha_2\right)}{\phi_0 r^2}n^i+\frac{G \bar m_2}{\phi_0 r^2 c^2}n^i \left[-\left(1-\bar\alpha_1\bar\alpha_2\right)v_1^2-2\left(v_2^2-2\v{v}_1\cdot\v{v}_2\right)\vphantom{\frac{G m_2}{r^2}}\right.\\
&\left.+\frac{3}{2}\left(1+\bar\alpha_1\bar\alpha_2\right)\left(\v{v}_2\cdot\v{n}\right)^2+\left(4+4\bar\alpha_1\bar\alpha_2-\bar\alpha'_1 \bar\alpha_2^2\right)\frac{G\bar m_2}{\phi_0 r}+\left(\left(5+\bar\alpha_1\bar\alpha_2\right)\left(1+\bar\alpha_1\bar\alpha_2\right)-\bar\alpha'_2\bar\alpha_1^2\right)\frac{G \bar m_1}{\phi_0 r}\right]\\
&+\frac{G \bar m_2}{\phi_0 r^2 c^2}(v_1-v_2)^i\left[4\left(\v{v}_1\cdot\v{n}\right)-\left(3-\bar\alpha_1\bar\alpha_2\right)\left(\v{v}_2\cdot\v{n}\right)\right],\label{eq:PNEOM1}
\end{split}\\
a^i_{2\,\text{(PN)}}&=\left(1\rightleftharpoons 2\right),\label{eq:PNEOM2}
\end{align}
\end{widetext}
where we have introduced the shorthand
\begin{align}
\bar m_i&\equiv m_i(\phi_0),\label{eq:Mbar}\\
\bar\alpha_i&\equiv \mu_0\left(1-2\d{\log m_i}{\log \phi}\right)_{\phi=\phi_0},\label{eq:Alphabar}\\
\bar\alpha_i'&\equiv \frac{B \bar{\alpha}_i}{2 \mu_0}-4\mu_0^2\left(\dd{\log m_i}{(\log \phi)}\right)_{\phi=\phi_0}.\label{eq:AlphaPrimebar}
\end{align}

The apparent differences between Eqs. (\ref{eq:1PSEOM1})--(\ref{eq:1PSEOM2}) and Eqs. (\ref{eq:PNEOM1})--(\ref{eq:PNEOM2}) are simply artifacts of the different notations. The disparities stem from the presence in Eq. (\ref{eq:PNEOM1}) of higher-order terms from expansions like Eq. (\ref{eq:Mexpand}). These terms are absorbed into the definitions of $m_A$ and $\alpha_A$ in the PD expansion [see Eq. (\ref{eq:ShorthandDef})]. We emphasize the differences between these two notations because the analytic model proposed in Ref. \cite{Palenzuela2014} directly adapted the equations of motion written as in Eqs. (\ref{eq:PNEOM1}) and (\ref{eq:PNEOM2}). Beyond post-Newtonian order, we expect a greater proportion of the corresponding terms in each notation to differ.

For a generic resummation scheme, the 1PD Eqs. (\ref{eq:1PSEOM1}) and (\ref{eq:1PSEOM2}) are not solutions to the Euler-Lagrange equations for any Fokker Lagrangian (a Lagrangian dependent solely on the the positions and velocities of the two bodies). A simple calculation reveals that the equations of motion can be integrated back to such a Lagrangian only when no resummation is performed, i.e. when $m^{(\text{PN})}$ is used.\footnote{This result contradicts the assertion of Ref. \cite{Palenzuela2014} that such a Lagrangian can be constructed by resumming (or not expanding) the scalar charge $\alpha$ in the PK Lagrangian of Ref. \cite{Damour1996b}.} The absence of a PD Fokker Lagrangian suggests that our model of DS requires the  two-body phase space to be augmented with additional degrees of freedom besides the bodies' positions and velocities, such as the scalar field $\xi$. We conjecture that any other extension to the PN formalism to incorporate DS will also require new, dynamical degrees of freedom.

\section{Structure of the far-zone fields}\label{sec:FZfields}

Having solved the dynamics of the binary, we now shift our attention to observables that can be extracted from the asymptotic geometry of the system. Our interest in this type of quantity is twofold. First, such objects encode all information needed to estimate GW signals (e.g. the waveform and its phase evolution estimated from the Bondi mass and flux). Second, there are several gauge-invariant quantities defined asymptotically that are easily computed in numerical relativity (e.g. the ADM mass and angular momentum) and thus can be used to directly check the validity of our model.  For simplicity, in this work we restrict our attention to the scalar mass, a coordinate independent measure of a spacetime's scalarization. We define the scalar mass at retarded time $\tau$ as
\begin{align}
M_S(\tau)&\equiv-\frac{c^2}{8\pi G}\oint_{\substack{|\v{x}|\rightarrow \infty\\ t-|\v{x}|=\tau}}\delta^{i j} \partial_i \phi \,d S_j, \label{eq:ScalarMassDef}\\
&=-\frac{\phi_0 c^2}{8\pi G}\oint_{\substack{|\v{x}|\rightarrow \infty\\ t-|\v{x}|=\tau}}\delta^{i j} \partial_i \Psi \,d S_j,\label{eq:ScalarMassDef2}
\end{align}
where $S_j$ is the surface-area element in flat space. We leave the other useful quantities described above for future work.

Calculating the scalar mass requires knowledge of the scalar field at a distance $|\v{x}|=R\gg \mathcal{R}$ (recall that $\mathcal{R}$ is the boundary of the near zone). As in the near zone, we will recycle the tools used to determine the scalar field in the radiation zone from previous PN calculations. We summarize this calculation for a generic stress-energy tensor below; for more detail, see Refs. \cite{Lang2014a,Lang2014b}

At the order at which we work, the scalar field at null infinity receives contributions from both the near and radiation zones, which we denote as $\Psi_\mathcal{N}$ and $\Psi_{\mathcal{C}-\mathcal{N}}$, respectively. We compute each piece separately, dropping any terms dependent on $\mathcal{R}$, which we assume will cancel when the pieces are combined (as was done in Sec. \ref{sec:NZfields}).
\subsection{Near-zone contribution to the scalar field}
The contribution to the scalar field at the point $(t,\v{x})$ in the radiation zone from points $(t',\v{x}')$ in the near zone is found by expanding the integral expression given in Eq. (\ref{eq:VarphiDef}) in powers of $|\v{x}'|/R$ 
\begin{align}
\Psi_\mathcal{N}&=\sum_{m=0}^\infty\frac{2G}{c^{2+m}} \frac{1}{m!} \frac{\partial^m}{\partial t^m}\int_{\mathcal{M}'} \tau_s(\tau,\v{x}')\frac{(\uv{N}\cdot\v{x}')}{R}d^3 x',\\
&=\frac{2G}{c^2}\sum_{m=0}^\infty \frac{(-1)^m}{m!} \partial_{k_1}\cdots\partial_{k_m}\left(\frac{1}{R}\mathcal{I}_s^{k_1\cdots k_m}(\tau)\right),\label{eq:ScalarFieldFar}
\end{align}
where $\uv{N}\equiv\v{x}/R$, $\mathcal{M}'$ is the intersection of the near zone with a hypersurface of constant retarded time $\tau=t-R$, and we have introduced the scalar multipole moments
\begin{align}
\mathcal{I}_s^{k_1\cdots k_m}(\tau)\equiv\int_{\mathcal{M}'} \tau_s(\tau,\v{x})x^{k_1}\cdots x^{k_m}d^3 x. \label{eq:ScalarMultipole}
\end{align}
We note that the terms that fall off faster that $R^{-1}$ in Eq. (\ref{eq:ScalarFieldFar}) will not contribute to the scalar mass; dropping these terms, the remaining piece of the scalar field is given by
\begin{align}
\Psi_\mathcal{N}=\sum_{m=0}^\infty\frac{2G}{Rc^{2+m}} \frac{1}{m!} \hat{N}^{k_1}\cdots\hat{N}^{k_m} \frac{d^m}{dt^m}\mathcal{I}_s^{k_1\cdots k_m}(\tau),\label{eq:ScalarFieldFar2}
\end{align}
We also note that only terms with even parity (with respect to inversions of $\v{x}$) will contribute to the scalar mass. These are the terms in Eq. (\ref{eq:ScalarFieldFar2}) with even $m$.

The source $\tau_s$ in the near zone is needed at higher order than what was given in Eq. (\ref{eq:TauSExpansion}) to calculate the 1PD scalar mass
\begin{widetext}
\begin{align}
\begin{split}
\tau_s=&\frac{{\mu_0}^2}{\phi_0}\left[\sigma_s+2\frac{G}{\phi_0 c^2} \sigma_s U+\frac{G (B-2{\mu_0}^2)}{\phi_0 c^2} \sigma_s U_s \vphantom{\frac{1}{2 \pi}}-\frac{1}{8 \pi} \frac{G (B+4{\mu_0}^2)}{\phi_0 c^2}(\nabla U_s)^2\right]\label{eq:TauSExpansionMore}\\
&+\frac{G \mu_0^2(1+\mu_0^2)}{\phi_0^2 c^4}\sigma_s\left\{\frac{G}{\phi_0}\left[2U^2-(2B-4\mu_0^2)(U U_s +\Phi^s_2)-12\mu_0^2\Phi_{2s}\right]-(4\Phi_1-\ddot{X})\right.\\
&\qquad\left.+\frac{G(B^2-10 B \mu_0^2+8\mu_0^2)}{4\phi_0}U_s^2+\frac{G(B-8\mu_0^2)(B-2\mu_0^2)}{2\phi_0}\Phi^s_{2s}+\frac{B-2\mu_0^2}{2}\ddot{X}_s\right\}\\
&-\frac{G \mu_0^2}{8 \pi \phi_0^2 c^4}\left\{8 U \ddot{U}_s +16 V^j \partial_j\dot{U}_s+8\Phi^{ij}_1\partial_i \partial_j U_s-(B+4\mu_0^2)(\dot{U}_s^2-\grad U_s\cdot \grad \ddot{X}_s)+\frac{G\mu_0^2(6B+8\mu_0^2)}{\phi_0} U_s \left(\grad U_s\right)^2\vphantom{\frac{G}{\phi_0}}\right.\\
&\qquad\left.-\frac{G}{\phi_0}\left[-4(B+4\mu_0^2)\grad U_s\cdot \grad \Phi^s_2-8 P^{i j}_2 \partial_i \partial_j U_s-8\mu_0^2 P^{i j}_{2s} \partial_i \partial_j U_s\right]+\frac{G(B-8\mu_0^2)(B+4\mu_0^2)}{\phi_0}\grad U_s\cdot\grad\Phi^s_{2s}\right\},
\end{split}
\end{align}
\end{widetext}
where, in addition to the potentials introduced in Sec. \ref{sec:NZfields}, we define
\begin{align}
\Phi_1\equiv&\int\frac{\sigma^{ii}(t,\v{x}')}{|\v{x}-\v{x}'|} d^3 x',\\
\Phi^{i j}_1\equiv&\int\frac{\sigma^{ij}(t,\v{x}')}{|\v{x}-\v{x}'|} d^3 x',\\
P^{i j}_2\equiv&\frac{1}{4\pi}\int\frac{\partial_i U(t,\v{x}') \partial_j U(t,\v{x}')}{|\v{x}-\v{x}'|} d^3 x',\label{eq:P2potential}\\
P^{i j}_{2s}\equiv&\frac{1}{4\pi}\int\frac{\partial_i U_s(t,x') \partial_j U_s(t,x')}{|\v{x}-\v{x}'|} d^3 x'.\label{eq:P2spotential}
\end{align}
\subsection{Radiation-zone contribution to the scalar field}
We rewrite the integral in Eq. (\ref{eq:VarphiDef}) in a more useful way when working far from the system
\begin{align}
\Psi=\frac{2G}{c^2}\int\frac{\tau_s(R'+\tau',\v{x}')\delta(t'-t+|\v{x}-\v{x}'|-R')}{|\v{x}-\v{x}'|} d^4 x',
\end{align}
where $R'=|\v{x}'|$ and $\tau'=t'-R'$. Thus, the contribution to the scalar field from the radiation zone (i.e. $R'>\mathcal{R}$) is given by
\begin{widetext}
\begin{align}
\begin{split}
\Psi_{\mathcal{C}-\mathcal{N}}=&\frac{2G}{c^2}\int_{\tau-2\mathcal{R}}^{\tau} d\tau' \int_0^{2\pi}d\phi\int_{1-\upsilon}^1 \frac{\tau_s(\tau'+R',\v{x}')}{t-\tau'-\uv{N}'\cdot \v{x}} (R')^2 d (\cos \theta')+\frac{2G}{c^2}\int_{-\infty}^{\tau-2\mathcal{R}}\oint\frac{\tau_s(\tau'+R',\v{x}')}{t-\tau'-\uv{N}'\cdot \v{x}} (R')^2 d^2 \Omega' \label{eq:FarZoneContributionMessy}
\end{split}
\end{align}
where $\upsilon=(\tau-\tau')(2R-2\mathcal{R}+\tau-\tau')/(2R\mathcal{R})$ and\ $\uv{N}'=\v{x}'/R'$. The source $\tau_s$ takes a different form in the radiation zone than in Eq. (\ref{eq:TauSExpansionMore}). To the order at which we work, the source in the radiation zone is given by
\begin{align}
\tau_s&=-\frac{B+4 \mu_0^2}{32 \pi G \mu_0^2}\left[c^2(\grad \Psi)^2-\dot{\Psi}^2\right]-\frac{1}{8\pi G}N\ddot{\Psi}.\label{eq:TauSFar}
\end{align}
The stress-energy tensor does not appear in this expression (under the guise of $\sigma$ or $\sigma_s$) because the radiation zone does not contain any matter. In computing the source $\tau_s$, we can ignore the radiation-zone contribution to the scalar field, as the corresponding contributions to the source will enter at beyond the order that we work. Thus, we use the scalar field as given in Eq. (\ref{eq:ScalarFieldFar}); the metric field $N$ can be expanded in a similar way. At this order, only the monopole and dipole pieces of these fields appear in $\tau_s$.
\begin{align}
N=& \frac{4G}{c^2}\frac{\mathcal{I}}{R}+\cdots,\\
\Psi=&\frac{2G}{c^2}\frac{\mathcal{I}_s}{R}-\frac{2G}{c^2}\partial_i\left(\frac{\mathcal{I}_s^i}{R}\right)+\cdots,
\end{align}
where the mass monopole moment $\mathcal{I}$ is defined as in Eq.  (\ref{eq:ScalarMultipole}) with $\tau^{0 0}$. Plugging these expressions into Eq. (\ref{eq:TauSFar}), we find
\begin{align}
\begin{split}
\tau_s=&-\frac{G(B+4 \mu_0^2)}{2 \pi \mu_0^2 c^2}\left(\frac{\mathcal{I}_s\dot{\mathcal{I}}_s}{R^3 c}+\frac{(\mathcal{I}_s)^2}{R^4}\right)-\frac{G}{\pi c^4}\frac{\mathcal{I}\ddot{\mathcal{I}}_s}{R^2}\\
&-\frac{G(B+4\mu_0^2)}{\pi c^2}\left(\frac{\mathcal{I}_s \ddot{\mathcal{I}}^j_s}{R^3 c^2}+\frac{2\mathcal{I}_s \dot{\mathcal{I}}^j_s}{R^4 c}+\frac{2\mathcal{I}_s \mathcal{I}^j_s}{R^5}+\frac{\dot{\mathcal{I}}_s \dot{\mathcal{I}}^j_s}{R^3 c^2}+\frac{\dot{\mathcal{I}}_s \mathcal{I}^j_s}{R^4 c}\right)\hat{N}^j-\frac{G}{\pi c^4}\left(\frac{\mathcal{I}\dddot{\mathcal{I}}^j_s}{R^2 c}+\frac{\mathcal{I}\ddot{\mathcal{I}}^j_s}{R^3}\right)\hat{N}^j.\label{eq:TauSFarMoments}
\end{split}
\end{align}
where we've used the fact that the moments are functions of retarded time, so that $\partial_j \mathcal{I}_s^j=-\dot{\mathcal{I}}_s^j \hat{N}^j/c$. 
The first line of Eq. (\ref{eq:TauSFarMoments}) contains the lowest order terms, which enter at $c^{-3}$ order relative to the leading contribution to $\Psi_\mathcal{N}$ from the near zone, while the second line contains terms that are suppressed by one additional factor of $c$.

We note that all of the terms in $\tau_s$ in the radiation zone take the form
\begin{align}
\tau_s(l,n)=\frac{1}{4\pi}\frac{f(\tau)}{R^n}\hat{N}^{k_1}\cdots \hat{N}^{k_l}.
\end{align}
With this information, each corresponding term in Eq. (\ref{eq:FarZoneContributionMessy}) can be rewritten as
\begin{align}
\begin{split}
\Psi_{\mathcal{C}-\mathcal{N}}(l,n)=&\frac{2G}{Rc^2}\hat{N}^{k_1}\cdots \hat{N}^{k_l}\left[\int_0^\mathcal{R}f(\tau-2s)A(s,R)ds+\int_\mathcal{R}^\infty f(\tau-2s)B(s,R)ds\right],\label{eq:lnPsiFZ}
\end{split}
\end{align}
with
\begin{align}
A(s,R)\equiv&\int_\mathcal{R}^{R+s}\frac{P_l(\Lambda)}{p^{n-1}}dp,\\
B(s,R)\equiv&\int_s^{R+s}\frac{P_l(\Lambda)}{p^{n-1}}dp,\\
\Lambda\equiv&\frac{R+2s}{R}-\frac{2s(R+s)}{Rp},
\end{align}
and where $P_l(\Lambda)$ are Legendre polynomials.

Given Eq. (\ref{eq:TauSFarMoments}), we see that $l=0,1$ and $n=2-5$ integrals contribute to the scalar field at this order. However, by inspection, the $l=1$ terms have odd parity, and thus will not contribute to the scalar mass. The $l=0$ contributions [in the notation of Eq. (\ref{eq:lnPsiFZ})] are given by
\begin{align}
\Psi_{\mathcal{C}-\mathcal{N}}(0,2)=&-\frac{4G^2}{Rc^6}\int^{\tau}_{-\infty} du\left(\log\left(R+\frac{\tau}{2}-\frac{u}{2}\right)\left[\mathcal{I} \ddot{\mathcal{I}}_s\right]_u-\log\left(\mathcal{R}+\frac{\tau}{2}-\frac{u}{2}\right)\left[\mathcal{I} \ddot{\mathcal{I}}_s\right]_{u-2\mathcal{R}}\right)-\log\mathcal{R}\int_{\tau-2\mathcal{R}}^\tau du \left[\mathcal{I}\ddot{\mathcal{I}}_s\right]_u,\label{eq:FarZoneLNContribution1}\\
\Psi_{\mathcal{C}-\mathcal{N}}(0,3)=&\frac{2G^2}{Rc^5}\frac{B+4\mu_0^2}{\mu_0^2}\left(\frac{\mathcal{I}_s^2(\tau)}{2R}-\frac{\mathcal{I}_s^2(\tau)}{2\mathcal{R}}-\int_{-\infty}^\tau du\left(\frac{\left[\mathcal{I}_s^2\right]_u}{(2R+\tau-u)^2}-\frac{\left[\mathcal{I}_s^2\right]_{u-2\mathcal{R}}}{(2\mathcal{R}+\tau-u)^2}\right)\right),\label{eq:FarZoneLNContribution2}\\
\Psi_{\mathcal{C}-\mathcal{N}}(0,4)=&\frac{4G^2}{Rc^4}\frac{B+4\mu_0^2}{\mu_0^2}\left(\int_{-\infty}^\tau du \left(\frac{\left[\mathcal{I}_s^2\right]_u}{(2R+\tau-u)^2}-\frac{\left[\mathcal{I}_s^2\right]_{u-2\mathcal{R}}}{(2\mathcal{R}+\tau-u)^2}\right)-\frac{1}{(2\mathcal{R})^2}\int_{\tau-2\mathcal{R}}^\tau du \left[\mathcal{I}^2_s\right]_u\right),\label{eq:FarZoneLNContribution3}
\end{align}
%\end{widetext}
where we have used the shorthand $[fg]_x=f(x)g(x)$. Nearly all of these terms are hereditary, i.e. depend on the full history of the system up to the retarded time $\tau$. The one exception is the first term in Eq. (\ref{eq:FarZoneLNContribution2}), but this term falls off too quickly with $R$ to contribute to the scalar mass.

\section{Two-body Scalar mass}\label{sec:ScalarMass}
Having expressed the scalar field in the radiation zone entirely in terms of the (even) scalar multipole moments, we now specialize to an inspiraling binary system. Plugging the potentials for a two-body system (Appendix \ref{sec:1PSpotentials}) into Eq. (\ref{eq:TauSExpansionMore}), we integrate to find the scalar moments. Integrals containing $\sigma_s$  can be evaluated directly as they contain delta functions at the worldlines of the bodies. The remaining terms are integrated by parts, using techniques analogous to those outlined in Sec. III of Ref. \cite{Lang2014b}. The multipoles needed to compute the scalar mass at 1PD order are given by
%\begin{widetext}
\begin{align}
\begin{split}
\mathcal{I}_s=&\frac{\mu_0 m_1 \alpha_1}{\phi_0}\left\{1-\frac{v_1^2}{2c^2}-\frac{G m_2}{\phi_0 rc^2}\left(1+\mu_0 \alpha_2\right)-\frac{v_1^4}{8 c^4}+\frac{G m_2}{\phi_0 r c^2}\left[\left(\frac{2 \mu_0(1-\alpha_2 \mu_0)-B \alpha_2}{4\mu_0}\right) \frac{v_1^2}{c^2}-\frac{3}{2}\frac{\left(\v{v}_2\cdot\v{n}\right)^2}{c^2}\right.\right.\\
&\left.\left.+\left(\frac{B \alpha_2 -8 \mu_0+6\alpha_2\mu_0^2}{4 \mu_0 }\right)\frac{\left(\v{v}_1\cdot\v{n}\right)^2}{c^2}+\left(\frac{\alpha_2(B+4\mu_0^2)}{4 \mu_0 }\right)\frac{\left(\v{v}_1\cdot\v{v}_2\right)}{c^2} -\left(\frac{B \alpha_2-16\mu_0+4\alpha_2 \mu_0^2}{4 \mu_0}\right)\frac{\left(\v{v}_1\cdot\v{n}\right)\left(\v{v}_2\cdot\v{n}\right)}{c^2}\right]\right.\\
&\left.+\frac{G^2 m_1 m_2}{\phi_0^2 r^2 c^4}\left[-\frac{1}{2}+\mu_0 \alpha_1-\frac{(B-6 \mu_0^2)\alpha_2}{4\mu_0}-\frac{(B+6-6\mu_0^2)\alpha_1\alpha_2}{4}-\frac{(B+2\mu_0^2)\alpha_1\alpha_2^2}{4\mu_0}\right]+\frac{G^2 m_2^2}{2 \phi_0^2 r^2 c^4}\right.\\
&\left.-\frac{G m_2}{\phi_0 c^4}\left[\frac{B \alpha_2-4\mu_0(1-\alpha_2 \mu_0)}{2 \mu_0}\right]\left(\v{a}_1\cdot\v{n}\right)\right\}+ \left(1\rightleftharpoons 2\right),
\end{split}\\
\mathcal{I}^{i j}_s=&\frac{\mu_0 m_1 \alpha_1 x_1^i x_1^j}{\phi_0}\left[1-\frac{v_1^2}{2 c^2}-\frac{G m_2}{\phi_0 rc^2}\left(1+\mu_0 \alpha_2\right)\right]+\frac{G m_1 m_2 \alpha_1 \alpha_2(B+4\mu_0^2)r}{4 \phi_0^2 \mu_0 c^2}\delta^{ij}+\left(1\rightleftharpoons 2\right),\\
\mathcal{I}^{i j k l}_s=&\frac{ \mu_0 m_1 \alpha_1 x_1^i x_1^j x_1^k x_1^l}{\phi_0 }+\left(1\rightleftharpoons 2\right).
\end{align}
We evaluate Eq. (\ref{eq:ScalarFieldFar2}) with these moments to compute the near zone contribution to the scalar field. Before proceeding, we briefly detail how time derivatives of the masses $m_i$ and scalar charges $\alpha_i$ are handled. Recall that the dependence of each body's mass (and scalar charge) on the local scalar field is decomposed into a resummed and expanded piece, represented by its dependence on $\xi$ and $\phi$, respectively. Thus, the derivative of the mass would be given by
\begin{align}
\d{m_A}{t}&=\pd{m_A}{\phi}v_A^\mu \partial_\mu \phi+\pd{m_A}{\xi}v_A^\mu \partial_\mu \xi,
\end{align}
where $v^\mu_A=u^\mu_A/u^0_A$. To reinforce that the fields $\phi$ and $\xi$ really represent the same physical scalar, we relate the two through Eq. (\ref{eq:JordanConstraint}). Thus (assuming differentiability), their gradients along each worldline are related as
\begin{align}
u^\mu_A\partial_\mu \xi=\d{F}{\phi}u^\mu_A\partial_\mu \phi. \label{eq:GradRelation}
\end{align}
In truth, because we expand only $\phi$ and not $\xi$, Eqs. (\ref{eq:JordanConstraint}) and (\ref{eq:GradRelation}) only hold in an approximate sense [e.g. up to 1PD order when using Eq. (\ref{eq:Xi1PS})]. Nevertheless, one finds that 
\begin{align}
\d{m_A}{t}&=\frac{D m_A}{D \phi} v^\mu \partial_\mu \phi+\Ord\left(\frac{1}{c^4}\right).
\end{align}
Because the time dependence of the mass enters only through the scalar field (whose leading order term is constant), its derivative is suppressed by an additional factor of $c^{-2}$ more than dimensional analysis would suggest, i.e. $\dot{m}/m\sim c^{-2}$. This suppression greatly simplifies our calculation of the scalar field.

Equipped with the scalar moments and a prescription for differentiating with respect to time, we calculate the near-zone contribution to the scalar field of a binary system
\begin{align}
\Psi_\mathcal{N}=&\Psi_\mathcal{N}^{(-1)}+\Psi_\mathcal{N}^{(0)}+\Psi_\mathcal{N}^{(1)},\label{eq:NearZoneFieldTot}
\end{align}
with
\begin{align}
\Psi_\mathcal{N}^{(-1)}=&\frac{2G\mu_0 m_1 \alpha_1 }{\phi_0 Rc^2 }+ \left(1\rightleftharpoons 2\right)\\
\Psi_\mathcal{N}^{(0)}=&\frac{2G\mu_0 m_1 \alpha_1 }{\phi_0 Rc^2 }\left\{-\frac{v_1^2}{2c^2}+\frac{(\uv{N}\cdot\v{v}_1)^2}{c^2}-\frac{G m_2}{\phi_0 r c^2}\left(1+\mu_0\alpha_2+(1+\alpha_1\alpha_2)\frac{(\uv{N}\cdot \v{x}_1)^2-(\uv{N}\cdot \v{x}_1)(\uv{N}\cdot \v{x}_2)}{r^2}\right)\right\}+ \left(1\rightleftharpoons 2\right)\\
\begin{split}
\Psi_\mathcal{N}^{(1)}=&\frac{2G\mu_0 m_1 \alpha_1 }{\phi_0 Rc^2}\left\{-\frac{v_1^4}{8c^4}+\frac{G m_2}{\phi_0 r c^2}\left[\left(\frac{1+\mu_0\alpha_2}{2}\right)\frac{v_1^2}{c^2}-\left(\frac{4-\mu_0\alpha_2}{2}\right)\frac{(\v{v}_1\cdot\v{n})^2}{c^2}-\frac{3(\v{v}_2\cdot\v{n})^2}{2c^2}+\frac{4(\v{v}_1\cdot\v{n})(\v{v}_2\cdot \v{n})}{c^2}\right]\right.\\
&\left.-\frac{G^2 m_1 m_2}{\phi_0 r^2 c^4}\left[\frac{1}{2}-\mu_0 \alpha_1-\frac{5\mu_0 \alpha_2}{2}+\frac{(6+B-6\mu_0^2)\alpha_1 \alpha_2}{4}+2 \alpha_2^2-\frac{\mu_0 \alpha_1 \alpha_2^2}{2}\right]-\frac{3 G^2 m_2^2}{2\phi_0^2 r^2 c^4}\right.\\
&\left.-\left(\frac{v_1^2}{2c^2}+\frac{G m_2(1+\mu_0\alpha_2)}{\phi_0 r c^2}\right)\frac{(\v{v}_1\cdot\uv{N})^2}{c^2}+\frac{G m_2}{\phi_0 r c^2}\left(-4 (\v{v}_1\cdot\v{n})+(3-\alpha_1\alpha_2)(\v{v}_2\cdot\v{n})\right)\frac{(\v{x}_1\cdot\uv{N})(\v{v}_2\cdot\uv{N})}{r c}\right.\\
&\left.+\frac{G m_2}{\phi_0 r c^2}\left[\left(8-\left(2\mu_0+\frac{B}{\mu_0}+4 \mu_0 \phi_0 \frac{D (\log m_1 \alpha_1)}{D \phi}\right)\alpha_2+2\alpha_1\alpha_2\right)\frac{(\v{v}_1\cdot\v{n})}{c}\right.\right.\\
&\qquad\left.\left.-\left(5-\left(2\mu_0+\frac{B}{\mu_0}+4 \mu_0 \phi_0 \frac{D (\log m_1 \alpha_1)}{D \phi}\right)\alpha_2-\alpha_1\alpha_2\right)\frac{(\v{v}_2\cdot\v{n})}{c}\right]\frac{(\v{x}_1\cdot\uv{N})(\v{v}_1\cdot\uv{N})}{r c}\right.\\
&\left.+\frac{G m_2}{\phi_0 r c^2}\left[\frac{4 \mu_0^2(\v{v}_1\cdot\v{n})}{c}-\frac{4 \mu_0^2(\v{v}_2\cdot\v{n})}{c}\right]\frac{(\v{x}_2\cdot\uv{N})(\v{v}_2\cdot\uv{N})}{r c}\right.\\
&\left.+\frac{G m_2}{\phi_0 r c^2}\left[-\frac{(1-3\alpha_1\alpha_2)v_1^2}{2c^2}-\frac{2v_2^2}{c^2}+\frac{4(\v{v}_1\cdot\v{v}_2)}{c^2}+\frac{3(1+\alpha_1\alpha_2)(\v{v}_2\cdot\v{n})^2}{2c^2}\right.\right.\\
&\qquad\left.\left.+\frac{G m_1(1+\alpha_1\alpha_2)(5+\mu_0\alpha_1)}{\phi_0 r c^2}+\frac{G m_2(1+\alpha_1\alpha_2)(5+\mu_0\alpha_2)}{\phi_0 r c^2}\right]\frac{(\v{x}_1\cdot\uv{N})(\v{n}\cdot\uv{N})}{r}\right.\\
&\left.+\frac{G m_2}{\phi_0 r c^2}\left[\frac{1}{2}\left(2-\left(\mu_0+\frac{B}{2\mu_0}+2 \phi_0\mu_0\frac{D(\log m_1 \alpha_1)}{D \phi}\right)\alpha_2+\alpha_1\alpha_2\right)\frac{v_1^2}{c^2}\right.\right.\\
&\qquad\left.\left.+\frac{1}{2}\left(1-\left(\mu_0+\frac{B}{2\mu_0}+2 \phi_0\mu_0\frac{D(\log m_1 \alpha_1)}{D \phi}\right)\alpha_2\right)\frac{v_2^2}{c^2}\right.\right.\\
&\qquad\left.\left.-\frac{1}{2}\left(3-\left(2\mu_0+\frac{B}{\mu_0}+4 \phi_0\mu_0\frac{D(\log m_1 \alpha_1)}{D \phi}\right)\alpha_2+\alpha_1\alpha_2\right)\frac{(\v{v}_1\cdot\v{v}_2)}{c^2}\right.\right.\\
&\qquad\left.\left.-\frac{3}{2}\left(2-\left(\mu_0+\frac{B}{2\mu_0}+2 \phi_0\mu_0\frac{D(\log m_1 \alpha_1)}{D \phi}\right)\alpha_2+\alpha_1\alpha_2\right)\frac{(\v{v}_1\cdot\v{n})^2}{c^2}\right.\right.\\
&\qquad\left.\left.-\frac{3}{2}\left(1-\left(\mu_0+\frac{B}{2\mu_0}+2 \phi_0\mu_0\frac{D(\log m_1 \alpha_1)}{D \phi}\right)\alpha_2\right)\frac{(\v{v}_2\cdot\v{n})^2}{c^2}\right.\right.\\
&\qquad\left.\left.+\frac{3}{2}\left(3-\left(2\mu_0+\frac{B}{\mu_0}+4 \phi_0\mu_0\frac{D(\log m_1 \alpha_1)}{D \phi}\right)\alpha_2+\alpha_1\alpha_2\right)\frac{(\v{v}_1\cdot\v{n})(\v{v}_2\cdot\v{n})}{c^2}\right.\right.\\
&\qquad\left.\left.-\frac{1}{2}\left(2+3\alpha_1\alpha_2+2\mu_0^2\alpha_2^2+\alpha_1^2\alpha_2^2 -\left(\mu_0+\frac{B}{2\mu_0}+2 \phi_0\mu_0\frac{D(\log m_1 \alpha_1)}{D \phi}\right)(1+\alpha_1\alpha_2)\alpha_2\right)\frac{Gm_2}{\phi_0 r c^2}\right.\right.\\
&\qquad\left.\left.-\frac{1}{2}\left(1+\alpha_1\alpha_2+2\mu_0^2\alpha_2^2 -\left(\mu_0+\frac{B}{2\mu_0}+2 \phi_0\mu_0\frac{D(\log m_1 \alpha_1)}{D \phi}\right)(1+\alpha_1\alpha_2)\alpha_2\right)\frac{Gm_1}{\phi_0 r c^2}\right]\frac{(\v{x}_1\cdot\uv{N})^2}{r^2}\right.\\
&+\left.\frac{G m_2}{\phi_0 r c^2}\left[\frac{\mu_0^2 v_1^2}{c^2}+\frac{\mu_0^2 v_2^2}{c^2}-\frac{2 \mu_0^2(\v{v}_1\cdot\v{v}_2)}{c^2}-\frac{3 \mu_0^2(\v{v}_1\cdot\v{n})^2}{c^2}-\frac{3 \mu_0^2(\v{v}_2\cdot\v{n})^2}{c^2}+\frac{6 \mu_0^2(\v{v}_1\cdot\v{n})(\v{v}_2\cdot\v{n}))}{c^2}\right.\right.\\
&\qquad\left.\left.-\frac{G \mu_0^2 m_2}{\phi_0 r c^2}-\frac{G \mu_0^2 m_1}{\phi_0 r c^2}\right]\frac{(\v{x}_2\cdot\uv{N})^2}{r^2}+\frac{(\v{v}_1\cdot\uv{N})^4}{c^4}-\frac{2 G m_2(1+\alpha_1\alpha_2)}{\phi_0 r c^2}\frac{(\v{x}_1\cdot \uv{N})^2(\v{v}_1\cdot\uv{N})^2}{r^2 c^2}\right.\\
&\left.-\frac{6 G m_2(1+\alpha_1\alpha_2)}{\phi_0 r c^2}\frac{(\v{x}_1\cdot \uv{N})(\v{v}_1\cdot\uv{N})^2(\v{n}\cdot\uv{N})}{r c^2}+\frac{2 G m_2(1+\alpha_1\alpha_2)}{\phi_0 r c^2}\frac{(\v{x}_1\cdot \uv{N})^2(\v{v}_1\cdot\uv{N})(\v{v}_2\cdot\uv{N})}{r^2 c^2}\right.\\
&\left.+\frac{6 G m_2(1+\alpha_1\alpha_2)}{\phi_0 r c^2}\left[\frac{(\v{v}_1\cdot\v{n})}{c}-\frac{(\v{v}_2\cdot\v{n})}{c}\right]\frac{(\v{x}_1\cdot \uv{N})^2(\v{v}_1\cdot\uv{N})(\v{n}\cdot\uv{N})}{r^2 c}\right.\\
&\left.+\frac{G m_2(1+\alpha_1\alpha_2)}{\phi_0 r c^2}\left[\frac{(\v{v}_1\cdot\v{n})}{c}-\frac{(\v{v}_2\cdot\v{n})}{c}\right]\frac{(\v{x}_1\cdot \uv{N})^3(\v{v}_1\cdot\uv{N})}{r^3 c}\right.\\
&\left.-\frac{G m_2(1+\alpha_1\alpha_2)}{\phi_0 r c^2}\left[\frac{(\v{v}_1\cdot\v{n})}{c}-\frac{(\v{v}_2\cdot\v{n})}{c}\right]\frac{(\v{x}_1\cdot \uv{N})^3(\v{v}_2\cdot\uv{N})}{r^3 c}+\frac{3G^2 m_2^2 (1+\alpha_1\alpha_2)^2}{2\phi_0^2 r^2 c^4}\frac{(\v{x}_1\cdot\uv{N})^2(\v{n}\cdot\uv{N})^2}{r^2}\right.\\
&\left.\frac{G m_2(1+\alpha_1\alpha_2)}{\phi_0 r c^2}\left[\frac{v_1^2}{2c^2}+\frac{v_2^2}{2c^2}-\frac{(\v{v}_1\cdot\v{v}_2)}{c^2}-\frac{5(\v{v}_1\cdot\v{n})^2}{2c^2}-\frac{5(\v{v}_2\cdot\v{n})^2}{2c^2}+\frac{5(\v{v}_1\cdot\v{n})(\v{v}_2\cdot\v{n})}{c^2}\right.\right.\\
&\qquad\left.\left.-\frac{G m_1(1+\alpha_1 \alpha_2)}{3 \phi_0 r c^2}-\frac{G m_2(1+\alpha_1 \alpha_2)}{3 \phi_0 r c^2}\right]\frac{(\v{x}_1\cdot\uv{N})^3(\v{n}\cdot\uv{N})}{r^3}\right\}+ \left(1\rightleftharpoons 2\right),\label{eq:1PDScalarMass}
\end{split}
\end{align}
\end{widetext}
where we have dropped the pieces that do not contribute to the scalar mass and have used Eqs. (\ref{eq:NewtonianEOM1}) and (\ref{eq:NewtonianEOM2}) to eliminate the bodies' accelerations.

To the order at which we work, the radiation-zone contribution to scalar mass is zero. The scalar monopole $\mathcal{I}_s$ is the only multipole that enters in Eqs. (\ref{eq:FarZoneLNContribution1})--(\ref{eq:FarZoneLNContribution3}); as discussed above, at leading order, the monopole is constant in time. This insight allows one to trivially evaluate these hereditary integrals. The non-zero terms either  depend on the arbitrarily chosen boundary $\mathcal{R}$ (and thus are canceled by near-zone contributions to the scalar field) or fall off too quickly with $R$ to contribute to the scalar mass.

Computing the scalar mass from the scalar field given in Eqs. (\ref{eq:NearZoneFieldTot})--(\ref{eq:1PDScalarMass}) is most easily done in the center of mass frame. However, we cannot compute the exact transformation to this frame in the PD formalism without first calculating the total momentum of the system.\footnote{In the PN formalism, the transformation to the center of mass frame is derived by forcing the total momentum of the binary system to vanish. The momentum is difficult to calculate within the PD approach because the equations of motion cannot be derived from a Lagrangian dependent solely on the particles' positions and velocities. Thus, the exact transformation to the center of mass frame remains unknown.} Instead, we consider frames in which the two bodies' positions are related by $\v{x}_1\propto -\v{x}_2$. Without dissipative effects, we expect the center of mass frame to satisfy this criterion.

Furthermore, we restrict our attention to binary systems undergoing circular motion. Neutron-star binaries are expected to radiate away any eccentricity relatively early in their evolution, long before they would be detectable by ground-based experiments like LIGO, thereby justifying this approximation.

We plug the expression for the scalar field in Eq. (\ref{eq:NearZoneFieldTot}) into Eq. (\ref{eq:ScalarMassDef2}) to obtain the scalar mass. This surface integral can be computed easily using the standard angular coordinates $(\theta,\phi)$ on the coordinate sphere of radius $R$. The scalar mass takes the exact same form as the scalar field with the $\uv{N}$-dependent terms replaced by the geometric quantities derived below
\begin{widetext}
\begin{align}
-\oint_{R\rightarrow \infty} \partial_i\left(\frac{f(\theta,\phi)}{R} \right)dS_i=\int f(\theta,\phi) d(\cos\theta)d\phi,
\end{align}
and
\begin{align}
\int (\uv{N} \cdot \v{x}_A)(\uv{N} \cdot \v{x}_B) d(\cos\theta)d\phi=&\frac{4 \pi}{3}\tilde{\gamma}_{AB} x_A x_B,\\
\int (\uv{N} \cdot \v{v}_A)(\uv{N} \cdot \v{v}_B) d(\cos\theta)d\phi=&\frac{4 \pi}{3}\tilde{\gamma}_{AB} v_A v_B,\\
\int (\uv{N} \cdot \v{x}_A)(\uv{N} \cdot \v{x}_B)(\uv{N} \cdot \v{x}_C)(\uv{N} \cdot \v{x}_D) d(\cos\theta)d\phi=&\frac{4 \pi}{5}\tilde{\gamma}_{AB}\tilde{\gamma}_{CD} x_A x_B x_C x_D,\\
\int (\uv{N} \cdot \v{v}_A)(\uv{N} \cdot \v{v}_B)(\uv{N} \cdot \v{v}_C)(\uv{N} \cdot \v{v}_D) d(\cos\theta)d\phi=&\frac{4 \pi}{5}\tilde{\gamma}_{AB}\tilde{\gamma}_{CD} v_A v_B v_C v_D,\\
\int (\uv{N} \cdot \v{x}_A)(\uv{N} \cdot \v{x}_B)(\uv{N} \cdot \v{v}_C)(\uv{N} \cdot \v{v}_D) d(\cos\theta)d\phi=&\frac{4 \pi}{15}\tilde{\gamma}_{AB}\tilde{\gamma}_{CD} x_A x_B v_C v_D,
\end{align}
where we have defined
\begin{align}
\tilde{\gamma}_{AB}\equiv\begin{cases}1, &\text{if }A=B\\-1,&\text{if }A\ne B\end{cases}.
\end{align}
The scalar mass is given by
\begin{align}
M_S&=\frac{m_1\alpha_1 \mu_0}{\phi_0}\left[1-\frac{v_1^2}{6 c^2}-\frac{G m_2}{\phi_0 r c^2}\left(1+\mu_0 \alpha_2 +\left(\frac{1+\alpha_1\alpha_2}{3}\right)\frac{r_1}{r}\right)\right]+\Big[\text{1PD}\Big]+ \left(1\rightleftharpoons 2\right),\label{eq:ScalarMass1PS}
\end{align}
\end{widetext}
where the 1PD terms are represented only schematically for the sake of compactness.
\section{Validity of the Post-Dickean expansion} \label{sec:Discussion}

\begin{figure*}
\includegraphics[width=\columnwidth,clip=true, trim=0 5 0 0]{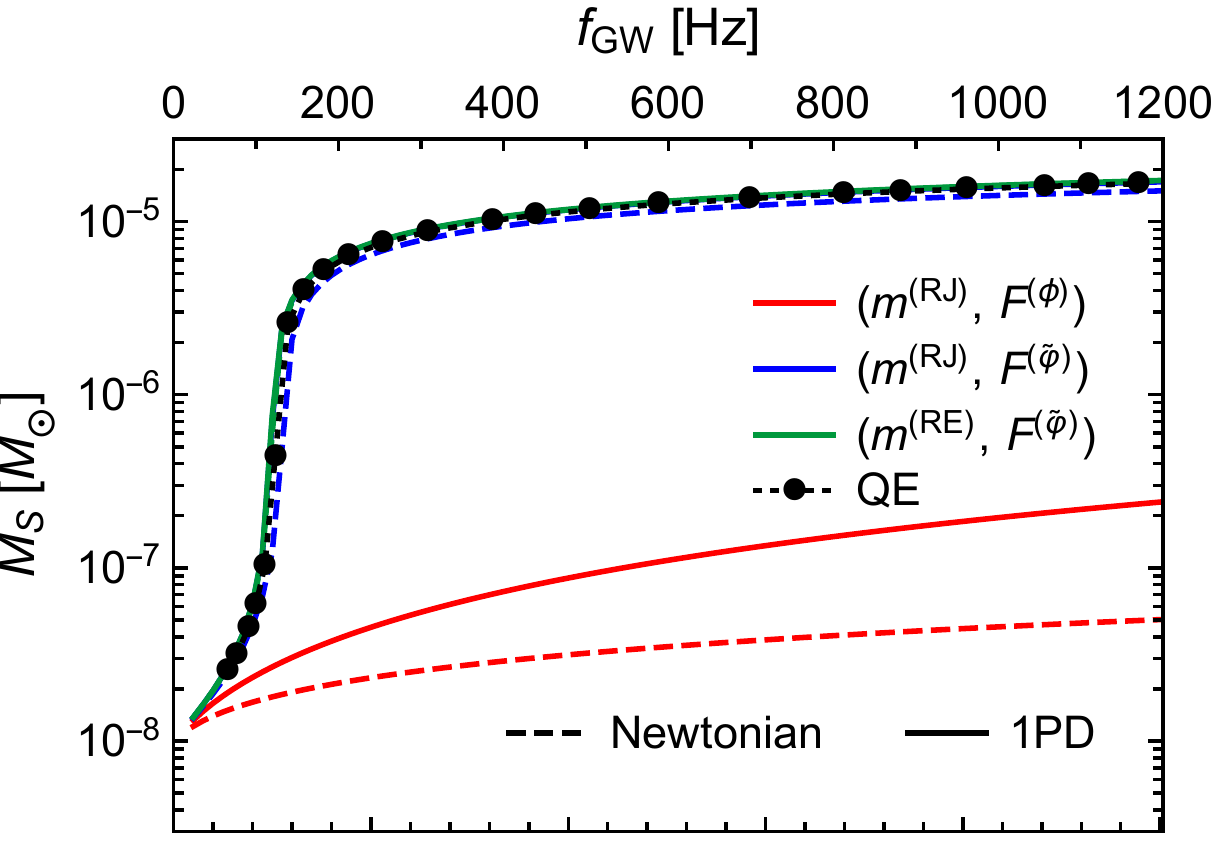}
\hspace{.75\columnsep}
\includegraphics[width=\columnwidth,clip=true, trim=0 5 0 0]{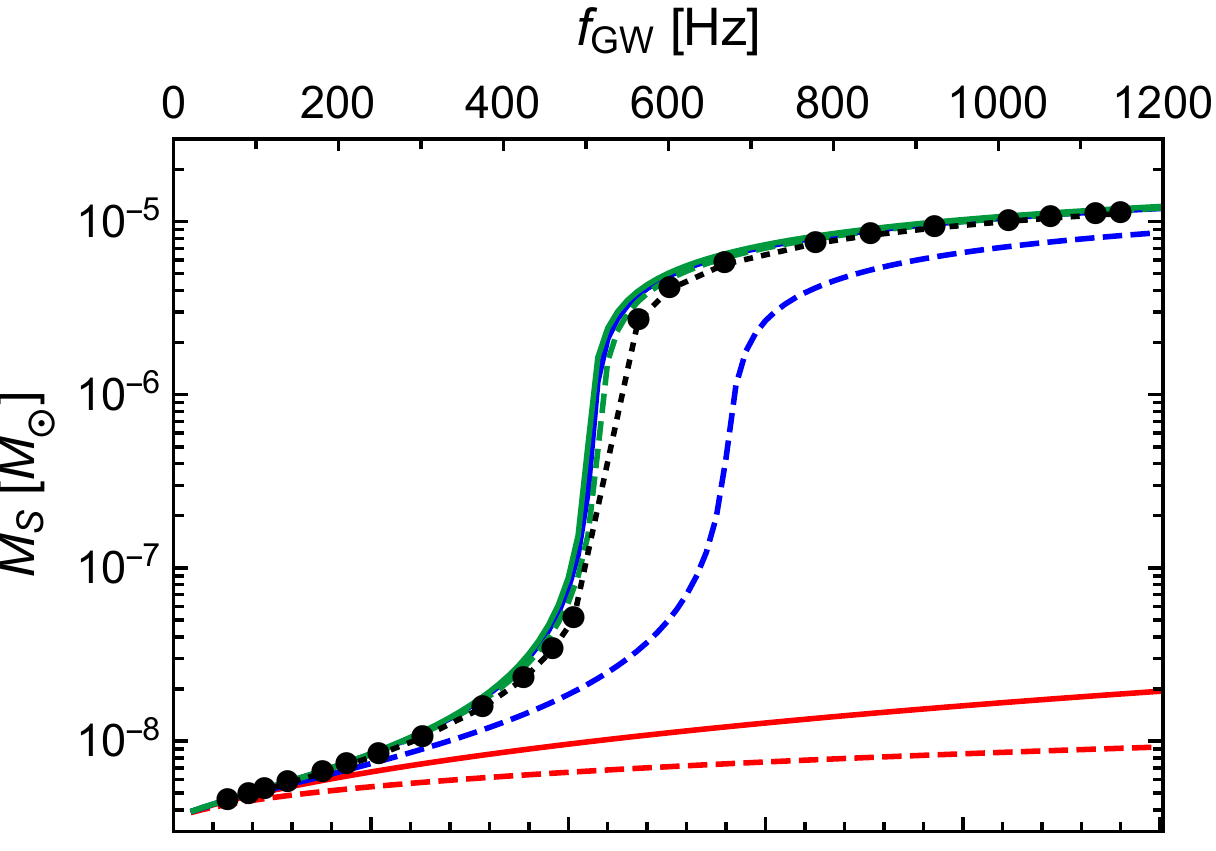}\\
\includegraphics[width=\columnwidth,clip=true, trim=0 0 0 0]{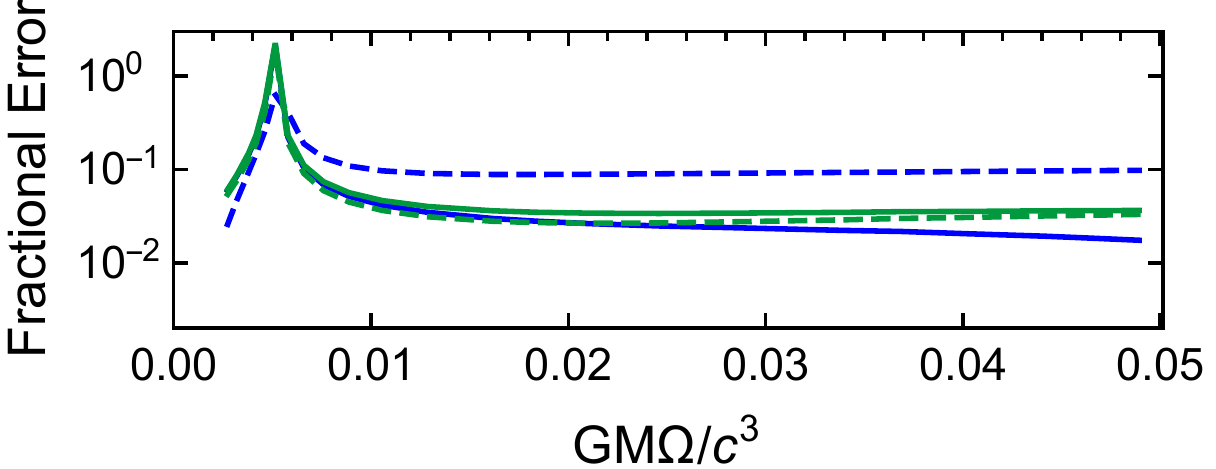}
\hspace{.75\columnsep}
\includegraphics[width=\columnwidth,clip=true, trim=0 0 0 0]{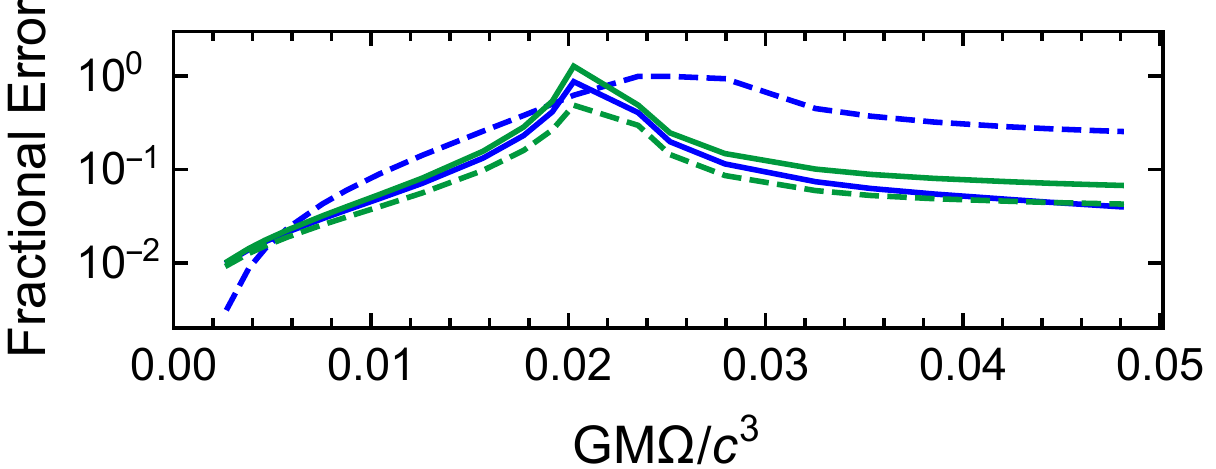}
\caption{Scalar mass of a $(1.35+1.35) M_\odot$ neutron-star binary system on a circular orbit as a function of the orbital angular frequency and gravitational wave frequency $(f_\text{GW}=\Omega/\pi)$. The scalar mass is computed at Newtonian (dashed) and 1PD (solid) order for resummation schemes listed in Table \ref{table:ResummationSchemes}. We also plot the quasi-equilibrium configuration calculations (QE) reported in Ref. \cite{Taniguchi2014} (dotted). The bottom panels depict the magnitude of the fractional error between the PD and quasi-equilibrium results. We use the APR4 equation of state with (left) $B=9, \,\tilde\varphi_0=3.33\times 10^{-11}$ and (right) $B=8.4, \,\tilde\varphi_0=3.45\times 10^{-11}$.}\label{fig:ResummationComparison}
\end{figure*}

The PD expansion was motivated through analogy: spontaneous and dynamical scalarization are suspected to arise from similar mechanisms, and so the analytic techniques applied to the former (resummation of expansions in $Gm/Rc^2$) should also be used with latter (partial resummation of expansions in $Gm/rc^2$). While such reasoning seems plausible, ultimately, the validity of our model can only be checked via comparison with high-precision numerical calculations. In absence of long numerical-relativity simulations of DS, we compare the PD approximation to recent quasi-equilibrium configuration calculations. We also closely examine the differences between the PD approximation and the analytic model proposed in Ref. \cite{Palenzuela2014} for completeness

\subsection{Quasi-equilibrium configurations}

The scalar mass of an equal-mass binary system was calculated along sequences of quasi-equilibrium configurations in Ref. \cite{Taniguchi2014}. Inherent to these calculations is the assumption of a conformally flat and stationary spacetime; physically, each configuration represents a binary system following a circular orbit. Despite neglecting the loss of energy and angular momentum through the emission of gravitational radiation, this setup is believed to closely resemble the adiabatic inspiral of a neutron-star binary system. Systematic errors enter these quasi-equilibrium calculations through the physical assumptions made above and imperfect numerical convergence, particularly at higher frequencies. At present, the magnitude of these errors is not well-understood.

We compare the PD predictions of the scalar mass with these numerical results to validate the accuracy of the model.  Figure \ref{fig:ResummationComparison} depicts the scalar mass as a function of orbital frequency $\Omega$ for a (1.35+1.35) $M_\odot$ binary system, where the PD corrections to Kepler's third law for an equal mass system (derived from the equations of motion)
\begin{align}
\begin{split}
\Omega^2=&\frac{G M(1+\alpha^2)}{r^3 \phi_0}-\frac{G^2 M^2(1+\alpha^2)(11+2\mu_0 \alpha+\alpha^2 )}{4 r^4 \phi_0^2 c^2},\label{eq:Kepler}
\end{split}
\end{align}
are used to replace the $r$-dependence in Eq. (\ref{eq:ScalarMass1PS}), and where $M=m_1+m_2$ and $\alpha=\alpha_1=\alpha_2$. The scalar mass is computed at Newtonian (dashed) and 1PD (solid) order; note that the former calculation is done consistently at Newtonian order [e.g. only the first term in Eq. (\ref{eq:Kepler}) is used]. We employ the APR4 equation of state, for which the allowed range of theory parameters in which DS can occur is spanned by $B\in[8,9]$ (see Ref. \cite{Shibata2014} for more detail). From this range, we focus on the cases $B=9$ and $B=8.4$, corresponding to the choices $\tilde{\varphi}_0=3.33\times10^{-11}$ and $\tilde{\varphi}_0=3.45\times10^{-11}$ considered in Ref. \cite{Taniguchi2014}. For all PD calculations, we use a Newton-Raphson method to numerically solve Eqs. (\ref{eq:Xi1PS}) and (\ref{eq:Xi1PS2}) to within a fractional error of $10^{-7}$.

Recall that the PD expansion encodes a flexibility in ``what to resum'' in the choice of $m(\phi,\xi)$ and $F(\phi)$. We compare each combination of the choices in Table \ref{table:ResummationSchemes} in Fig. \ref{fig:ResummationComparison}, denoting each resummation scheme by the pair $(m,F)$. The scalar mass estimated with the $(m^{(\text{RJ})},F^{(\phi)})$ and $(m^{(\text{RE})},F^{(\phi)})$ resummation schemes differ by only $\sim 0.01 \%$; to improve legibility, we only plot the former (in red).

The two most important features depicted in Fig. \ref{fig:ResummationComparison} that we hope to recover with our model are the frequency at which DS occurs $\Omega_\text{DS}$ and the magnitude of the scalar mass after scalarization. We extract the onset of DS from the figure using the fitting procedure detailed in Ref. \cite{Taniguchi2014}; these values are given in Table \ref{table:DSFreq}. One finds that the scalar mass $M_S$ can be well approximated by
\begin{align}
\left(1+\left(\frac{M_S}{M \mu_0}\right)^2\right)^{10/3}=\begin{cases}1,&\text{if }\Omega<\Omega_\text{DS}\\\
a_0+a_1 x,&\text{if }\Omega>\Omega_\text{DS}\end{cases}
\end{align}
where $x\equiv \left(G M \Omega/c^3\right)^{2/3}$. We determine the coefficients $a_0$ and $a_1$ by fitting the high frequency part of the curves in Fig. \ref{fig:ResummationComparison} and then find $\Omega_\text{DS}$ from the intersection of this linear function with 1.

The 1PD predictions for both the location and magnitude of scalarization match the results of Ref. \cite{Taniguchi2014} at the $\lesssim10\%$ level for the choice $F^{(\tilde\varphi)}$. (Note that the peaks in the relative error seen in the bottom panels of Fig. \ref{fig:ResummationComparison} stem from the slight misalignment of the scalar mass predictions at the sharp onset of DS.) Interestingly, for systems that scalarize later in the inspiral (i.e. smaller values of $B$), the Newtonian order prediction in the ($m^{(\text{RE})},F^{(\tilde\varphi)}$) scheme agrees more closely with the numerical results. Without a more comprehensive study of various resummation schemes or the PD expansion at higher order, it is difficult to say whether this agreement is coincidental.

The choice of $m(\phi,\xi)$ seems to have little effect on the scalar mass predictions of the PD model. The two resummation schemes with $F^{(\phi)}$ are essentially indistinguishable, while the schemes with  $F^{(\tilde\varphi)}$ appear to converge to within a few percent at 1PD order.

\begin{table}[t]\centering
\caption{Orbital angular frequency and gravitational wave frequency at which dynamical scalarization occurs $(f^\text{GW}=\Omega/\pi)$ for the systems considered in Fig. \ref{fig:ResummationComparison}. Only resummation schemes with the choice $F^{(\tilde\varphi)}$ produce DS. For comparison, we list the results of the quasi-equilibrium configuration calculations (QE) of Ref. \cite{Taniguchi2014}.}\label{table:DSFreq}
\begin{ruledtabular}
\begin{tabular}{C{.1\columnwidth} C{.22\columnwidth} C{.2\columnwidth}C{.22\columnwidth}C{.15\columnwidth}} 
$B$&Model&Order&$G M \Omega_{\text{DS}}/c^3$&$f^\text{GW}_\text{DS}$ [Hz]\\
\midrule
9.0&$(m^{(\text{RJ})},F^{(\tilde\varphi)})$&Newtonian&0.0044&106\\
9.0&$(m^{(\text{RJ})},F^{(\tilde\varphi)})$&1PD&0.0047&112\\
9.0&$(m^{(\text{RE})},F^{(\tilde\varphi)})$&Newtonian&0.0052&124\\
9.0&$(m^{(\text{RE})},F^{(\tilde\varphi)})$&1PD&0.0051&122\\
9.0&QE&\-------&0.0051&123\\
\midrule
8.4&$(m^{(\text{RJ})},F^{(\tilde\varphi)})$&Newtonian&0.0282&674\\
8.4&$(m^{(\text{RJ})},F^{(\tilde\varphi)})$&1PD&0.0212&508\\
8.4&$(m^{(\text{RE})},F^{(\tilde\varphi)})$&Newtonian&0.0217&520\\
8.4&$(m^{(\text{RE})},F^{(\tilde\varphi)})$&1PD&0.0212&508\\
8.4&QE&\-------&0.0223&534
\end{tabular}
\end{ruledtabular}
\end{table}

\begin{figure}
\includegraphics[width=\columnwidth, clip=true, trim=0 0 0 17]{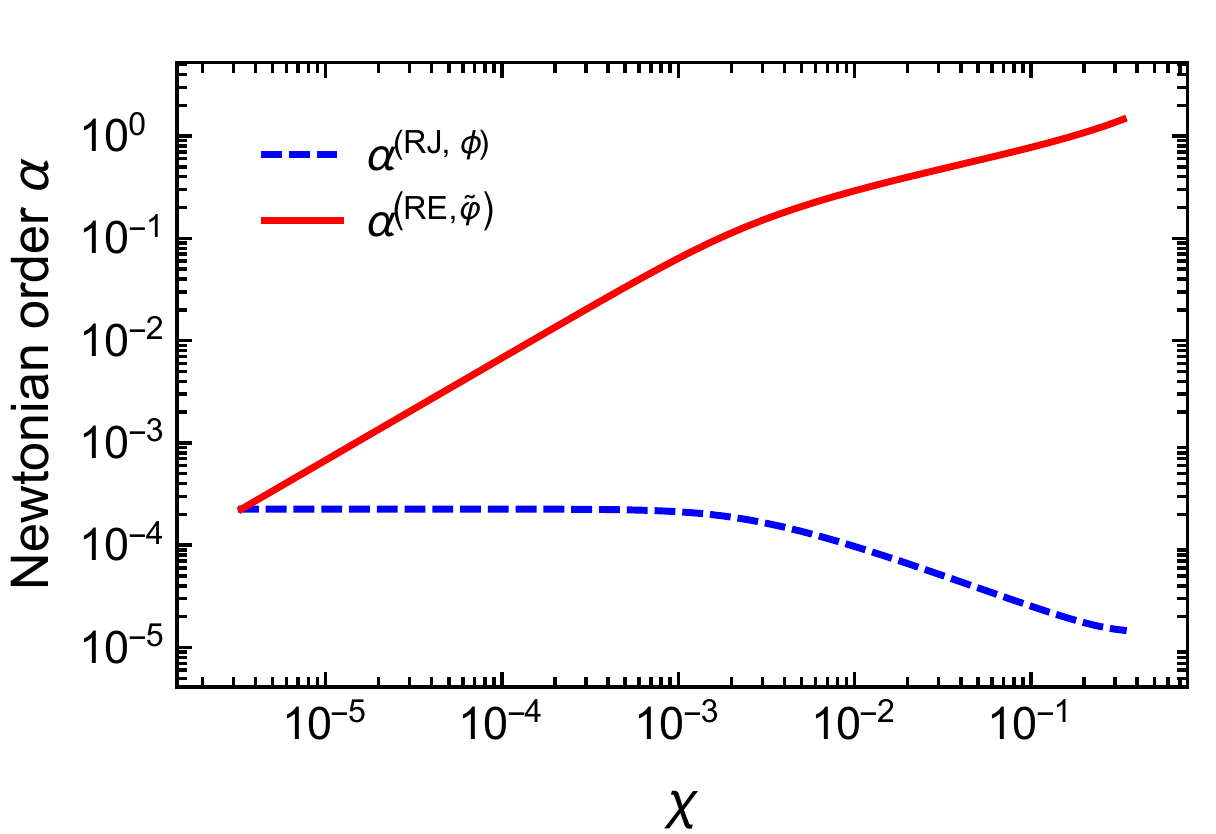}
\caption{Newtonian order contribution to the scalar charge $\alpha$ of each neutron star in a $(1.35+1.35) M_\odot$ binary system as a function of the auxiliary field $\chi$ in the $(m^{(\text{RJ})},F^{(\phi)})$ and $(m^{(\text{RE})},F^{(\tilde\varphi)})$ resummation schemes. We use the APR4 equation of state with $B=9,\, \tilde\varphi_0=3.33\times 10^{-11}$.}\label{fig:ChargeComparisonPlot}
\end{figure}

On the other hand, the choice of $F(\phi)$ drastically alters the growth of the scalar mass. Of the two options presented in Table \ref{table:ResummationSchemes}, only $F^{(\tilde\varphi)}$ reproduces the sharp transition consistent with dynamical scalarization. The significance of the choice of $F$ can be seen by studying the behavior of the scalar charge $\alpha(\phi,\xi)$. Because the definition of $\xi$ relies on the choice of resummation scheme [see  Eq. (\ref{eq:JordanConstraint})], we invert this definition and instead consider the dependence of the charge on an auxiliary field $\chi$ that is the same in all resummation schemes, defined as
\begin{align}
\chi\equiv\sqrt{\frac{2 \log (F^{-1}(\xi))}{B}}=\begin{cases}\sqrt{\frac{ 2 \log \xi}{B}},&\text{if } F(\phi)=\phi\\
\xi,&\text{if } F(\phi)=\sqrt{\frac{2 \log \phi}{B}}\end{cases}\label{eq:Chi}
\end{align}
Figure \ref{fig:ChargeComparisonPlot} shows the leading order piece of the scalar charge $\alpha$ in the  $(m^{(\text{RJ})},F^{(\phi)})$ and $(m^{(\text{RE})},F^{(\tilde\varphi)})$ resummation schemes given in Eqs. (\ref{eq:NewtonianAlpha1}) and (\ref{eq:NewtonianAlpha2}). The resummed scalar charges in each scheme agree at $\chi=\tilde\varphi_0$, but they scale as
\begin{align}
\alpha^{(\text{RJ},\phi)}&\sim \d{\log m}{\xi}\sim \d{\log m}{\chi}\frac{1}{B \chi} e^{-B \chi^2/2}\sim \d{\log m}{\chi}\frac{1}{\chi},\label{eq:ChargeScale1}\\
\alpha^{(\text{RE},\tilde\varphi)}&\sim \d{\log m^{(E)}}{\xi}\sim \d{\log m}{\chi}-\frac{B \chi}{2}\sim  \d{\log m}{\chi},\label{eq:ChargeScale2}
\end{align}
where we have used the fact that $\chi\ll 1$.

Without the additional factor of $\chi^{-1}$, the scalar charge in the $(m^{(\text{RE})},F^{(\tilde\varphi)})$ scheme grows with the local scalar field (the red curve in Fig. \ref{fig:ChargeComparisonPlot}). This trend enables a positive feedback loop that ultimately emulates DS \cite{Palenzuela2014}. Intuitively, an increase in the field $\chi$ at one body increases its charge $\alpha$, which, in turn, increases the field $\chi$ at the other body (and so on). No such feedback is possible within the $(m^{(\text{RJ})},F^{(\phi)})$ resummation scheme because $\alpha$ does not increase with greater $\xi$.

\subsection{Earlier analytic models}\label{sec:Analytic}

\begin{figure*}[t]
\includegraphics[width=\columnwidth,clip=true, trim=0 5 0 0]{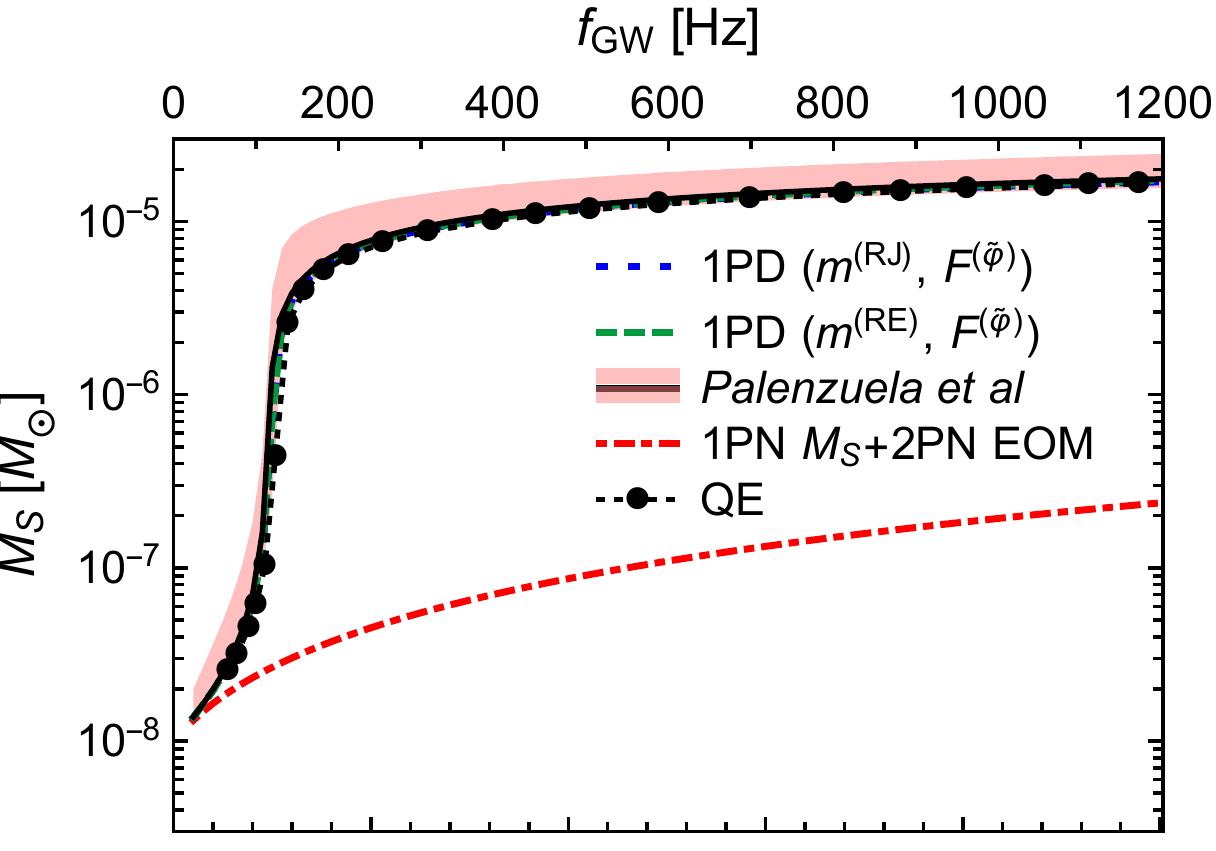}
\hspace{.75\columnsep}
\includegraphics[width=\columnwidth,clip=true, trim=0 5 0 0]{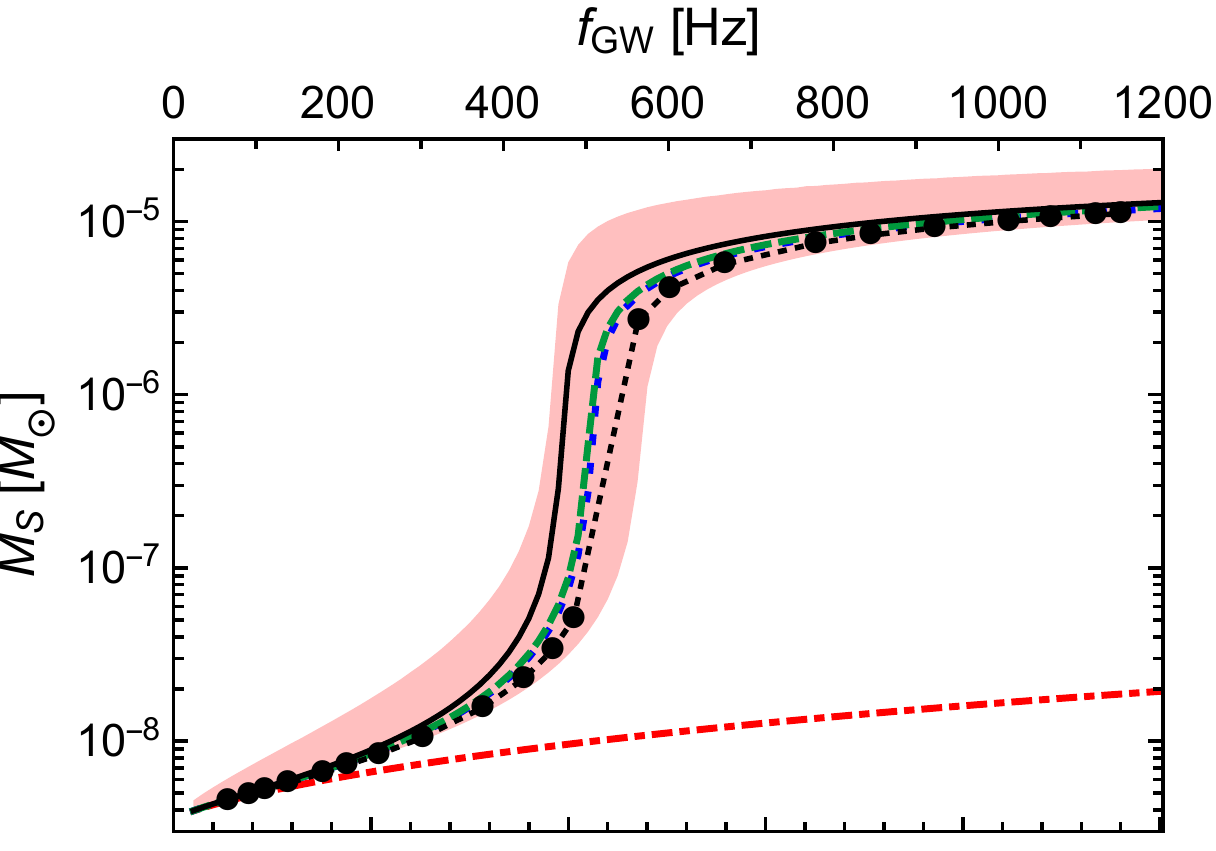}\\
\includegraphics[width=\columnwidth,clip=true, trim=0 0 0 0]{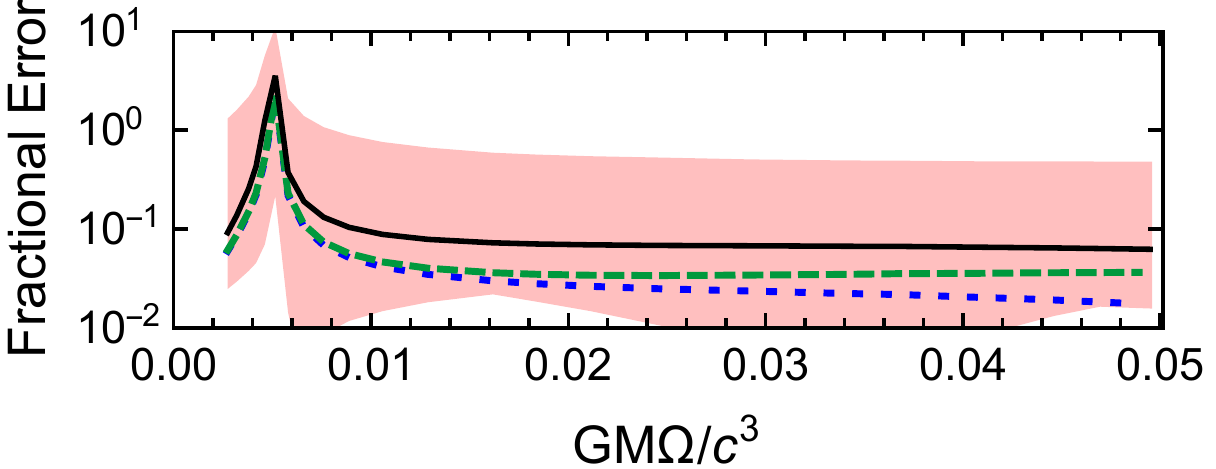}
\hspace{.75\columnsep}
\includegraphics[width=\columnwidth,clip=true, trim=0 0 0 0]{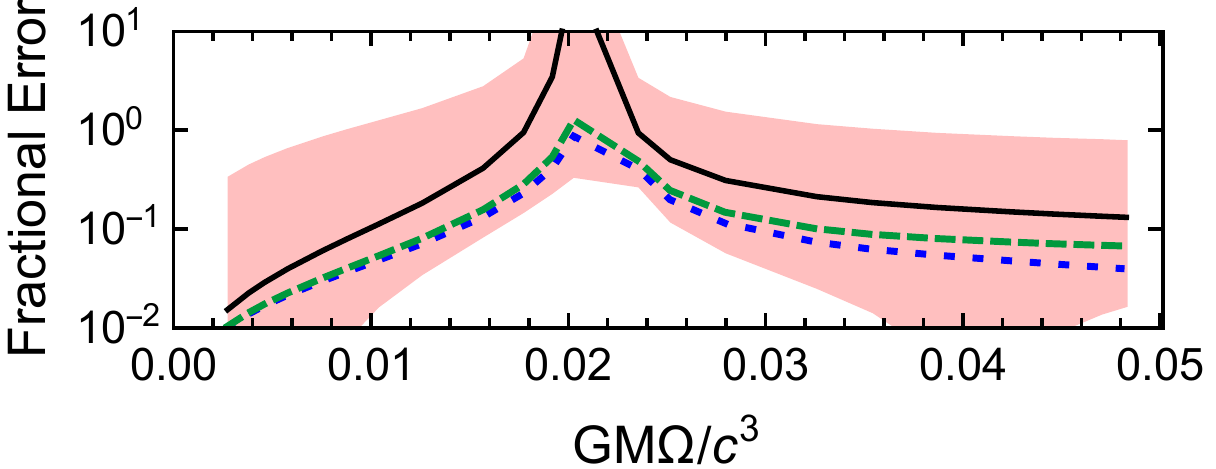}
\caption{Scalar mass as a function of orbital frequency $\Omega$ and gravitational wave frequency $f_\text{GW}$ of the binary system depicted in Fig. \ref{fig:ResummationComparison}. The post-Dickean curves are calculated at 1PD order with the resummation schemes using $F^{(\tilde\varphi)}$. The model proposed in Ref. \cite{Palenzuela2014} is plotted in black alongside the variations that we develop in Appendix \ref{sec:Palenzuela}, which are collectively depicted by the pink region. For comparison, we plot the 1PN scalar mass (red) computed using the results of Ref. \cite{Lang2014a} (we use the 2PN equations of motion of Ref. \cite{Mirshekari2013} to restrict to circular orbits). The bottom panels depict the magnitude of the fractional error between the models and the quasi-equilibrium configurations (QE) of Ref. \cite{Taniguchi2014}.}\label{fig:ScalarMassPlot1}
\end{figure*}

The first analytic model of DS was proposed in Ref. \cite{Palenzuela2014}. This model used the 2.5PN equations of motion computed in Ref. \cite{Mirshekari2013}, but altered the coefficients using a feedback mechanism designed to mimic DS. To 1PN order, these modified equations of motion are given in Eqs. (\ref{eq:PNEOM1})--(\ref{eq:Alphabar}) but with the 
important difference that $\bar m_i$ and $\bar \alpha_i$ are evaluated at an enhanced field value $\varphi_B$ instead of at $\phi_0$. To determine $\varphi_B$ the authors numerically solved the Newtonian order relations
\begin{align}
\varphi_B^{(1)}&=\tilde\varphi_0+\frac{G \bar m_2(\varphi_B^{(2)})\bar\alpha_2(\varphi_B^{(2)})}{\phi_0 r c^2},\label{eq:PalenzuelaVarphi}\\
\varphi_B^{(2)}&=\left(1\rightleftharpoons 2\right).\label{eq:PalenzuelaVarphi2}
\end{align}
Additionally, the authors explicitly set the derivatives of the scalar charge $\bar\alpha'$, $\bar  \alpha''$ to zero.

As reported in Ref. \cite{Palenzuela2014}, this model captures dynamical scalarization and produces results (qualitatively) consistent with numerical-relativity simulations. The model is easily implemented because it directly augments the PN results of Ref. \cite{Mirshekari2013} with Eqs. (\ref{eq:PalenzuelaVarphi}) and (\ref{eq:PalenzuelaVarphi2}). However, mixing the Newtonian order feedback mechanism with higher-order equations of motion produces technical ambiguities in the model; we address these uncertainties in greater detail in Appendix \ref{sec:Palenzuela}.

Comparing Eq. (\ref{eq:NewtonianXi}) with Eq. (\ref{eq:PalenzuelaVarphi}), we immediately see that our PD approach recovers the feedback mechanism of Ref. \cite{Palenzuela2014} at Newtonian order with the resummation schemes that use $F^{(\tilde\varphi)}$. Similarly, comparing Eq. (\ref{eq:NewtonianEOM1}) with Eq. (\ref{eq:PNEOM1}), we see that the equations of motion for the binary system agree at Newtonian order with those of Ref. \cite{Palenzuela2014} provided we also use $m^{(\text{RJ})}$.

Disparities arise between the two formalisms beyond Newtonian order. For the same resummation scheme adopted above, the auxiliary field given in Eq. (\ref{eq:Xi1PS}) is the natural extension of the feedback model of Ref. \cite{Palenzuela2014} to higher order. Beyond the difference between $\xi$ and $\varphi_B$, the equations of motion of each approach [Eq. (\ref{eq:1PSEOM1}) and Eq. (\ref{eq:PNEOM1})] differ only in the terms proportional to $r^{-2}$ (recall that $m_i$ and $\alpha_i$ receive PD corrections as discussed in Appendix \ref{sec:Charge}). However, as discussed at the end of Sec. \ref{sec:EOM}, we expect a greater proportion of terms in the model of Ref. \cite{Palenzuela2014} to disagree with the PD equations of motion beyond post-Newtonian order.

To provide some context of the PD expansion's place relative to previous models, the scalar mass predicted by each of the analytic approximations discussed above is plotted in Fig. \ref{fig:ScalarMassPlot1}. As discussed in Sec \ref{sec:DS}, the unaltered PN approximation (denoted in red) does not reproduce DS, giving a scalar mass orders of magnitude smaller than numerical predictions. In contrast, the PD approximation (blue and green) agrees with the quasi-equilibrium calculations (dotted black) reported in Ref. \cite{Taniguchi2014} at the level of $\lesssim10\% $ when equipped with the proper resummation scheme. This level of accuracy is comparable to that achieved by the analytic model proposed in Ref. \cite{Palenzuela2014} (solid black). In addition, the technical ambiguities found in this earlier model (see Appendix \ref{sec:Palenzuela}) generate some systematic uncertainty in its predictions. As a rough estimate of this uncertainty, we denote with the pink region the range of values spanned by all of the alternatives considered in Figs. \ref{fig:PalenzuelaScalarMass} and \ref{fig:PalenzuelaScalarMass1PN}. The PD formalism alleviates this issue by resumming the PN approximation in a mathematically consistent way, albeit with a freedom in the exact choice of quantities to resum.
\section{Conclusions}\label{sec:Conclusions}

In this paper, we proposed the post-Dickean expansion, a new model of dynamical scalarization constructed by resumming the post-Newtonian expansion. The motivation for this approach stems from the success of previous analytic treatments of spontaneous scalarization, a phenomenon suspected to be closely related to DS. By appropriating tools from recent PN calculations \cite{Mirshekari2013,Lang2014a,Lang2014b}, we derived the equations of motion and the scalar mass (a measure of scalarization) of a binary system at post-Newtonian order. Comparisons with recent numerical results \cite{Taniguchi2014} indicate that our new formalism captures DS accurately. The PD model exactly coincides with the analytic model introduced in Ref. \cite{Palenzuela2014} at leading order, but the ambiguities that arise at higher order in that earlier work are avoided with the PD approach because of its more rigorous and self-consistent formulation.

While this work establishes a framework for modeling DS, several further steps remain before it can be used to generate waveforms needed to test GR with GW detectors. Fortunately, most of these remaining calculations are straightforward, albeit lengthy. The waveform was recently computed to 2PN order in Ref. \cite{Lang2014b, Marsat2016}.\footnote{Advanced LIGO is most sensitive to a GW's transverse-traceless polarizations, for which the 2PN calculation was done. An additional transverse ``breathing'' mode would accompany the signal; this third polarization is determined by $\Psi$ and has only been computed to 1.5PN order \cite{Lang2014a}.} Similarly to what was done in Secs. \ref{sec:NZfields} and \ref{sec:FZfields}, the PD waveform can be calculated in precisely the same way as the PN result with a slightly modified stress-energy tensor. To reach the 2PD accuracy, one would also need to derive the equations of motion at that order. Again, all of the necessary steps have been completed for the PN calculation \cite{Mirshekari2013}, so one can simply recycle that work with a new stress-energy tensor to produce the corresponding PD result.

The evolution of a binary system directly impacts the GW signal it produces. Thus, in conjunction with the waveform calculation sketched above, one would need to estimate the phase evolution of a binary in the PD formalism. One approach, analogous to what was done in Refs. \cite{Palenzuela2014,Sampson2014a}, would be to directly integrate the equations of motion. However, earlier surveys of PN models in GR indicate that such a procedure can produce unreliable waveforms \cite{Buonanno2003}.
Instead, a better approximation can be found by balancing the change in the (conservative) binding energy and the radiated flux far from the system. The flux was computed to 1PN order in Ref. \cite{Lang2014a}; this calculation could be redone in the PD expansion with a modified stress-energy tensor.

Unfortunately, the PD binding energy cannot be easily recomputed with existing PN work. To date, this energy has been calculated in the PN approach by integrating the (conservative) equations of motion to produce a Lagrangian and performing a Legendre transformation. However, as discussed at the end of Sec. \ref{sec:EOM}, no such Lagrangian exists for the PD equations of motion because of the presence of the auxiliary field $\xi$. Without this shortcut, one would need to instead calculate the ADM energy at spatial infinity. To our knowledge, the full asymptotic metric has not been computed at spatial infinity to any PN order for the class of ST theories we consider. In principle, the 1PD energy could be estimated at null infinity because the system is fully conservative up to that order, but the results of Ref. \cite{Lang2014b} would have to be considerably extended, as the author computed only the traceless piece of the asymptotic metric. A more systematic approach should mimic the PN calculation of the ADM Hamiltonian in GR \cite{Schafer1985}, in which all of the gravitational degrees are integrated out, leaving an energy dependent only on each body's position, momentum, and local scalar field $\xi$.

Besides the litany of PN results that need to be recomputed in the PD formalism to produce waveforms, the model could offer a better physical understanding of DS. Surprisingly, we found that the PD predictions were largely independent of the choice $m(\phi,\xi)$ in the resummation scheme. While this result needs to be confirmed with a more comprehensive survey of possible schemes, the dependence of our formalism on the sole function $F(\phi)$ suggests that DS could be modeled with a single effective potential for the scalar charge at the level of the action. An analogous method was employed in Ref. \cite{Hinderer2016}, in which the quadrupole modes of a neutron star were promoted to dynamical field variables governed by an effective potential to model their response to the tidal fields produced by a companion black hole. This procedure could be adopted for dynamical scalarization, where each body's scalar monopole (i.e. scalar charge) dynamically responds to the monopolar scalar field sourced by the companion star \cite{Steinhoff}. This investigation could offer a more intuitive view of DS as a non-linear phenomenon.

\section*{Acknowledgements}
We are grateful to Gilles Esposito-Far\`{e}se, Jan Steinhoff, Enrico Barausse, and Nicol\`{a}s Yunes for useful discussions and comments. N.S. and partially A.B. acknowledge support from NSF Grant No. PHY-1208881.  A.B. also acknowledges partial support from NASA Grant NNX12AN10G. N.S. thanks the Max Planck Institute for Gravitational Physics for its hospitality during the completion of this work.

\appendix

\section{Post-Dickean expansion of mass and scalar charge}\label{sec:Charge}
The exact form of the mass $m$ and the scalar charge $\alpha$ defined in Eq. (\ref{eq:ShorthandDef}) depend on how the mass is resummed, that is, on our choice of $m(\phi,\xi)$ and $F(\phi)$. For example, using the expressions in $m^{(\text{RJ})}$ and $F^{(\phi)}$ given in Table \ref{table:ResummationSchemes}, the mass and scalar charge are given at 1PD order by 
\begin{widetext}
\begin{align}
m_A^{(\text{RJ},\phi)}=&m_A(\xi),\label{eq:NewtonianM1}\\
\begin{split}
\alpha_A^{(\text{RJ},\phi)}=&\mu_0\left(1-2 \phi_0 \d{\log m_A}{\xi}\right)+\frac{B\mu_0}{2}\left(1-2 \phi_0 \d{\log m_A}{\xi}\right)\left(1-2 \phi_0 \d{\log m_B}{\xi}\right)\frac{G m_B}{\phi_0 r c^2}\\
&-4{\mu_0}^3\left(\d{\log m_A}{\xi}\right)\left(1-2 \phi_0 \d{\log m_B}{\xi}\right)\frac{G m_B}{ r c^2}+\Ord\left(\frac{1}{c^4}\right),\label{eq:NewtonianAlpha1}
\end{split}
\end{align}
where $A\neq B$, while the choice of $m^{(\text{RE})}$ and $F^{(\tilde\varphi)}$ gives
\begin{align}
m_A^{(\text{RE},\varphi)}&=\sqrt{\phi_0}m_A^{(E)}(\xi)\left(1+\frac{ G m_B^{(E)} \mu_0 \alpha_B}{\sqrt{\phi_0} r c^2}\right)+\Ord\left(\frac{1}{c^4}\right),\label{eq:NewtonianM2}\\
\alpha_A^{(\text{RE},\varphi)}&=-\d{\log m_A^{(E)}(\xi)}{\xi} \label{eq:NewtonianAlpha2}.
\end{align}
\end{widetext}
Note that the expressions in Eqs. (\ref{eq:NewtonianAlpha1}) and (\ref{eq:NewtonianM2}) receive higher-order corrections, while Eqs. (\ref{eq:NewtonianM1}) and (\ref{eq:NewtonianAlpha2}) are exact. In general, whenever the function $m(\phi,\xi)$ can be factored into
\begin{align}
m(\phi,\xi)=m_\phi(\phi)m_\xi(\xi),
\end{align}
the quantities
\begin{gather}
\tilde{m}(\xi)\equiv m_\xi(\xi),\\
q\equiv -\d{\log m_\xi}{\xi}=\left(\d{\log m_\phi}{\phi}-\frac{D\log m(\phi,\xi)}{D \phi}\right)\bigg/\d{F}{\phi},
\end{gather}
are exact at all orders in the PD expansion. These quantities represent the resummed piece of the mass and scalar charge; for the resummation schemes defined in Table \ref{table:ResummationSchemes}, these quantities are listed in Table \ref{table:Charge}. When using a particular resummation scheme, it is most convenient to work with these variables instead of $m$ and $\alpha$ so as to avoid the additional bookkeeping required to track the PD corrections to the mass and scalar charge.
\begin{table}[h]\centering
\caption{Resummed piece of the mass $m$ and scalar charge $\alpha$ for the resummation schemes given in Table \ref{table:ResummationSchemes}. We denote the differential operator $\frac{D}{D \phi}$ with the abbreviation $D$. }\label{table:Charge}
\begin{ruledtabular}
\begin{tabular}{C{.15\columnwidth} C{.15\columnwidth} C{.15\columnwidth} C{.45\columnwidth}}
\multicolumn{2}{c}{Resummation Scheme}&\multicolumn{1}{c}{$\tilde{m}(\xi)$}&\multicolumn{1}{c}{$q(\xi)$}\\
\xmulticolumn{2}{c}{Resummation Scheme}&\xmulticolumn{1}{c}{r}&\xmulticolumn{1}{c}{r}\\
$m^{(\rule{7pt}{.5pt})}$&$F^{(\rule{7pt}{.5pt})}$&&\\
\midrule
RJ&$\phi$&$m$& $-D \log m$\\
RJ&$\tilde\varphi$&$m$& $-2\phi\left(\frac{B \log \phi}{2}\right)^{1/2} D\log m$\\
RE&$\phi$&$m^{(E)}$& $\frac{1}{2\phi}-D\log m$\\
RE&$\tilde\varphi$&$m^{(E)}$& $\left(\frac{B \log \phi}{2}\right)^{1/2} \left(1-2 \phi D\,\log m \right)$\\
\end{tabular}
\end{ruledtabular}
\end{table}
\section{Two-body potentials at post-Dickean order}\label{sec:1PSpotentials}
The sources defined in Eqs. (\ref{eq:SigmaDefstart})--(\ref{eq:SigmaDefend}) computed at 1PD order are
\begin{widetext}
\begin{align}
\sigma&=m_1\left(1+\frac{3 v_1^2}{2 c^2}-\frac{G m_2(1-5 \mu_0 \alpha_2)}{\phi_0 rc^2}\right)\delta^{(3)}(x-x_1)+\Ord\left(\frac{1}{c^4}\right)+\left(1\rightleftharpoons 2\right),\\
\sigma_s&=\frac{m_1 \alpha_1}{\mu_0}\left(1-\frac{v_1^2}{2c^2}-\frac{G m_2(6\mu_0+B\alpha_2-6\mu_0^2 \alpha_2)}{2 \mu_0 \phi_0 r c^2}\right)\delta^{(3)}(x-x_1)+\Ord\left(\frac{1}{c^4}\right)+\left(1\rightleftharpoons 2\right),\\
\sigma^i&=\frac{m_1 v_1^i}{c} \delta^{(3)}(x-x_1)+\Ord\left(\frac{1}{c^3}\right)+\left(1\rightleftharpoons 2\right),\\
\sigma^{ii}&= \frac{m_1 v_1^2}{c^2}\delta^{(3)}(x-x_1)+\Ord\left(\frac{1}{c^4}\right)+\left(1\rightleftharpoons 2\right),
\end{align}
where $r=|\v{x}_1-\v{x}_2| $ and we have suppressed the expansions in $m_i$ and $\alpha_i$ using the notation of Eq. (\ref{eq:ShorthandDef}).
Hence, the two body potentials needed to compute the equations of motion and scalar mass in Secs. \ref{sec:EOM} and \ref{sec:ScalarMass} are given by
\begin{align}
U&\equiv \int \frac{\sigma(t,\v{x}')}{|\v{x}-\v{x}'|} d^3 x'= \frac{m_1}{r_1}\left(1+\frac{3 v_1^2}{2 c^2}-\frac{G m_2(1-5 \mu_0 \alpha_2)}{\phi_0 r c^2}\right)+\Ord\left(\frac{1}{c^4}\right)+\left(1\rightleftharpoons 2\right),\\
U_s&\equiv \int \frac{\sigma_s(t,\v{x}')}{|\v{x}-\v{x}'|} d^3 x'=\frac{m_1 \alpha_1}{\mu_0 r_1}\left(1-\frac{v_1^2}{2c^2}-\frac{G m_2(6\mu_0+B\alpha_2-6\mu_0^2 \alpha_2)}{2 \mu_0 \phi_0 r c^2 }\right)+\Ord\left(\frac{1}{c^4}\right)+\left(1\rightleftharpoons 2\right),\\
\begin{split}
M_s&\equiv\int \sigma_s(t,\v{x}')d^3 x'={\mu_0}^{-1} m_1 \alpha_1+\Ord\left(\frac{1}{c^2}\right)+\left(1\rightleftharpoons 2\right),\\
\dot{M}_s&=\Ord\left(\frac{1}{c^3}\right),
\end{split}\\
V^i&\equiv\int\frac{\sigma^i(t,\v{x}')}{|\v{x}-\v{x}'|} d^3 x'=\frac{m_1}{r_1}\frac{v_1^i}{c}+\Ord\left(\frac{1}{c^3}\right)+\left(1\rightleftharpoons 2\right),\\
V_s^i&\equiv\int\frac{\sigma_s(t,\v{x}')v'^i}{|\v{x}-\v{x}'|} d^3 x'=\frac{m_1\alpha_1}{\mu_0 r_1}\frac{v_1^i}{c}+\Ord\left(\frac{1}{c^3}\right)+\left(1\rightleftharpoons 2\right),\\
\Phi_1&\equiv\int\frac{\sigma^{ii}(t,\v{x}')}{|\v{x}-\v{x}'|} d^3 x'=\frac{m_1}{r_1}\frac{v_1^2}{c^2}+\Ord\left(\frac{1}{c^3}\right)+\left(1\rightleftharpoons 2\right),\\
\Phi^{i j}_1&\equiv\int\frac{\sigma^{ij}(t,\v{x}')}{|\v{x}-\v{x}'|} d^3 x'=\frac{m_1}{r_1}\frac{v_1^i v_1^j}{c^2}+\Ord\left(\frac{1}{c^3}\right)+\left(1\rightleftharpoons 2\right),\\
\Phi^s_2&\equiv\int\frac{\sigma_s(t,\v{x}')U(t,\v{x}')}{|\v{x}-\v{x}'|} d^3 x'=\frac{m_1 m_2 \alpha_1}{\mu_0 r_1 r}+\Ord\left(\frac{1}{c^2}\right)+\left(1\rightleftharpoons 2\right),\\
\Phi_{2s}&\equiv\int\frac{\sigma(t,\v{x}')U_s(t,\v{x}')}{|\v{x}-\v{x}'|} d^3 x'=\frac{m_1 m_2 \alpha_2}{\mu_0 r_1 r}+\Ord\left(\frac{1}{c^2}\right)+\left(1\rightleftharpoons 2\right),\\
\Phi^s_{2s}&\equiv\int\frac{\sigma_s(t,\v{x}')U_s(t,\v{x}')}{|\v{x}-\v{x}'|} d^3 x'=\frac{m_1 m_2 \alpha_1 \alpha_2}{{\mu_0}^2r_1 r}+\Ord\left(\frac{1}{c^2}\right)+\left(1\rightleftharpoons 2\right),\\
\begin{split}
X&\equiv\int \sigma(t,\v{x}') |\v{x}-\v{x}'| d^3 x'=m_1 r_1+\Ord\left(\frac{1}{c^2}\right)+\left(1\rightleftharpoons 2\right),\\
\ddot{X}&=\dd{m_1}{t}r_1+2\d{m_1}{t}\d{r_1}{t}+m_1\dd{r_1}{t}+\Ord\left(\frac{1}{c^2}\right)+\left(1\rightleftharpoons 2\right),\\
&=m_1\left(\v{a}_1\cdot \v{n}_1+\frac{v_1^2}{r_1}-\frac{(\v{v}_1\cdot \v{n}_1)^2}{r_1}\right)+\Ord\left(\frac{1}{c^2}\right)+\left(1\rightleftharpoons 2\right),
\end{split}\\
\begin{split}
X_s&\equiv\int \sigma_s(t,\v{x}') |\v{x}-\v{x}'| d^3 x'={\mu_0}^{-1}m_1 \alpha_1 r_1+\Ord\left(\frac{1}{c^2}\right)+\left(1\rightleftharpoons 2\right).\\
\ddot{X}_s&={\mu_0}^{-1}\left[\dd{(m_1 \alpha_1)}{t}r_1+2\d{(m_1\alpha_1)}{t}\d{r_1}{t}+m_1\alpha_1\dd{r_1}{t}\right]+\left(1\rightleftharpoons 2\right),\\
&=\frac{m_1\alpha_1}{\mu_0}\left(\v{a}_1\cdot \v{n}_1+\frac{v_1^2}{r_1}-\frac{(\v{v}_1\cdot \v{n}_1)^2}{r_1}\right)+\Ord\left(\frac{1}{c^2}\right)+\left(1\rightleftharpoons 2\right),\label{eq:1PSsolEnd}
\end{split}
\end{align}
where the time derivatives of the masses and scalar charges are pushed to higher PD order because
\begin{align}
\d{m_A}{t}&=\frac{D m_A}{D \phi}v_A^\mu \partial_\mu \phi(x_A)\sim\Ord\left(\frac{1}{c^3}\right),
\end{align}
where $v_A^\mu\equiv u_A^\mu/u_A^0$.
\end{widetext}

\section{A closer look at the model of \textit{Palenzuela et. al.}}\label{sec:Palenzuela}

\begin{figure*}[t]
\includegraphics[width=\columnwidth,clip=true, trim=0 5 0 0]{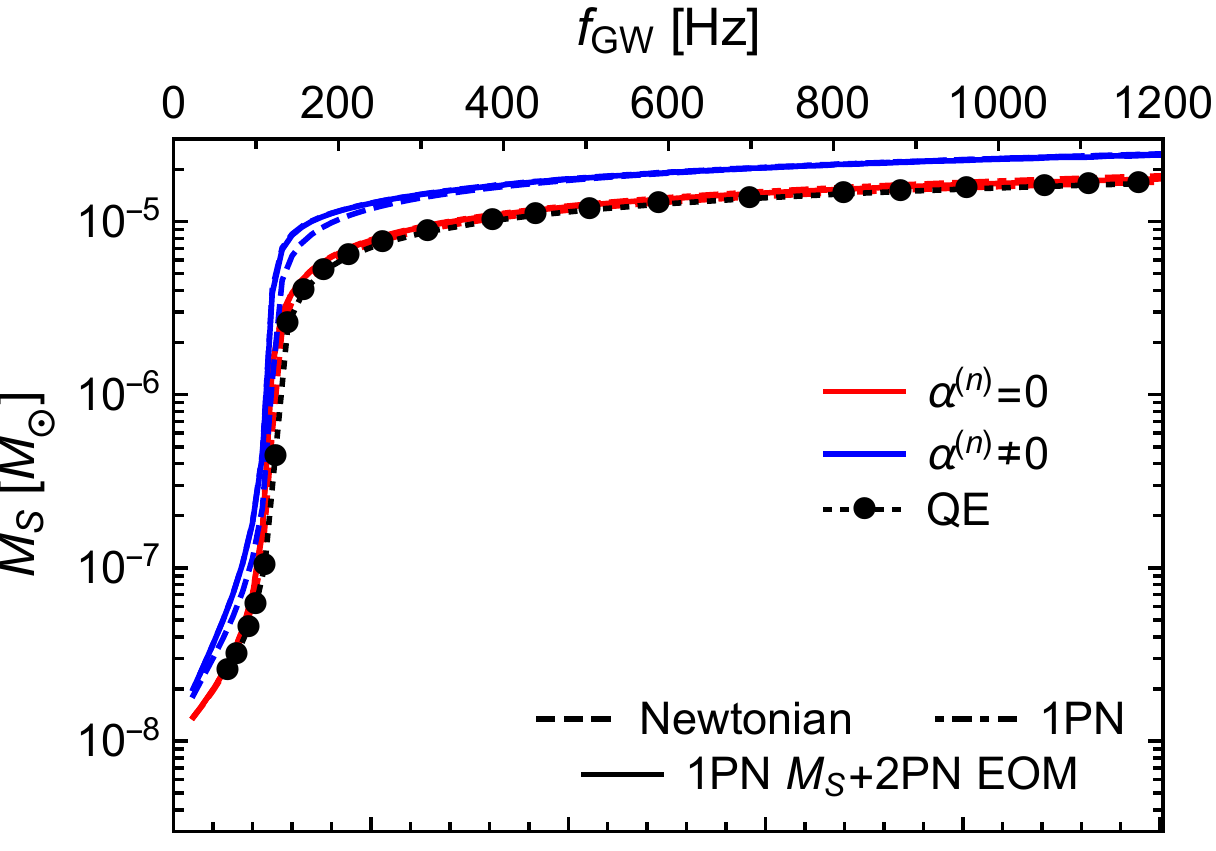} 
\hspace{.75\columnsep}
\includegraphics[width=\columnwidth,clip=true, trim=0 5 0 0]{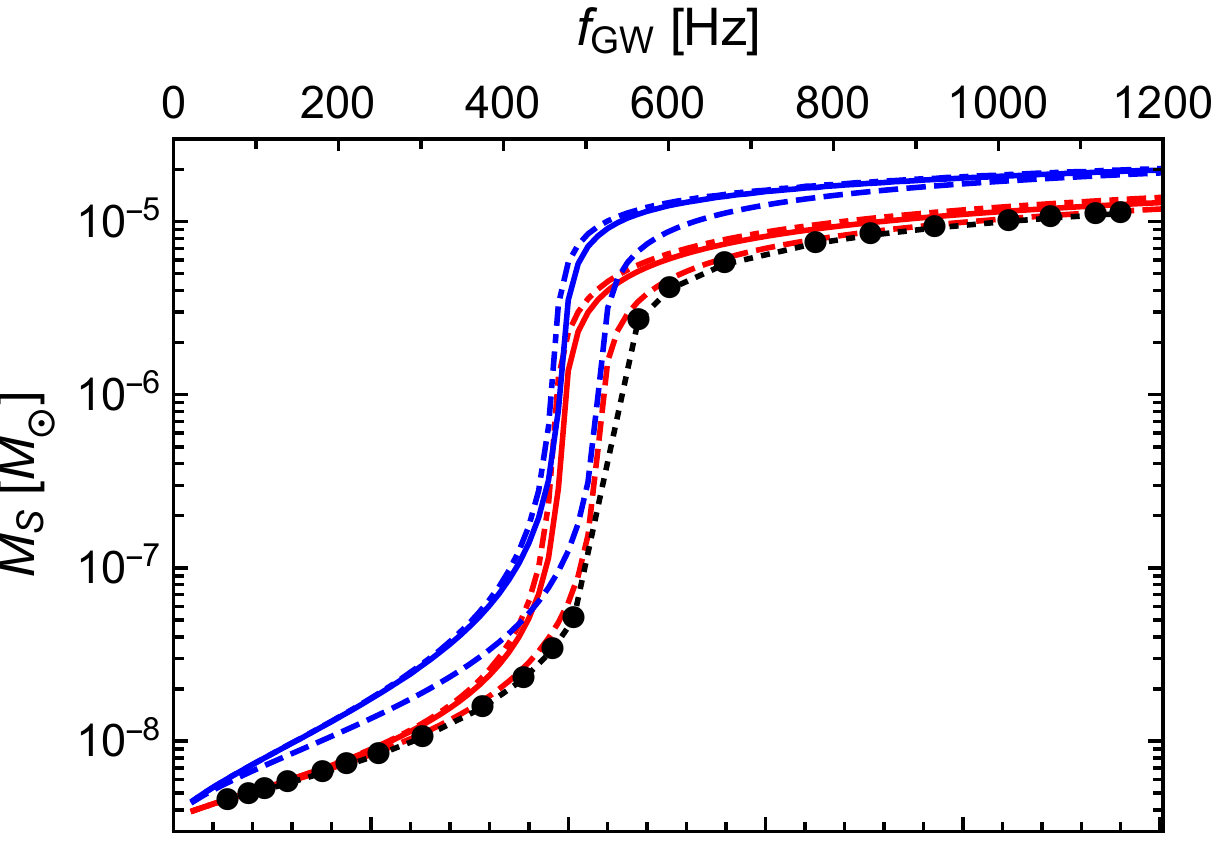}\\
\includegraphics[width=\columnwidth,clip=true, trim=0 0 0 0]{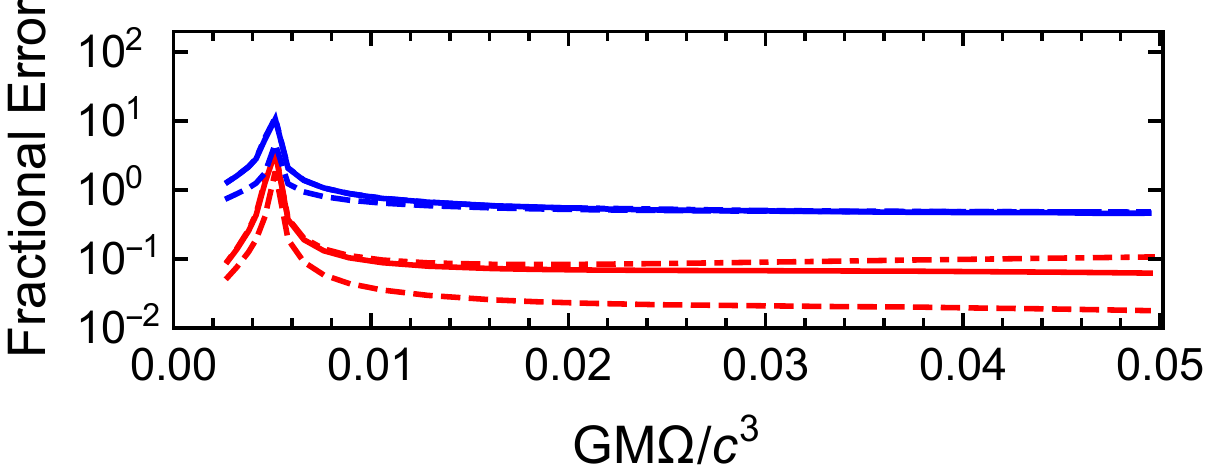} 
\hspace{.75\columnsep}
\includegraphics[width=\columnwidth,clip=true, trim=0 0 0 0]{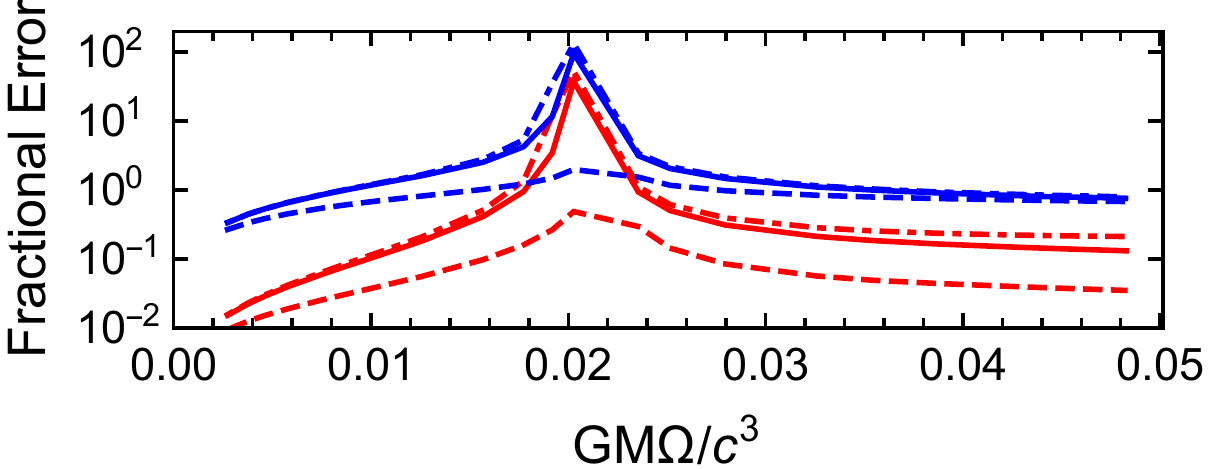}
\caption{Scalar mass predicted by the model of Ref. \cite{Palenzuela2014} for a $(1.35+1.35) M_\odot$ neutron-star binary system on a circular orbit as a function of the orbital frequency and gravitational wave frequency. The scalar mass at Newtonian and 1PN order computed without (with) the derivatives of scalar charge is plotted in red (blue) using the Newtonian, 1PN, and 2PN equations of motion (dashed, dot-dashed, and solid lines, respectively) and the Newtonian order feedback mechanism given in Eq. (\ref{eq:PalenzuelaVarphi}). We also plot the quasi-equilibrium configurations (QE) reported in Ref. \cite{Taniguchi2014} (dotted). The bottom panels depict the magnitude of the fractional error between the PD and quasi-equilibrium results. We use the APR4 equation of state with (left) $B=9, \,\tilde\varphi_0=3.33\times 10^{-11}$ and (right) $B=8.4, \,\tilde\varphi_0=3.45\times 10^{-11}$.}\label{fig:PalenzuelaScalarMass}
\end{figure*}

Another analytic model of dynamical scalarization was proposed in Ref. \cite{Palenzuela2014}. This model augments the PN equations of motion with a feedback mechanism that simulates the non-perturbative growth of the scalar field around each body (see Sec. \ref{sec:Analytic} for more detail). This prescription is uniquely defined when working at leading order but becomes ambiguous when extended to higher PN orders. The construction given in Ref. \cite{Palenzuela2014} uses the 2.5PN equations of motion (given in Ref. \cite{Mirshekari2013}) and a Newtonian order feedback mechanism [Eqs. (\ref{eq:PalenzuelaVarphi}) and (\ref{eq:PalenzuelaVarphi2})]. The authors also set to zero all derivatives of the scalar charge [the first of which is given in Eq. (\ref{eq:AlphaPrimebar})].

While this particular set of choices leads to predictions consistent with numerical-relativity, we would like to explore other realizations of this model for two reasons. First, we want to understand the impact of these algorithmic decisions; if a particular choice greatly impacts the model's performance, understanding its physical significance is important. Second, we would like to track the changes to the model at each order so as to check the best way to improve the results of Ref. \cite{Palenzuela2014} with future PN calculations. We address these two concerns by investigating the effects of including derivatives of the scalar charge and using a higher-order feedback mechanism. The authors of Ref. \cite{Palenzuela2014} briefly mention these two modifications and argue that they do not significantly impact the model; we expand on this discussion here, offering a precise, quantitative description of their effects.

\textit{Including derivatives of the scalar charge:} The derivatives of the scalar charge enter this model through the equations of motion [see Eqs. (\ref{eq:PNEOM1}) and (\ref{eq:PNEOM2})] and the feedback mechanism [see Eqs. (\ref{eq:PalenzuelaVarphi1PN}) and (\ref{eq:PalenzuelaVarphi1PN2})] beginning at 1PN order. The decision to set these derivatives to zero was made in Ref. \cite{Palenzuela2014} to ensure that $\bar{m}(\phi)$ and $\bar{\alpha}(\phi)$ were evaluated at each star rather than expanded about the background value $\phi_0$. However, this procedure is problematic, as simply setting the derivatives of $\bar{\alpha}$ to zero does not properly resum these expansions. For example, $\bar{\alpha}$ itself appears in every term of the Einstein-frame mass [analogous to Eq. (\ref{eq:Mexpand})]
\begin{align}
\bar{m}^{(E)}(\tilde\varphi)=&m^{(E)}(\tilde \varphi_0)\left[\vphantom{\frac{1}{2}}1+\bar{\alpha} \Delta \right. \nonumber \\
& \left.-\frac{1}{2}(\bar{\alpha}^2-\bar{\alpha}')\Delta^2+\cdots\right],\label{eq:PNChargeExpansion}
\end{align}
where $\Delta\equiv (\tilde\varphi-\tilde\varphi_0)$. Thus, the higher-order terms remaining after eliminating derivatives of $\bar\alpha$ are effectively double counted by the feedback model.

\begin{figure*}[t]
\includegraphics[width=\columnwidth,clip=true, trim=0 5 0 0]{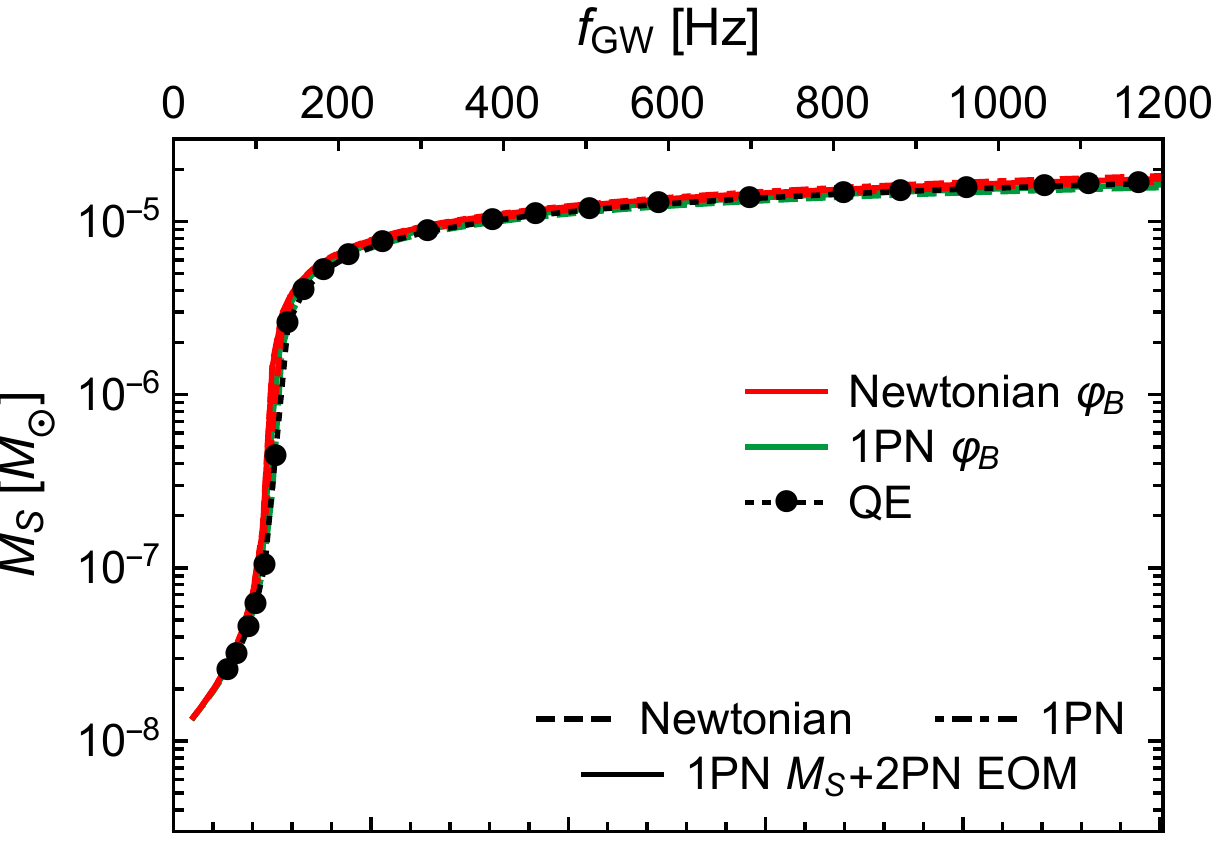} 
\hspace{.75\columnsep}
\includegraphics[width=\columnwidth,clip=true, trim=0 5 0 0]{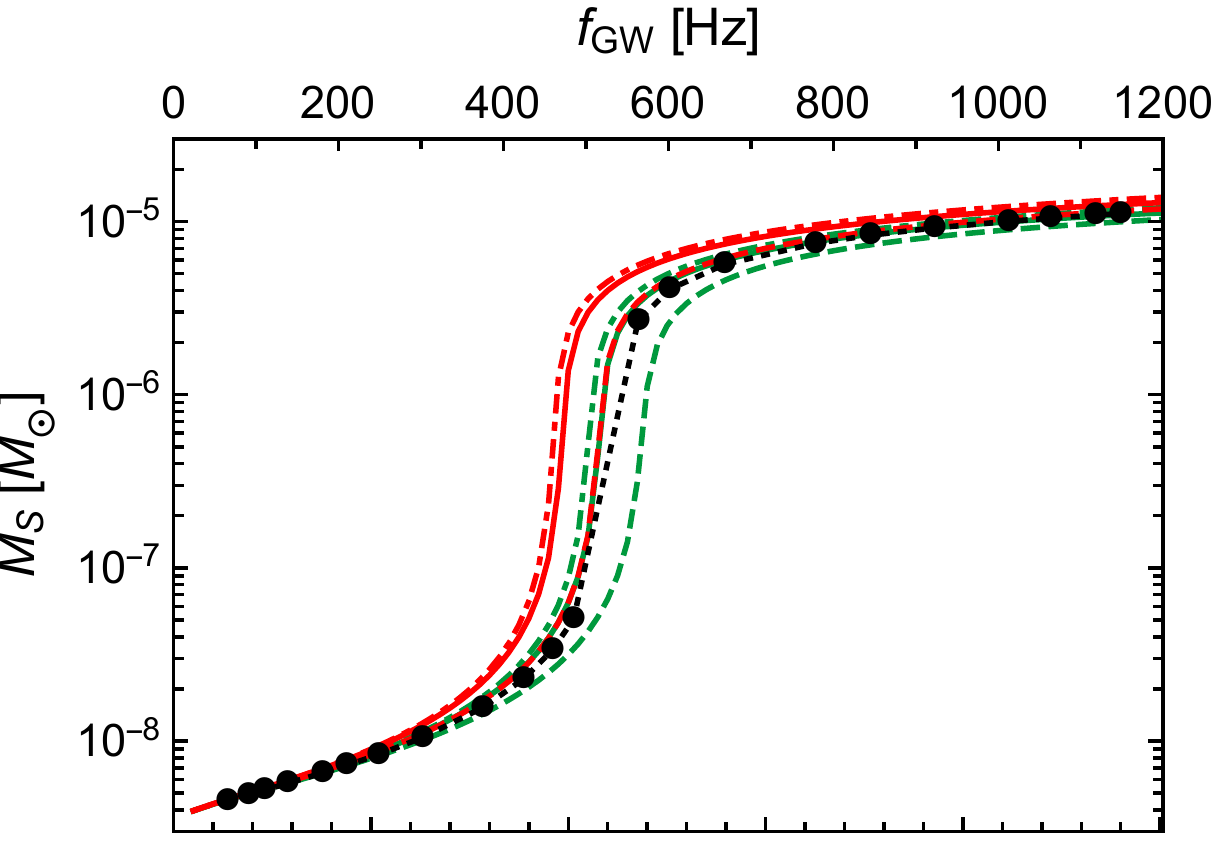}\\
\includegraphics[width=\columnwidth,clip=true, trim=0 0 0 0]{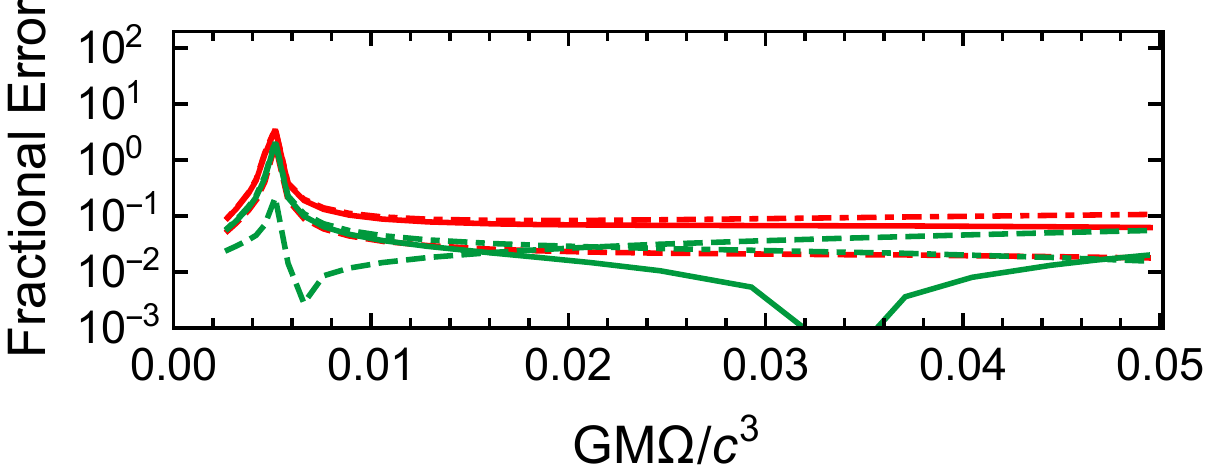} 
\hspace{.75\columnsep}
\includegraphics[width=\columnwidth,clip=true, trim=0 0 0 0]{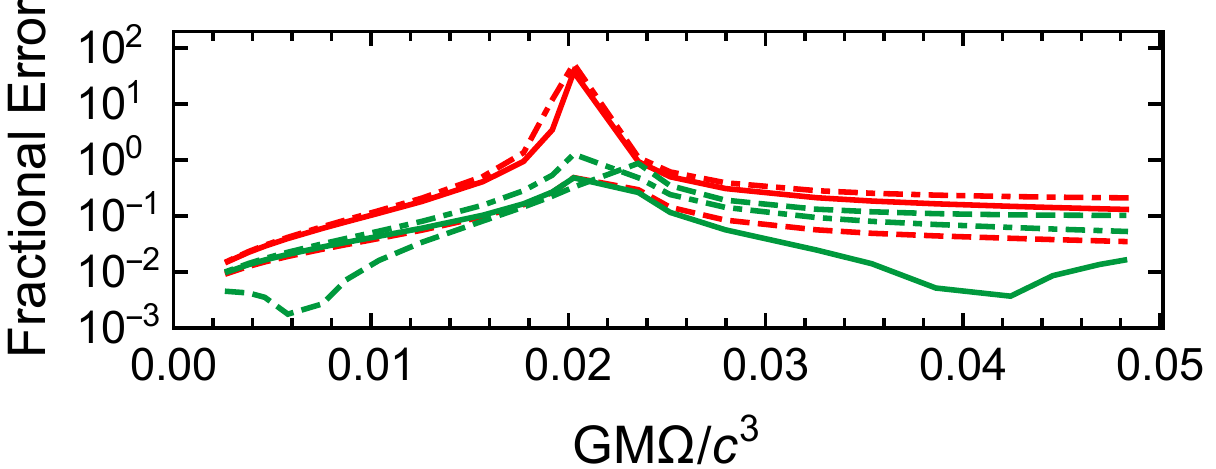}
\caption{Same as Fig. \ref{fig:PalenzuelaScalarMass} but also including the predictions of the model of Ref. \cite{Palenzuela2014} computed using the 1PN extension of the feedback model (green) given in Eq. (\ref{eq:PalenzuelaVarphi1PN}). The derivatives of the scalar charge were dropped in computing all of the plotted curves.}\label{fig:PalenzuelaScalarMass1PN}
\end{figure*}

Even without a mathematically rigorous motivation, the choice to drop the derivatives of the charge still yields predictions in qualitative agreement with numerical relativity. However, there are many other equally valid ways to alter the coefficients in expansions like that of Eq. (\ref{eq:PNChargeExpansion}) --- for example, the coefficients containing $\bar{\alpha}'$ could be halved rather than set to zero. To provide some bound on the effect of these choices, in Fig. \ref{fig:PalenzuelaScalarMass} we compare the total scalar mass predicted by the model of Ref. \cite{Palenzuela2014} when all derivatives of the scalar charge are dropped (red) and when all are kept (blue). We restrict to circular orbits using the Newtonian, 1PN, and 2PN equations of motion (dashed, dot-dashed, and solid lines, respectively) --- the exact prescription used in Ref. \cite{Palenzuela2014} is the solid, red curve.

The inclusion of these terms increases the scalar mass by approximately $20-50\%$ both before and after scalarization. This result is consistent with our previous observation that higher-order terms left in the PN expansion should produce extraneous contributions when the mass and charge are resummed. In addition, we note that the model of Ref. \cite{Palenzuela2014} overestimates the scalar mass compared to the PD approximation at the same order. Combining these two observations, we argue that double counting can become a significant issue when using a simple feedback mechanism like that of Ref. \cite{Palenzuela2014}, and that while simply dropping particular terms from the PN expansion can help remedy these issues, it is not the ideal solution. Instead, the corresponding resummation should be accounted for in a more systematic way, as is done with the PD approach.

\textit{Extending the feedback mechanism to 1PN order:} The feedback mechanism used in Ref. \cite{Palenzuela2014} contains only the leading order contributions to the scalar field despite being paired with the 2.5PN equations of motion computed.\footnote{The authors of Ref. \cite{Palenzuela2014} considered the effect of adding to this feedback mechanism the order $\Ord(1/r^2)$ terms from the field felt by a static test mass far from an isolated body. These terms were shown to have negligible impact on their model. Here, we consider the 1PN corrections to the scalar field felt by each body in a comparable-mass binary system, which comprise a more comprehensive set of $\Ord(1/r^2)$ corrections.} The impact of using approximants of such different order is unclear, but the mismatch may lead to certain unintuitive predictions. We consider instead using the natural 1PN extension of the feedback mechanism
\begin{align}
\begin{split}
\varphi_B^{(1)}=&\tilde\varphi_0+\frac{G \bar m_2 \bar\alpha_2}{\phi_0 r c^2}+\frac{G \bar m_2}{\phi_0 r c^4}\left[-\frac{1}{2}\bar\alpha_2(\v{v}_2\cdot\v{n})^2\right.\\
&\left.-\left(\frac{3}{2} \bar\alpha_2+\frac{3}{2}\bar\alpha_1\bar\alpha_2^2-\bar\alpha_1\bar\alpha'_2\right)\frac{G \bar m_1}{\phi_0 r}\right],\label{eq:PalenzuelaVarphi1PN}
\end{split}\\
\varphi_B^{(2)}=&\left(1\rightleftharpoons 2\right),\label{eq:PalenzuelaVarphi1PN2}
\end{align}
where $\bar m_i$, $\bar\alpha_i$, and $\bar\alpha'_i$ are evaluated at $\varphi_B^{(i)}$.

We compare the total scalar mass computed using the Newtonian (red) and 1PN feedback models (green) with equations of motion at Newtonian, 1PN, and 2PN order in Fig. \ref{fig:PalenzuelaScalarMass1PN}, making the additional choice to set all derivatives of the scalar charge to zero, as was done in Ref. \cite{Palenzuela2014}. The inclusion of higher-order effects in these two aspects of the model produces competing shifts in the predicted onset of DS: the choice of 1PN feedback system over the Newtonian system pushes this transition point to higher frequency, while the 1PN terms in the equations of motion push the transition to lower frequency. These two effects nearly cancel each other in such a way that the predictions when working consistently at Newtonian order (i.e. Newtonian order feedback and equations of motion) are very close to those when working consistently at 1PN order. We observe that working consistently at one order generally improves the agreement with the quasi-equilibrium configuration calculations of Ref. \cite{Taniguchi2014}. The most accurate model depicted in Fig. \ref{fig:PalenzuelaScalarMass1PN} uses the 1PN feedback model in conjunction with the 2PN equations of motion, but in line with the previous observation, we suspect that adding the 2PN corrections to Eqs. (\ref{eq:PalenzuelaVarphi1PN}) and (\ref{eq:PalenzuelaVarphi1PN2}) will improve these predictions; we leave the calculation and implementation of these higher-order terms for future work.

To recap, some of the technical aspects in the construction of the model proposed in Ref. \cite{Palenzuela2014} are ambiguous; the prescription for these options is only precisely specified when working at Newtonian order. These choices arise because the model splices a non-linear feedback mechanism on to independently computed PN equations of motion. We find that the model is most accurate when one uses a feedback mechanism and equations of motion of the same order and when one drops some of the higher-order terms in the PN expansions (e.g. the derivatives of the charge) to minimize double counting.

The PD formalism avoids these issues by performing a resummation of the post-Newtonian expansion at the level of the action. By carrying through this resummation consistently, a non-linear feedback mechanism analogous to Eqs. (\ref{eq:PalenzuelaVarphi1PN}) and (\ref{eq:PalenzuelaVarphi1PN2}) organically arises alongside the equations of motion. Thus, the PD model gives results at a consistent order while also avoiding double counting.

\bibliography{reference,DS_Resummed}

\begin{thebibliography}{64}%
\makeatletter
\providecommand \@ifxundefined [1]{%
 \@ifx{#1\undefined}
}%
\providecommand \@ifnum [1]{%
 \ifnum #1\expandafter \@firstoftwo
 \else \expandafter \@secondoftwo
 \fi
}%
\providecommand \@ifx [1]{%
 \ifx #1\expandafter \@firstoftwo
 \else \expandafter \@secondoftwo
 \fi
}%
\providecommand \natexlab [1]{#1}%
\providecommand \enquote  [1]{``#1''}%
\providecommand \bibnamefont  [1]{#1}%
\providecommand \bibfnamefont [1]{#1}%
\providecommand \citenamefont [1]{#1}%
\providecommand \href@noop [0]{\@secondoftwo}%
\providecommand \href [0]{\begingroup \@sanitize@url \@href}%
\providecommand \@href[1]{\@@startlink{#1}\@@href}%
\providecommand \@@href[1]{\endgroup#1\@@endlink}%
\providecommand \@sanitize@url [0]{\catcode `\\12\catcode `\$12\catcode
  `\&12\catcode `\#12\catcode `\^12\catcode `\_12\catcode `\%12\relax}%
\providecommand \@@startlink[1]{}%
\providecommand \@@endlink[0]{}%
\providecommand \url  [0]{\begingroup\@sanitize@url \@url }%
\providecommand \@url [1]{\endgroup\@href {#1}{\urlprefix }}%
\providecommand \urlprefix  [0]{URL }%
\providecommand \Eprint [0]{\href }%
\providecommand \doibase [0]{http://dx.doi.org/}%
\providecommand \selectlanguage [0]{\@gobble}%
\providecommand \bibinfo  [0]{\@secondoftwo}%
\providecommand \bibfield  [0]{\@secondoftwo}%
\providecommand \translation [1]{[#1]}%
\providecommand \BibitemOpen [0]{}%
\providecommand \bibitemStop [0]{}%
\providecommand \bibitemNoStop [0]{.\EOS\space}%
\providecommand \EOS [0]{\spacefactor3000\relax}%
\providecommand \BibitemShut  [1]{\csname bibitem#1\endcsname}%
\let\auto@bib@innerbib\@empty
%</preamble>
\bibitem [{\citenamefont {Abbott}\ \emph
  {et~al.}(2016{\natexlab{a}})\citenamefont {Abbott} \emph
  {et~al.}}]{detection}%
  \BibitemOpen
  \bibfield  {author} {\bibinfo {author} {\bibfnamefont {B.~P.}\ \bibnamefont
  {Abbott}} \emph {et~al.} (\bibinfo {collaboration} {LIGO Scientific,
  Virgo}),\ }\href {\doibase 10.1103/PhysRevLett.116.061102} {\bibfield
  {journal} {\bibinfo  {journal} {Phys.\ Rev.\ Lett.}\ }\textbf {\bibinfo
  {volume} {116}},\ \bibinfo {pages} {061102} (\bibinfo {year}
  {2016}{\natexlab{a}})}\BibitemShut {NoStop}%
%%CITATION = ARXIV:1602.03837;%%
\bibitem [{\citenamefont {Will}(2014)}]{Will2014}%
  \BibitemOpen
  \bibfield  {author} {\bibinfo {author} {\bibfnamefont {C.~M.}\ \bibnamefont
  {Will}},\ }\href {http://www.livingreviews.org/lrr-2014-4} {\bibfield
  {journal} {\bibinfo  {journal} {Living Rev.~Rel.}\ }\textbf {\bibinfo
  {volume} {17}} (\bibinfo {year} {2014})}\BibitemShut {NoStop}%
\bibitem [{\citenamefont {Wex}(2014)}]{Wex2014}%
  \BibitemOpen
  \bibfield  {author} {\bibinfo {author} {\bibfnamefont {N.}~\bibnamefont
  {Wex}},\ }\href@noop {} {\  (\bibinfo {year} {2014})},\ \Eprint
  {http://arxiv.org/abs/1402.5594} {arXiv:1402.5594 [gr-qc]} \BibitemShut
  {NoStop}%
\bibitem [{\citenamefont {Berti}\ \emph {et~al.}(2015)\citenamefont {Berti}
  \emph {et~al.}}]{Berti2015}%
  \BibitemOpen
  \bibfield  {author} {\bibinfo {author} {\bibfnamefont {E.}~\bibnamefont
  {Berti}} \emph {et~al.},\ }\href {\doibase 10.1088/0264-9381/32/24/243001}
  {\bibfield  {journal} {\bibinfo  {journal} {Class.\ Quantum Grav.}\ }\textbf
  {\bibinfo {volume} {32}},\ \bibinfo {pages} {243001} (\bibinfo {year}
  {2015})}\BibitemShut {NoStop}%
\bibitem [{\citenamefont {Aasi}\ \emph {et~al.}(2013)\citenamefont {Aasi} \emph
  {et~al.}}]{Aasi2013}%
  \BibitemOpen
  \bibfield  {author} {\bibinfo {author} {\bibfnamefont {J.}~\bibnamefont
  {Aasi}} \emph {et~al.} (\bibinfo {collaboration} {LIGO Scientific, Virgo}),\
  }\href@noop {} {\  (\bibinfo {year} {2013})},\ \Eprint
  {http://arxiv.org/abs/1304.0670} {arXiv:1304.0670 [gr-qc]} \BibitemShut
  {NoStop}%
%%CITATION = ARXIV:1304.0670;%%
\bibitem [{\citenamefont {Abbott}\ \emph
  {et~al.}(2016{\natexlab{b}})\citenamefont {Abbott} \emph
  {et~al.}}]{Abbott2016}%
  \BibitemOpen
  \bibfield  {author} {\bibinfo {author} {\bibfnamefont {B.~P.}\ \bibnamefont
  {Abbott}} \emph {et~al.} (\bibinfo {collaboration} {LIGO Scientific,
  Virgo}),\ }\href@noop {} {\  (\bibinfo {year} {2016}{\natexlab{b}})},\
  \Eprint {http://arxiv.org/abs/1602.03842} {arXiv:1602.03842 [gr-qc]}
  \BibitemShut {NoStop}%
\bibitem [{\citenamefont {Abbott}\ \emph
  {et~al.}(2016{\natexlab{c}})\citenamefont {Abbott} \emph
  {et~al.}}]{Abbott2016b}%
  \BibitemOpen
  \bibfield  {author} {\bibinfo {author} {\bibfnamefont {B.~P.}\ \bibnamefont
  {Abbott}} \emph {et~al.} (\bibinfo {collaboration} {LIGO Scientific,
  Virgo}),\ }\href@noop {} {\  (\bibinfo {year} {2016}{\natexlab{c}})},\
  \Eprint {http://arxiv.org/abs/1602.03838} {arXiv:1602.03838 [gr-qc]}
  \BibitemShut {NoStop}%
\bibitem [{\citenamefont {Acernese}\ \emph {et~al.}(2015)\citenamefont
  {Acernese} \emph {et~al.}}]{aVIRGO}%
  \BibitemOpen
  \bibfield  {author} {\bibinfo {author} {\bibfnamefont {F.}~\bibnamefont
  {Acernese}} \emph {et~al.},\ }\href
  {http://stacks.iop.org/0264-9381/32/i=2/a=024001} {\bibfield  {journal}
  {\bibinfo  {journal} {Class.\ Quantum Grav.}\ }\textbf {\bibinfo {volume}
  {32}},\ \bibinfo {pages} {024001} (\bibinfo {year} {2015})}\BibitemShut
  {NoStop}%
\bibitem [{\citenamefont {Aso}\ \emph {et~al.}(2013)\citenamefont {Aso},
  \citenamefont {Michimura}, \citenamefont {Somiya}, \citenamefont {Ando},
  \citenamefont {Miyakawa}, \citenamefont {Sekiguchi}, \citenamefont
  {Tatsumi},\ and\ \citenamefont {Yamamoto}}]{KAGRA}%
  \BibitemOpen
  \bibfield  {author} {\bibinfo {author} {\bibfnamefont {Y.}~\bibnamefont
  {Aso}}, \bibinfo {author} {\bibfnamefont {Y.}~\bibnamefont {Michimura}},
  \bibinfo {author} {\bibfnamefont {K.}~\bibnamefont {Somiya}}, \bibinfo
  {author} {\bibfnamefont {M.}~\bibnamefont {Ando}}, \bibinfo {author}
  {\bibfnamefont {O.}~\bibnamefont {Miyakawa}}, \bibinfo {author}
  {\bibfnamefont {T.}~\bibnamefont {Sekiguchi}}, \bibinfo {author}
  {\bibfnamefont {D.}~\bibnamefont {Tatsumi}}, \ and\ \bibinfo {author}
  {\bibfnamefont {H.}~\bibnamefont {Yamamoto}} (\bibinfo {collaboration} {The
  KAGRA Collaboration}),\ }\href {\doibase 10.1103/PhysRevD.88.043007}
  {\bibfield  {journal} {\bibinfo  {journal} {Phys.\ Rev.\ D}\ }\textbf
  {\bibinfo {volume} {88}},\ \bibinfo {pages} {043007} (\bibinfo {year}
  {2013})}\BibitemShut {NoStop}%
\bibitem [{\citenamefont {Yunes}\ and\ \citenamefont
  {Siemens}(2013)}]{YunesLR}%
  \BibitemOpen
  \bibfield  {author} {\bibinfo {author} {\bibfnamefont {N.}~\bibnamefont
  {Yunes}}\ and\ \bibinfo {author} {\bibfnamefont {X.}~\bibnamefont
  {Siemens}},\ }\href {http://www.livingreviews.org/lrr-2013-9} {\bibfield
  {journal} {\bibinfo  {journal} {Living Rev.~Rel.}\ }\textbf {\bibinfo
  {volume} {16}} (\bibinfo {year} {2013})}\BibitemShut {NoStop}%
\bibitem [{\citenamefont {Gair}\ \emph {et~al.}(2013)\citenamefont {Gair},
  \citenamefont {Vallisneri}, \citenamefont {Larson},\ and\ \citenamefont
  {Baker}}]{GairLR}%
  \BibitemOpen
  \bibfield  {author} {\bibinfo {author} {\bibfnamefont {J.~R.}\ \bibnamefont
  {Gair}}, \bibinfo {author} {\bibfnamefont {M.}~\bibnamefont {Vallisneri}},
  \bibinfo {author} {\bibfnamefont {S.~L.}\ \bibnamefont {Larson}}, \ and\
  \bibinfo {author} {\bibfnamefont {J.~G.}\ \bibnamefont {Baker}},\ }\href
  {http://www.livingreviews.org/lrr-2013-7} {\bibfield  {journal} {\bibinfo
  {journal} {Living Rev.~Rel.}\ }\textbf {\bibinfo {volume} {16}} (\bibinfo
  {year} {2013})}\BibitemShut {NoStop}%
\bibitem [{\citenamefont {Arun}\ \emph {et~al.}(2006)\citenamefont {Arun},
  \citenamefont {Iyer}, \citenamefont {Qusailah},\ and\ \citenamefont
  {Sathyaprakash}}]{Arun2006b}%
  \BibitemOpen
  \bibfield  {author} {\bibinfo {author} {\bibfnamefont {K.~G.}\ \bibnamefont
  {Arun}}, \bibinfo {author} {\bibfnamefont {B.~R.}\ \bibnamefont {Iyer}},
  \bibinfo {author} {\bibfnamefont {M.~S.~S.}\ \bibnamefont {Qusailah}}, \ and\
  \bibinfo {author} {\bibfnamefont {B.~S.}\ \bibnamefont {Sathyaprakash}},\
  }\href {\doibase 10.1103/PhysRevD.74.024006} {\bibfield  {journal} {\bibinfo
  {journal} {Phys.\ Rev.\ D}\ }\textbf {\bibinfo {volume} {74}},\ \bibinfo
  {pages} {024006} (\bibinfo {year} {2006})}\BibitemShut {NoStop}%
%%CITATION = GR-QC/0604067;%%
\bibitem [{\citenamefont {Yunes}\ and\ \citenamefont
  {Pretorius}(2009)}]{Yunes2009}%
  \BibitemOpen
  \bibfield  {author} {\bibinfo {author} {\bibfnamefont {N.}~\bibnamefont
  {Yunes}}\ and\ \bibinfo {author} {\bibfnamefont {F.}~\bibnamefont
  {Pretorius}},\ }\href {\doibase 10.1103/PhysRevD.80.122003} {\bibfield
  {journal} {\bibinfo  {journal} {Phys.\ Rev.\ D}\ }\textbf {\bibinfo {volume}
  {80}},\ \bibinfo {pages} {122003} (\bibinfo {year} {2009})}\BibitemShut
  {NoStop}%
%%CITATION = ARXIV:0909.3328;%%
\bibitem [{\citenamefont {Del~Pozzo}\ \emph {et~al.}(2011)\citenamefont
  {Del~Pozzo}, \citenamefont {Veitch},\ and\ \citenamefont
  {Vecchio}}]{DelPozzo2011}%
  \BibitemOpen
  \bibfield  {author} {\bibinfo {author} {\bibfnamefont {W.}~\bibnamefont
  {Del~Pozzo}}, \bibinfo {author} {\bibfnamefont {J.}~\bibnamefont {Veitch}}, \
  and\ \bibinfo {author} {\bibfnamefont {A.}~\bibnamefont {Vecchio}},\ }\href
  {\doibase 10.1103/PhysRevD.83.082002} {\bibfield  {journal} {\bibinfo
  {journal} {Phys. Rev. D}\ }\textbf {\bibinfo {volume} {83}},\ \bibinfo
  {pages} {082002} (\bibinfo {year} {2011})}\BibitemShut {NoStop}%
\bibitem [{\citenamefont {Cornish}\ \emph {et~al.}(2011)\citenamefont
  {Cornish}, \citenamefont {Sampson}, \citenamefont {Yunes},\ and\
  \citenamefont {Pretorius}}]{Cornish2011}%
  \BibitemOpen
  \bibfield  {author} {\bibinfo {author} {\bibfnamefont {N.}~\bibnamefont
  {Cornish}}, \bibinfo {author} {\bibfnamefont {L.}~\bibnamefont {Sampson}},
  \bibinfo {author} {\bibfnamefont {N.}~\bibnamefont {Yunes}}, \ and\ \bibinfo
  {author} {\bibfnamefont {F.}~\bibnamefont {Pretorius}},\ }\href {\doibase
  10.1103/PhysRevD.84.062003} {\bibfield  {journal} {\bibinfo  {journal} {Phys.
  Rev. D}\ }\textbf {\bibinfo {volume} {84}},\ \bibinfo {pages} {062003}
  (\bibinfo {year} {2011})}\BibitemShut {NoStop}%
\bibitem [{\citenamefont {Li}\ \emph {et~al.}(2012{\natexlab{a}})\citenamefont
  {Li}, \citenamefont {Del~Pozzo}, \citenamefont {Vitale}, \citenamefont {Van
  Den~Broeck}, \citenamefont {Agathos}, \citenamefont {Veitch}, \citenamefont
  {Grover}, \citenamefont {Sidery}, \citenamefont {Sturani},\ and\
  \citenamefont {Vecchio}}]{Li2012a}%
  \BibitemOpen
  \bibfield  {author} {\bibinfo {author} {\bibfnamefont {T.~G.~F.}\
  \bibnamefont {Li}}, \bibinfo {author} {\bibfnamefont {W.}~\bibnamefont
  {Del~Pozzo}}, \bibinfo {author} {\bibfnamefont {S.}~\bibnamefont {Vitale}},
  \bibinfo {author} {\bibfnamefont {C.}~\bibnamefont {Van Den~Broeck}},
  \bibinfo {author} {\bibfnamefont {M.}~\bibnamefont {Agathos}}, \bibinfo
  {author} {\bibfnamefont {J.}~\bibnamefont {Veitch}}, \bibinfo {author}
  {\bibfnamefont {K.}~\bibnamefont {Grover}}, \bibinfo {author} {\bibfnamefont
  {T.}~\bibnamefont {Sidery}}, \bibinfo {author} {\bibfnamefont
  {R.}~\bibnamefont {Sturani}}, \ and\ \bibinfo {author} {\bibfnamefont
  {A.}~\bibnamefont {Vecchio}},\ }\href {\doibase 10.1103/PhysRevD.85.082003}
  {\bibfield  {journal} {\bibinfo  {journal} {Phys. Rev. D}\ }\textbf {\bibinfo
  {volume} {85}},\ \bibinfo {pages} {082003} (\bibinfo {year}
  {2012}{\natexlab{a}})}\BibitemShut {NoStop}%
\bibitem [{\citenamefont {Li}\ \emph {et~al.}(2012{\natexlab{b}})\citenamefont
  {Li}, \citenamefont {Del~Pozzo}, \citenamefont {Vitale}, \citenamefont {Van
  Den~Broeck}, \citenamefont {Agathos}, \citenamefont {Veitch}, \citenamefont
  {Grover}, \citenamefont {Sidery}, \citenamefont {Sturani},\ and\
  \citenamefont {Vecchio}}]{Li2012b}%
  \BibitemOpen
  \bibfield  {author} {\bibinfo {author} {\bibfnamefont {T.~G.~F.}\
  \bibnamefont {Li}}, \bibinfo {author} {\bibfnamefont {W.}~\bibnamefont
  {Del~Pozzo}}, \bibinfo {author} {\bibfnamefont {S.}~\bibnamefont {Vitale}},
  \bibinfo {author} {\bibfnamefont {C.}~\bibnamefont {Van Den~Broeck}},
  \bibinfo {author} {\bibfnamefont {M.}~\bibnamefont {Agathos}}, \bibinfo
  {author} {\bibfnamefont {J.}~\bibnamefont {Veitch}}, \bibinfo {author}
  {\bibfnamefont {K.}~\bibnamefont {Grover}}, \bibinfo {author} {\bibfnamefont
  {T.}~\bibnamefont {Sidery}}, \bibinfo {author} {\bibfnamefont
  {R.}~\bibnamefont {Sturani}}, \ and\ \bibinfo {author} {\bibfnamefont
  {A.}~\bibnamefont {Vecchio}},\ }\href
  {http://stacks.iop.org/1742-6596/363/i=1/a=012028} {\bibfield  {journal}
  {\bibinfo  {journal} {Journal of Physics: Conference Series}\ }\textbf
  {\bibinfo {volume} {363}},\ \bibinfo {pages} {012028} (\bibinfo {year}
  {2012}{\natexlab{b}})}\BibitemShut {NoStop}%
\bibitem [{\citenamefont {Sampson}\ \emph {et~al.}(2013)\citenamefont
  {Sampson}, \citenamefont {Cornish},\ and\ \citenamefont
  {Yunes}}]{Sampson2013b}%
  \BibitemOpen
  \bibfield  {author} {\bibinfo {author} {\bibfnamefont {L.}~\bibnamefont
  {Sampson}}, \bibinfo {author} {\bibfnamefont {N.}~\bibnamefont {Cornish}}, \
  and\ \bibinfo {author} {\bibfnamefont {N.}~\bibnamefont {Yunes}},\ }\href
  {\doibase 10.1103/PhysRevD.87.102001} {\bibfield  {journal} {\bibinfo
  {journal} {Phys.\ Rev.\ D}\ }\textbf {\bibinfo {volume} {87}},\ \bibinfo
  {pages} {102001} (\bibinfo {year} {2013})}\BibitemShut {NoStop}%
%%CITATION = ARXIV:1303.1185;%%
\bibitem [{\citenamefont {Sampson}\ \emph
  {et~al.}(2014{\natexlab{a}})\citenamefont {Sampson}, \citenamefont
  {Cornish},\ and\ \citenamefont {Yunes}}]{Sampson2013}%
  \BibitemOpen
  \bibfield  {author} {\bibinfo {author} {\bibfnamefont {L.}~\bibnamefont
  {Sampson}}, \bibinfo {author} {\bibfnamefont {N.}~\bibnamefont {Cornish}}, \
  and\ \bibinfo {author} {\bibfnamefont {N.}~\bibnamefont {Yunes}},\ }\href
  {\doibase 10.1103/PhysRevD.89.064037} {\bibfield  {journal} {\bibinfo
  {journal} {Phys.\ Rev.\ D}\ }\textbf {\bibinfo {volume} {89}},\ \bibinfo
  {pages} {064037} (\bibinfo {year} {2014}{\natexlab{a}})}\BibitemShut
  {NoStop}%
%%CITATION = ARXIV:1311.4898;%%
\bibitem [{\citenamefont {Abbott}\ \emph
  {et~al.}(2016{\natexlab{d}})\citenamefont {Abbott} \emph
  {et~al.}}]{testingGR}%
  \BibitemOpen
  \bibfield  {author} {\bibinfo {author} {\bibfnamefont {B.~P.}\ \bibnamefont
  {Abbott}} \emph {et~al.} (\bibinfo {collaboration} {LIGO Scientific,
  Virgo}),\ }\href@noop {} {\  (\bibinfo {year} {2016}{\natexlab{d}})},\
  \Eprint {http://arxiv.org/abs/1602.03841} {arXiv:1602.03841 [gr-qc]}
  \BibitemShut {NoStop}%
%%CITATION = ARXIV:1602.03841;%%
\bibitem [{\citenamefont {Will}(1998)}]{Will1997}%
  \BibitemOpen
  \bibfield  {author} {\bibinfo {author} {\bibfnamefont {C.~M.}\ \bibnamefont
  {Will}},\ }\href {\doibase 10.1103/PhysRevD.57.2061} {\bibfield  {journal}
  {\bibinfo  {journal} {Phys.\ Rev.\ D}\ }\textbf {\bibinfo {volume} {57}},\
  \bibinfo {pages} {2061} (\bibinfo {year} {1998})}\BibitemShut {NoStop}%
\bibitem [{\citenamefont {Barausse}\ \emph {et~al.}(2013)\citenamefont
  {Barausse}, \citenamefont {Palenzuela}, \citenamefont {Ponce},\ and\
  \citenamefont {Lehner}}]{Barausse2013}%
  \BibitemOpen
  \bibfield  {author} {\bibinfo {author} {\bibfnamefont {E.}~\bibnamefont
  {Barausse}}, \bibinfo {author} {\bibfnamefont {C.}~\bibnamefont
  {Palenzuela}}, \bibinfo {author} {\bibfnamefont {M.}~\bibnamefont {Ponce}}, \
  and\ \bibinfo {author} {\bibfnamefont {L.}~\bibnamefont {Lehner}},\ }\href
  {\doibase 10.1103/PhysRevD.87.081506} {\bibfield  {journal} {\bibinfo
  {journal} {Phys.\ Rev.\ D}\ }\textbf {\bibinfo {volume} {87}},\ \bibinfo
  {pages} {081506} (\bibinfo {year} {2013})}\BibitemShut {NoStop}%
%%CITATION = ARXIV:1212.5053;%%
\bibitem [{\citenamefont {Shibata}\ \emph {et~al.}(2014)\citenamefont
  {Shibata}, \citenamefont {Taniguchi}, \citenamefont {Okawa},\ and\
  \citenamefont {Buonanno}}]{Shibata2014}%
  \BibitemOpen
  \bibfield  {author} {\bibinfo {author} {\bibfnamefont {M.}~\bibnamefont
  {Shibata}}, \bibinfo {author} {\bibfnamefont {K.}~\bibnamefont {Taniguchi}},
  \bibinfo {author} {\bibfnamefont {H.}~\bibnamefont {Okawa}}, \ and\ \bibinfo
  {author} {\bibfnamefont {A.}~\bibnamefont {Buonanno}},\ }\href {\doibase
  10.1103/PhysRevD.89.084005} {\bibfield  {journal} {\bibinfo  {journal}
  {Phys.\ Rev.\ D}\ }\textbf {\bibinfo {volume} {89}},\ \bibinfo {pages}
  {084005} (\bibinfo {year} {2014})}\BibitemShut {NoStop}%
\bibitem [{\citenamefont {Palenzuela}\ \emph {et~al.}(2014)\citenamefont
  {Palenzuela}, \citenamefont {Barausse}, \citenamefont {Ponce},\ and\
  \citenamefont {Lehner}}]{Palenzuela2014}%
  \BibitemOpen
  \bibfield  {author} {\bibinfo {author} {\bibfnamefont {C.}~\bibnamefont
  {Palenzuela}}, \bibinfo {author} {\bibfnamefont {E.}~\bibnamefont
  {Barausse}}, \bibinfo {author} {\bibfnamefont {M.}~\bibnamefont {Ponce}}, \
  and\ \bibinfo {author} {\bibfnamefont {L.}~\bibnamefont {Lehner}},\ }\href
  {\doibase 10.1103/PhysRevD.89.044024} {\bibfield  {journal} {\bibinfo
  {journal} {Phys.\ Rev.\ D}\ }\textbf {\bibinfo {volume} {89}},\ \bibinfo
  {pages} {044024} (\bibinfo {year} {2014})}\BibitemShut {NoStop}%
%%CITATION = ARXIV:1310.4481;%%
\bibitem [{\citenamefont {Sampson}\ \emph
  {et~al.}(2014{\natexlab{b}})\citenamefont {Sampson}, \citenamefont {Yunes},
  \citenamefont {Cornish}, \citenamefont {Ponce}, \citenamefont {Barausse},
  \citenamefont {Klein}, \citenamefont {Palenzuela},\ and\ \citenamefont
  {Lehner}}]{Sampson2014a}%
  \BibitemOpen
  \bibfield  {author} {\bibinfo {author} {\bibfnamefont {L.}~\bibnamefont
  {Sampson}}, \bibinfo {author} {\bibfnamefont {N.}~\bibnamefont {Yunes}},
  \bibinfo {author} {\bibfnamefont {N.}~\bibnamefont {Cornish}}, \bibinfo
  {author} {\bibfnamefont {M.}~\bibnamefont {Ponce}}, \bibinfo {author}
  {\bibfnamefont {E.}~\bibnamefont {Barausse}}, \bibinfo {author}
  {\bibfnamefont {A.}~\bibnamefont {Klein}}, \bibinfo {author} {\bibfnamefont
  {C.}~\bibnamefont {Palenzuela}}, \ and\ \bibinfo {author} {\bibfnamefont
  {L.}~\bibnamefont {Lehner}},\ }\href {\doibase 10.1103/PhysRevD.90.124091}
  {\bibfield  {journal} {\bibinfo  {journal} {Phys.\ Rev.\ D}\ }\textbf
  {\bibinfo {volume} {90}},\ \bibinfo {pages} {124091} (\bibinfo {year}
  {2014}{\natexlab{b}})}\BibitemShut {NoStop}%
%%CITATION = ARXIV:1407.7038;%%
\bibitem [{\citenamefont {Taniguchi}\ \emph {et~al.}(2015)\citenamefont
  {Taniguchi}, \citenamefont {Shibata},\ and\ \citenamefont
  {Buonanno}}]{Taniguchi2014}%
  \BibitemOpen
  \bibfield  {author} {\bibinfo {author} {\bibfnamefont {K.}~\bibnamefont
  {Taniguchi}}, \bibinfo {author} {\bibfnamefont {M.}~\bibnamefont {Shibata}},
  \ and\ \bibinfo {author} {\bibfnamefont {A.}~\bibnamefont {Buonanno}},\
  }\href {\doibase 10.1103/PhysRevD.91.024033} {\bibfield  {journal} {\bibinfo
  {journal} {Phys.\ Rev.\ D}\ }\textbf {\bibinfo {volume} {91}},\ \bibinfo
  {pages} {024033} (\bibinfo {year} {2015})}\BibitemShut {NoStop}%
%%CITATION = ARXIV:1410.0738;%%
\bibitem [{\citenamefont {Damour}\ and\ \citenamefont
  {Esposito-Far\`{e}se}(1992)}]{Damour1992}%
  \BibitemOpen
  \bibfield  {author} {\bibinfo {author} {\bibfnamefont {T.}~\bibnamefont
  {Damour}}\ and\ \bibinfo {author} {\bibfnamefont {G.}~\bibnamefont
  {Esposito-Far\`{e}se}},\ }\href {\doibase 10.1088/0264-9381/9/9/015}
  {\bibfield  {journal} {\bibinfo  {journal} {Class. Quant. Grav.}\ }\textbf
  {\bibinfo {volume} {9}},\ \bibinfo {pages} {2093} (\bibinfo {year}
  {1992})}\BibitemShut {NoStop}%
%%CITATION = CQGRD,9,2093;%%
\bibitem [{\citenamefont {Damour}\ and\ \citenamefont
  {Esposito-Far\`{e}se}(1993)}]{Esposito-Farese1993}%
  \BibitemOpen
  \bibfield  {author} {\bibinfo {author} {\bibfnamefont {T.}~\bibnamefont
  {Damour}}\ and\ \bibinfo {author} {\bibfnamefont {G.}~\bibnamefont
  {Esposito-Far\`{e}se}},\ }\href {\doibase 10.1103/PhysRevLett.70.2220}
  {\bibfield  {journal} {\bibinfo  {journal} {Phys. Rev. Lett.}\ }\textbf
  {\bibinfo {volume} {70}},\ \bibinfo {pages} {2220} (\bibinfo {year}
  {1993})}\BibitemShut {NoStop}%
%%CITATION = PRLTA,70,2220;%%
\bibitem [{\citenamefont {Damour}\ and\ \citenamefont
  {Nordtvedt}(1993{\natexlab{a}})}]{Damour1993}%
  \BibitemOpen
  \bibfield  {author} {\bibinfo {author} {\bibfnamefont {T.}~\bibnamefont
  {Damour}}\ and\ \bibinfo {author} {\bibfnamefont {K.}~\bibnamefont
  {Nordtvedt}},\ }\href {\doibase 10.1103/PhysRevLett.70.2217} {\bibfield
  {journal} {\bibinfo  {journal} {Phys.\ Rev.\ Lett.}\ }\textbf {\bibinfo
  {volume} {70}},\ \bibinfo {pages} {2217} (\bibinfo {year}
  {1993}{\natexlab{a}})}\BibitemShut {NoStop}%
\bibitem [{\citenamefont {Damour}\ and\ \citenamefont
  {Nordtvedt}(1993{\natexlab{b}})}]{Damour1993b}%
  \BibitemOpen
  \bibfield  {author} {\bibinfo {author} {\bibfnamefont {T.}~\bibnamefont
  {Damour}}\ and\ \bibinfo {author} {\bibfnamefont {K.}~\bibnamefont
  {Nordtvedt}},\ }\href {\doibase 10.1103/PhysRevD.48.3436} {\bibfield
  {journal} {\bibinfo  {journal} {Phys.\ Rev.\ D}\ }\textbf {\bibinfo {volume}
  {48}},\ \bibinfo {pages} {3436} (\bibinfo {year}
  {1993}{\natexlab{b}})}\BibitemShut {NoStop}%
\bibitem [{\citenamefont {Damour}\ and\ \citenamefont
  {Esposito-Far{\`{e}}se}(1996)}]{Damour1996}%
  \BibitemOpen
  \bibfield  {author} {\bibinfo {author} {\bibfnamefont {T.}~\bibnamefont
  {Damour}}\ and\ \bibinfo {author} {\bibfnamefont {G.}~\bibnamefont
  {Esposito-Far{\`{e}}se}},\ }\href {\doibase 10.1103/PhysRevD.54.1474}
  {\bibfield  {journal} {\bibinfo  {journal} {Phys.\ Rev.\ D}\ }\textbf
  {\bibinfo {volume} {54}},\ \bibinfo {pages} {1474} (\bibinfo {year}
  {1996})}\BibitemShut {NoStop}%
\bibitem [{\citenamefont {Ramazano\u{g}lu}\ and\ \citenamefont
  {Pretorius}(2016)}]{Ramazanoglu2016}%
  \BibitemOpen
  \bibfield  {author} {\bibinfo {author} {\bibfnamefont {F.~M.}\ \bibnamefont
  {Ramazano\u{g}lu}}\ and\ \bibinfo {author} {\bibfnamefont {F.}~\bibnamefont
  {Pretorius}},\ }\href@noop {} {\  (\bibinfo {year} {2016})},\ \Eprint
  {http://arxiv.org/abs/1601.07475} {arXiv:1601.07475 [gr-qc]} \BibitemShut
  {NoStop}%
%%CITATION = ARXIV:1601.07475;%%
\bibitem [{\citenamefont {DeDeo}\ and\ \citenamefont
  {Psaltis}(2003)}]{DeDeo2003}%
  \BibitemOpen
  \bibfield  {author} {\bibinfo {author} {\bibfnamefont {S.}~\bibnamefont
  {DeDeo}}\ and\ \bibinfo {author} {\bibfnamefont {D.}~\bibnamefont
  {Psaltis}},\ }\href {\doibase 10.1103/PhysRevLett.90.141101} {\bibfield
  {journal} {\bibinfo  {journal} {Phys. Rev. Lett.}\ }\textbf {\bibinfo
  {volume} {90}},\ \bibinfo {pages} {141101} (\bibinfo {year}
  {2003})}\BibitemShut {NoStop}%
%%CITATION = ASTRO-PH/0302095;%%
\bibitem [{\citenamefont {Sotani}\ and\ \citenamefont
  {Kokkotas}(2004)}]{Sotani2004}%
  \BibitemOpen
  \bibfield  {author} {\bibinfo {author} {\bibfnamefont {H.}~\bibnamefont
  {Sotani}}\ and\ \bibinfo {author} {\bibfnamefont {K.~D.}\ \bibnamefont
  {Kokkotas}},\ }\href {\doibase 10.1103/PhysRevD.70.084026} {\bibfield
  {journal} {\bibinfo  {journal} {Phys.\ Rev.\ D}\ }\textbf {\bibinfo {volume}
  {70}},\ \bibinfo {pages} {084026} (\bibinfo {year} {2004})}\BibitemShut
  {NoStop}%
%%CITATION = GR-QC/0409066;%%
\bibitem [{\citenamefont {Damour}\ and\ \citenamefont
  {Esposito-Far\`{e}se}(1998)}]{Damour1998}%
  \BibitemOpen
  \bibfield  {author} {\bibinfo {author} {\bibfnamefont {T.}~\bibnamefont
  {Damour}}\ and\ \bibinfo {author} {\bibfnamefont {G.}~\bibnamefont
  {Esposito-Far\`{e}se}},\ }\href
  {http://prd.aps.org/abstract/PRD/v58/i4/e042001} {\bibfield  {journal}
  {\bibinfo  {journal} {Phys.\ Rev.\ D}\ }\textbf {\bibinfo {volume} {58}},\
  \bibinfo {pages} {042001} (\bibinfo {year} {1998})}\BibitemShut {NoStop}%
\bibitem [{\citenamefont {DeDeo}\ and\ \citenamefont
  {Psaltis}(2004)}]{DeDeo2004}%
  \BibitemOpen
  \bibfield  {author} {\bibinfo {author} {\bibfnamefont {S.}~\bibnamefont
  {DeDeo}}\ and\ \bibinfo {author} {\bibfnamefont {D.}~\bibnamefont
  {Psaltis}},\ }\href@noop {} {\enquote {\bibinfo {title} {{Testing
  strong-field gravity with quasiperiodic oscillations}},}\ } (\bibinfo {year}
  {2004}),\ \bibinfo {note} {submitted to: Phys. Rev. D},\ \Eprint
  {http://arxiv.org/abs/astro-ph/0405067} {arXiv:astro-ph/0405067 [astro-ph]}
  \BibitemShut {NoStop}%
%%CITATION = ASTRO-PH/0405067;%%
\bibitem [{\citenamefont {Doneva}\ \emph {et~al.}(2014)\citenamefont {Doneva},
  \citenamefont {Yazadjiev}, \citenamefont {Stergioulas}, \citenamefont
  {Kokkotas},\ and\ \citenamefont {Athanasiadis}}]{Doneva2014}%
  \BibitemOpen
  \bibfield  {author} {\bibinfo {author} {\bibfnamefont {D.~D.}\ \bibnamefont
  {Doneva}}, \bibinfo {author} {\bibfnamefont {S.~S.}\ \bibnamefont
  {Yazadjiev}}, \bibinfo {author} {\bibfnamefont {N.}~\bibnamefont
  {Stergioulas}}, \bibinfo {author} {\bibfnamefont {K.~D.}\ \bibnamefont
  {Kokkotas}}, \ and\ \bibinfo {author} {\bibfnamefont {T.~M.}\ \bibnamefont
  {Athanasiadis}},\ }\href {\doibase 10.1103/PhysRevD.90.044004} {\bibfield
  {journal} {\bibinfo  {journal} {Phys.\ Rev.\ D}\ }\textbf {\bibinfo {volume}
  {90}},\ \bibinfo {pages} {044004} (\bibinfo {year} {2014})}\BibitemShut
  {NoStop}%
\bibitem [{\citenamefont {Freire}\ \emph {et~al.}(2012)\citenamefont {Freire},
  \citenamefont {Wex}, \citenamefont {Esposito-Far{\`{e}}se}, \citenamefont
  {Verbiest}, \citenamefont {Bailes}, \citenamefont {Jacoby}, \citenamefont
  {Kramer}, \citenamefont {Stairs}, \citenamefont {Antoniadis},\ and\
  \citenamefont {Janssen}}]{Freire2012}%
  \BibitemOpen
  \bibfield  {author} {\bibinfo {author} {\bibfnamefont {P.~C.~C.}\
  \bibnamefont {Freire}}, \bibinfo {author} {\bibfnamefont {N.}~\bibnamefont
  {Wex}}, \bibinfo {author} {\bibfnamefont {G.}~\bibnamefont
  {Esposito-Far{\`{e}}se}}, \bibinfo {author} {\bibfnamefont {J.~P.~W.}\
  \bibnamefont {Verbiest}}, \bibinfo {author} {\bibfnamefont {M.}~\bibnamefont
  {Bailes}}, \bibinfo {author} {\bibfnamefont {B.~A.}\ \bibnamefont {Jacoby}},
  \bibinfo {author} {\bibfnamefont {M.}~\bibnamefont {Kramer}}, \bibinfo
  {author} {\bibfnamefont {I.~H.}\ \bibnamefont {Stairs}}, \bibinfo {author}
  {\bibfnamefont {J.}~\bibnamefont {Antoniadis}}, \ and\ \bibinfo {author}
  {\bibfnamefont {G.~H.}\ \bibnamefont {Janssen}},\ }\href {\doibase
  10.1111/j.1365-2966.2012.21253.x} {\bibfield  {journal} {\bibinfo  {journal}
  {Mon.\ Not.\ Roy.\ Astr.\ Soc.}\ }\textbf {\bibinfo {volume} {423}},\
  \bibinfo {pages} {3328} (\bibinfo {year} {2012})}\BibitemShut {NoStop}%
\bibitem [{\citenamefont {Just}(1959)}]{Just}%
  \BibitemOpen
  \bibfield  {author} {\bibinfo {author} {\bibfnamefont {K.}~\bibnamefont
  {Just}},\ }\href@noop {} {\bibfield  {journal} {\bibinfo  {journal} {Z.
  Phys.}\ }\textbf {\bibinfo {volume} {157}} (\bibinfo {year}
  {1959})}\BibitemShut {NoStop}%
\bibitem [{\citenamefont {Read}\ \emph {et~al.}(2009)\citenamefont {Read},
  \citenamefont {Lackey}, \citenamefont {Owen},\ and\ \citenamefont
  {Friedman}}]{Read2009}%
  \BibitemOpen
  \bibfield  {author} {\bibinfo {author} {\bibfnamefont {J.~S.}\ \bibnamefont
  {Read}}, \bibinfo {author} {\bibfnamefont {B.~D.}\ \bibnamefont {Lackey}},
  \bibinfo {author} {\bibfnamefont {B.~J.}\ \bibnamefont {Owen}}, \ and\
  \bibinfo {author} {\bibfnamefont {J.~L.}\ \bibnamefont {Friedman}},\ }\href
  {\doibase 10.1103/PhysRevD.79.124032} {\bibfield  {journal} {\bibinfo
  {journal} {Phys.\ Rev.\ D}\ }\textbf {\bibinfo {volume} {79}},\ \bibinfo
  {pages} {124032} (\bibinfo {year} {2009})}\BibitemShut {NoStop}%
\bibitem [{\citenamefont {Akmal}\ \emph {et~al.}(1998)\citenamefont {Akmal},
  \citenamefont {Pandharipande},\ and\ \citenamefont {Ravenhall}}]{Akmal1998}%
  \BibitemOpen
  \bibfield  {author} {\bibinfo {author} {\bibfnamefont {A.}~\bibnamefont
  {Akmal}}, \bibinfo {author} {\bibfnamefont {V.~R.}\ \bibnamefont
  {Pandharipande}}, \ and\ \bibinfo {author} {\bibfnamefont {D.~G.}\
  \bibnamefont {Ravenhall}},\ }\href {\doibase 10.1103/PhysRevC.58.1804}
  {\bibfield  {journal} {\bibinfo  {journal} {Phys.\ Rev.\ C}\ }\textbf
  {\bibinfo {volume} {58}},\ \bibinfo {pages} {1804} (\bibinfo {year}
  {1998})}\BibitemShut {NoStop}%
\bibitem [{\citenamefont {Hawking}(1972)}]{Hawking1972}%
  \BibitemOpen
  \bibfield  {author} {\bibinfo {author} {\bibfnamefont {S.~W.}\ \bibnamefont
  {Hawking}},\ }\href {\doibase 10.1007/BF01877518} {\bibfield  {journal}
  {\bibinfo  {journal} {Commun. Math. Phys.}\ }\textbf {\bibinfo {volume}
  {25}},\ \bibinfo {pages} {167} (\bibinfo {year} {1972})}\BibitemShut
  {NoStop}%
%%CITATION = CMPHA,25,167;%%
\bibitem [{\citenamefont {Damour}\ and\ \citenamefont
  {Esposito-Far\`{e}se}(1996)}]{Damour1996b}%
  \BibitemOpen
  \bibfield  {author} {\bibinfo {author} {\bibfnamefont {T.}~\bibnamefont
  {Damour}}\ and\ \bibinfo {author} {\bibfnamefont {G.}~\bibnamefont
  {Esposito-Far\`{e}se}},\ }\href {\doibase 10.1103/PhysRevD.53.5541}
  {\bibfield  {journal} {\bibinfo  {journal} {Phys.\ Rev.\ D}\ }\textbf
  {\bibinfo {volume} {53}},\ \bibinfo {pages} {5541} (\bibinfo {year}
  {1996})}\BibitemShut {NoStop}%
%%CITATION = GR-QC/9506063;%%
\bibitem [{\citenamefont {Damour}\ and\ \citenamefont
  {Taylor}(1992)}]{Damour1992a}%
  \BibitemOpen
  \bibfield  {author} {\bibinfo {author} {\bibfnamefont {T.}~\bibnamefont
  {Damour}}\ and\ \bibinfo {author} {\bibfnamefont {J.~H.}\ \bibnamefont
  {Taylor}},\ }\href {http://link.aps.org/doi/10.1103/PhysRevD.45.1840}
  {\bibfield  {journal} {\bibinfo  {journal} {Phys.\ Rev.\ D}\ }\textbf
  {\bibinfo {volume} {45}},\ \bibinfo {pages} {1840} (\bibinfo {year}
  {1992})}\BibitemShut {NoStop}%
\bibitem [{\citenamefont {Lang}(2015)}]{Lang2014a}%
  \BibitemOpen
  \bibfield  {author} {\bibinfo {author} {\bibfnamefont {R.~N.}\ \bibnamefont
  {Lang}},\ }\href {\doibase 10.1103/PhysRevD.91.084027} {\bibfield  {journal}
  {\bibinfo  {journal} {Phys.\ Rev.\ D}\ }\textbf {\bibinfo {volume} {91}},\
  \bibinfo {pages} {084027} (\bibinfo {year} {2015})}\BibitemShut {NoStop}%
%%CITATION = ARXIV:1411.3073;%%
\bibitem [{\citenamefont {Mirshekari}\ and\ \citenamefont
  {Will}(2013)}]{Mirshekari2013}%
  \BibitemOpen
  \bibfield  {author} {\bibinfo {author} {\bibfnamefont {S.}~\bibnamefont
  {Mirshekari}}\ and\ \bibinfo {author} {\bibfnamefont {C.~M.}\ \bibnamefont
  {Will}},\ }\href {\doibase 10.1103/PhysRevD.87.084070} {\bibfield  {journal}
  {\bibinfo  {journal} {Phys.\ Rev.\ D}\ }\textbf {\bibinfo {volume} {87}},\
  \bibinfo {pages} {084070} (\bibinfo {year} {2013})}\BibitemShut {NoStop}%
%%CITATION = ARXIV:1301.4680;%%
\bibitem [{\citenamefont {Esposito-Far\`{e}se}(2011)}]{Esposito-Farese2011}%
  \BibitemOpen
  \bibfield  {author} {\bibinfo {author} {\bibfnamefont {G.}~\bibnamefont
  {Esposito-Far\`{e}se}},\ }\bibfield  {booktitle} {\emph {\bibinfo {booktitle}
  {{Mass and Motion in General Relativity}}},\ }\href {\doibase
  10.1007/978-90-481-3015-3_17} {\bibfield  {journal} {\bibinfo  {journal}
  {Fundam. Theor. Phys.}\ }\textbf {\bibinfo {volume} {162}},\ \bibinfo {pages}
  {461} (\bibinfo {year} {2011})}\BibitemShut {NoStop}%
%%CITATION = ARXIV:0905.2575;%%
\bibitem [{\citenamefont {Eardley}(1975)}]{Eardley1975}%
  \BibitemOpen
  \bibfield  {author} {\bibinfo {author} {\bibfnamefont {D.~M.}\ \bibnamefont
  {Eardley}},\ }\href {\doibase 10.1086/181744} {\bibfield  {journal} {\bibinfo
   {journal} {Astrophys.\ J.\ Lett.}\ }\textbf {\bibinfo {volume} {196}},\
  \bibinfo {pages} {L59} (\bibinfo {year} {1975})}\BibitemShut {NoStop}%
\bibitem [{\citenamefont {Scherrer}(1941)}]{Scherrer}%
  \BibitemOpen
  \bibfield  {author} {\bibinfo {author} {\bibfnamefont {W.}~\bibnamefont
  {Scherrer}},\ }\href@noop {} {\bibfield  {journal} {\bibinfo  {journal}
  {Verh. Schweiz. Naturf. Ges.}\ }\textbf {\bibinfo {volume} {121}},\ \bibinfo
  {pages} {86} (\bibinfo {year} {1941})}\BibitemShut {NoStop}%
\bibitem [{\citenamefont {Thiry}(1951)}]{Thiry}%
  \BibitemOpen
  \bibfield  {author} {\bibinfo {author} {\bibfnamefont {Y.}~\bibnamefont
  {Thiry}},\ }\href@noop {} {\bibfield  {journal} {\bibinfo  {journal} {J.
  Math. Pures et Appl.}\ }\textbf {\bibinfo {volume} {30}},\ \bibinfo {pages}
  {275} (\bibinfo {year} {1951})}\BibitemShut {NoStop}%
\bibitem [{\citenamefont {Jordan}(1955)}]{Jordan}%
  \BibitemOpen
  \bibfield  {author} {\bibinfo {author} {\bibfnamefont {P.}~\bibnamefont
  {Jordan}},\ }\href@noop {} {\emph {\bibinfo {title} {Schwerkraft und
  Weltall}}}\ (\bibinfo  {publisher} {F. Vieweg},\ \bibinfo {address}
  {Braunschweig},\ \bibinfo {year} {1955})\BibitemShut {NoStop}%
\bibitem [{\citenamefont {Fierz}(1956)}]{Fierz}%
  \BibitemOpen
  \bibfield  {author} {\bibinfo {author} {\bibfnamefont {M.}~\bibnamefont
  {Fierz}},\ }\href@noop {} {\bibfield  {journal} {\bibinfo  {journal} {Helv.
  Phys. Acta}\ }\textbf {\bibinfo {volume} {29}},\ \bibinfo {pages} {128}
  (\bibinfo {year} {1956})}\BibitemShut {NoStop}%
%%CITATION = HPACA,29,128;%%
\bibitem [{\citenamefont {Brans}\ and\ \citenamefont {Dicke}(1961)}]{Brans}%
  \BibitemOpen
  \bibfield  {author} {\bibinfo {author} {\bibfnamefont {C.}~\bibnamefont
  {Brans}}\ and\ \bibinfo {author} {\bibfnamefont {R.~H.}\ \bibnamefont
  {Dicke}},\ }\href {\doibase 10.1103/PhysRev.124.925} {\bibfield  {journal}
  {\bibinfo  {journal} {Phys.\ Rev.}\ }\textbf {\bibinfo {volume} {124}},\
  \bibinfo {pages} {925} (\bibinfo {year} {1961})}\BibitemShut {NoStop}%
%%CITATION = PHRVA,124,925;%%
\bibitem [{\citenamefont {Epstein}\ and\ \citenamefont
  {Wagoner}(1975)}]{Epstein1975}%
  \BibitemOpen
  \bibfield  {author} {\bibinfo {author} {\bibfnamefont {R.}~\bibnamefont
  {Epstein}}\ and\ \bibinfo {author} {\bibfnamefont {R.~V.}\ \bibnamefont
  {Wagoner}},\ }\href {\doibase 10.1086/153561} {\bibfield  {journal} {\bibinfo
   {journal} {Astrophys.\ J.}\ }\textbf {\bibinfo {volume} {197}},\ \bibinfo
  {pages} {717} (\bibinfo {year} {1975})}\BibitemShut {NoStop}%
\bibitem [{\citenamefont {Wagoner}\ and\ \citenamefont
  {Will}(1976)}]{Wagoner1976}%
  \BibitemOpen
  \bibfield  {author} {\bibinfo {author} {\bibfnamefont {R.~V.}\ \bibnamefont
  {Wagoner}}\ and\ \bibinfo {author} {\bibfnamefont {C.~M.}\ \bibnamefont
  {Will}},\ }\href {\doibase 10.1086/154886} {\bibfield  {journal} {\bibinfo
  {journal} {Astrophys.\ J.}\ }\textbf {\bibinfo {volume} {210}},\ \bibinfo
  {pages} {764} (\bibinfo {year} {1976})}\BibitemShut {NoStop}%
\bibitem [{\citenamefont {Will}\ and\ \citenamefont
  {Wiseman}(1996)}]{Will1996}%
  \BibitemOpen
  \bibfield  {author} {\bibinfo {author} {\bibfnamefont {C.~M.}\ \bibnamefont
  {Will}}\ and\ \bibinfo {author} {\bibfnamefont {A.~G.}\ \bibnamefont
  {Wiseman}},\ }\href {\doibase 10.1103/PhysRevD.54.4813} {\bibfield  {journal}
  {\bibinfo  {journal} {Phys.\ Rev.\ D}\ }\textbf {\bibinfo {volume} {54}},\
  \bibinfo {pages} {4813} (\bibinfo {year} {1996})}\BibitemShut {NoStop}%
%%CITATION = GR-QC/9608012;%%
\bibitem [{\citenamefont {Pati}\ and\ \citenamefont {Will}(2000)}]{Pati2000}%
  \BibitemOpen
  \bibfield  {author} {\bibinfo {author} {\bibfnamefont {M.~E.}\ \bibnamefont
  {Pati}}\ and\ \bibinfo {author} {\bibfnamefont {C.~M.}\ \bibnamefont
  {Will}},\ }\href {\doibase 10.1103/PhysRevD.62.124015} {\bibfield  {journal}
  {\bibinfo  {journal} {Phys.\ Rev.\ D}\ }\textbf {\bibinfo {volume} {62}},\
  \bibinfo {pages} {124015} (\bibinfo {year} {2000})}\BibitemShut {NoStop}%
%%CITATION = GR-QC/0007087;%%
\bibitem [{\citenamefont {Pati}\ and\ \citenamefont {Will}(2002)}]{Pati2002}%
  \BibitemOpen
  \bibfield  {author} {\bibinfo {author} {\bibfnamefont {M.~E.}\ \bibnamefont
  {Pati}}\ and\ \bibinfo {author} {\bibfnamefont {C.~M.}\ \bibnamefont
  {Will}},\ }\href {\doibase 10.1103/PhysRevD.65.104008} {\bibfield  {journal}
  {\bibinfo  {journal} {Phys.\ Rev.\ D}\ }\textbf {\bibinfo {volume} {65}},\
  \bibinfo {pages} {104008} (\bibinfo {year} {2002})}\BibitemShut {NoStop}%
\bibitem [{\citenamefont {Lang}(2014)}]{Lang2014b}%
  \BibitemOpen
  \bibfield  {author} {\bibinfo {author} {\bibfnamefont {R.~N.}\ \bibnamefont
  {Lang}},\ }\href {\doibase 10.1103/PhysRevD.89.084014} {\bibfield  {journal}
  {\bibinfo  {journal} {Phys.\ Rev.\ D}\ }\textbf {\bibinfo {volume} {89}},\
  \bibinfo {pages} {084014} (\bibinfo {year} {2014})}\BibitemShut {NoStop}%
%%CITATION = ARXIV:1310.3320;%%
\bibitem [{\citenamefont {Marsat}\ \emph {et~al.}(2016)\citenamefont {Marsat},
  \citenamefont {Sennett},\ and\ \citenamefont {Buonanno}}]{Marsat2016}%
  \BibitemOpen
  \bibfield  {author} {\bibinfo {author} {\bibfnamefont {S.}~\bibnamefont
  {Marsat}}, \bibinfo {author} {\bibfnamefont {N.}~\bibnamefont {Sennett}}, \
  and\ \bibinfo {author} {\bibfnamefont {A.}~\bibnamefont {Buonanno}},\
  }\href@noop {} {} (\bibinfo {year} {2016}),\ \bibinfo {note} {in
  preparation}\BibitemShut {NoStop}%
\bibitem [{\citenamefont {Buonanno}\ \emph {et~al.}(2003)\citenamefont
  {Buonanno}, \citenamefont {Chen},\ and\ \citenamefont
  {Vallisneri}}]{Buonanno2003}%
  \BibitemOpen
  \bibfield  {author} {\bibinfo {author} {\bibfnamefont {A.}~\bibnamefont
  {Buonanno}}, \bibinfo {author} {\bibfnamefont {Y.}~\bibnamefont {Chen}}, \
  and\ \bibinfo {author} {\bibfnamefont {M.}~\bibnamefont {Vallisneri}},\
  }\href {\doibase 10.1103/PhysRevD.67.024016} {\bibfield  {journal} {\bibinfo
  {journal} {Phys.\ Rev.\ D}\ }\textbf {\bibinfo {volume} {67}},\ \bibinfo
  {pages} {024016} (\bibinfo {year} {2003})}\BibitemShut {NoStop}%
\bibitem [{\citenamefont {Sch{\"{a}}fer}(1985)}]{Schafer1985}%
  \BibitemOpen
  \bibfield  {author} {\bibinfo {author} {\bibfnamefont {G.}~\bibnamefont
  {Sch{\"{a}}fer}},\ }\href {\doibase 10.1016/0003-4916(85)90337-9} {\bibfield
  {journal} {\bibinfo  {journal} {Annals of Physics}\ }\textbf {\bibinfo
  {volume} {161}},\ \bibinfo {pages} {81} (\bibinfo {year} {1985})}\BibitemShut
  {NoStop}%
\bibitem [{\citenamefont {Hinderer}\ \emph {et~al.}(2016)\citenamefont
  {Hinderer}, \citenamefont {Taracchini}, \citenamefont {Foucart},
  \citenamefont {Buonanno}, \citenamefont {Duez}, \citenamefont {Kidder},
  \citenamefont {Pfeiffer}, \citenamefont {Scheel}, \citenamefont {Hotokezaka},
  \citenamefont {Kyutoku}, \citenamefont {Shibata},\ and\ \citenamefont
  {Carpenter}}]{Hinderer2016}%
  \BibitemOpen
  \bibfield  {author} {\bibinfo {author} {\bibfnamefont {T.}~\bibnamefont
  {Hinderer}}, \bibinfo {author} {\bibfnamefont {A.}~\bibnamefont
  {Taracchini}}, \bibinfo {author} {\bibfnamefont {F.}~\bibnamefont {Foucart}},
  \bibinfo {author} {\bibfnamefont {A.}~\bibnamefont {Buonanno}}, \bibinfo
  {author} {\bibfnamefont {M.}~\bibnamefont {Duez}}, \bibinfo {author}
  {\bibfnamefont {L.~E.}\ \bibnamefont {Kidder}}, \bibinfo {author}
  {\bibfnamefont {H.~P.}\ \bibnamefont {Pfeiffer}}, \bibinfo {author}
  {\bibfnamefont {M.~A.}\ \bibnamefont {Scheel}}, \bibinfo {author}
  {\bibfnamefont {K.}~\bibnamefont {Hotokezaka}}, \bibinfo {author}
  {\bibfnamefont {K.}~\bibnamefont {Kyutoku}}, \bibinfo {author} {\bibfnamefont
  {M.}~\bibnamefont {Shibata}}, \ and\ \bibinfo {author} {\bibfnamefont
  {C.~W.}\ \bibnamefont {Carpenter}},\ }\href@noop {} {\  (\bibinfo {year}
  {2016})},\ \Eprint {http://arxiv.org/abs/1602.00599} {arXiv:1602.00599
  [gr-qc]} \BibitemShut {NoStop}%
\bibitem [{\citenamefont {Steinhoff}()}]{Steinhoff}%
  \BibitemOpen
  \bibfield  {author} {\bibinfo {author} {\bibfnamefont {J.}~\bibnamefont
  {Steinhoff}},\ }\href@noop {} {}\bibinfo {howpublished} {private
  communication}\BibitemShut {NoStop}%
\end{thebibliography}%
\end{document}